\documentclass[11pt]{article}
\usepackage{amsmath,amsfonts,amssymb,graphicx,epsfig,pdflscape,multirow}
\usepackage{arydshln,cite,enumitem}

\usepackage{amsthm}

\numberwithin{equation}{section}
\graphicspath{{./eps/}}
\usepackage{cite}
\usepackage[usenames]{color}
\usepackage{pstricks}
\usepackage{caption}
\usepackage[backref=false]{hyperref}
\usepackage[capitalise,noabbrev]{cleveref} 
\hypersetup{
colorlinks=true,
citecolor=red,
linkcolor=darkblue,
urlcolor=darkblue
}

\long\def\ignore#1{}
\definecolor{darkblue}{rgb}{0,0,.8}
\definecolor{red}{rgb}{1,0,0}
\definecolor{purple}{rgb}{1,0.4,1}
\definecolor{coloroflink}{rgb}{0.7,0,1}
\definecolor{pink}{rgb}{1,.7,.7}
\definecolor{lightblue}{rgb}{.61,.61,1}
\definecolor{midblue}{rgb}{.7,.7,1}
\definecolor{lightlightblue}{rgb}{.9,.9,1}
\definecolor{lightestblue}{rgb}{.96,.96,1}
\definecolor{lightpurple}{rgb}{1,.65,1}
\definecolor{darkgreen}{rgb}{0.180392, 0.545098, 0.341176}

\newtheoremstyle{smallcaps}{5pt}{5pt}{\itshape}{}{}{}{.5em}
{\scshape\thmname{#1}~\thmnumber{#2}.\thmnote{~\textnormal{(#3)}}}
\theoremstyle{smallcaps}

\numberwithin{equation}{section}

\newcommand{\nc}{\newcommand}
\nc{\bib}{\bibitem}
\nc{\be}{\begin{equation}}
\nc{\ee}{\end{equation}}
\nc{\Mod}{\textrm{\,mod\,}}
\nc{\Tt}{\textrm{t}}	

\nc{\ir}{\mathrm{i}}
\nc{\eE}{\mathsf{e}} 
\nc{\dd}{\mathrm{d}}   

\nc{\notelegant}{1.5pt}
\nc{\elegant}{1.25pt}
\nc{\moyen}{1.0pt}
\nc{\mince}{0.5pt}

\nc{\Tb}{\mbox{\boldmath $T$}}
\nc{\tl}{\mbox{$\mathsf {TL}$}}
\nc{\eptl}{\mathsf{\mathcal EPTL}}
\nc{\chit}{\protect\raisebox{0.25ex}{$\chi$}}

\def\facegrid#1#2{
\psframe[fillstyle=solid,fillcolor=lightlightblue,linewidth=0pt]#1#2
\psgrid[gridlabels=0pt,subgriddiv=1]#1#2}

\def\loopa{
\psframe[linewidth=.25pt](0,0)(1,1)
\psarc[linewidth=1.25pt,linecolor=blue](1,0){.5}{90}{180}
\psarc[linewidth=1.25pt,linecolor=blue](0,1){.5}{-90}{0}
}
\def\loopb{
\psframe[linewidth=.25pt](0,0)(1,1)
\psarc[linewidth=1.25pt,linecolor=blue](0,0){.5}{0}{90}
\psarc[linewidth=1.25pt,linecolor=blue](1,1){.5}{180}{270}
}

\def\twodefdown{
\psset{unit=0.4cm}
\begin{pspicture}[shift=-0.09cm](0,-0.5)(0.8,0)
\psline[linewidth=0.5pt]{-}(0,0)(0.8,0)
\psline[linecolor=blue,linewidth=\moyen]{-}(0.2,0)(0.2,-0.5)
\psline[linecolor=blue,linewidth=\moyen]{-}(0.6,0)(0.6,-0.5)
\end{pspicture}
}

\def\fourdefdown{
\psset{unit=0.4cm}
\begin{pspicture}[shift=-0.09cm](0,-0.5)(1.6,0)
\psline[linewidth=0.5pt]{-}(0,0)(1.6,0)
\psline[linecolor=blue,linewidth=\moyen]{-}(0.2,0)(0.2,-0.5)
\psline[linecolor=blue,linewidth=\moyen]{-}(0.6,0)(0.6,-0.5)
\psline[linecolor=blue,linewidth=\moyen]{-}(1.0,0)(1.0,-0.5)
\psline[linecolor=blue,linewidth=\moyen]{-}(1.4,0)(1.4,-0.5)
\end{pspicture}
}

\def\arcupin{
\psset{unit=0.4cm}
\begin{pspicture}[shift=0.015cm](0,0)(0.8,0.4)
\psline[linewidth=0.5pt]{-}(0,0)(0.8,0)
\psarc[linecolor=blue,linewidth=\moyen]{-}(0.4,0){0.2}{0}{180}
\end{pspicture}
}

\def\arcupout{
\psset{unit=0.4cm}
\begin{pspicture}[shift=0.015cm](0,0)(0.8,0.4)
\psline[linewidth=0.5pt]{-}(0,0)(0.8,0)
\psarc[linecolor=blue,linewidth=\moyen]{-}(0,0){0.2}{0}{90}
\psarc[linecolor=blue,linewidth=\moyen]{-}(0.8,0){0.2}{90}{180}
\end{pspicture}
}

\def\arcsupone{
\psset{unit=0.4cm}
\begin{pspicture}[shift=0.015cm](0,0)(1.6,0.4)
\psline[linewidth=0.5pt]{-}(0,0)(1.6,0)
\psarc[linecolor=blue,linewidth=\moyen]{-}(0.4,0){0.2}{0}{180}
\psarc[linecolor=blue,linewidth=\moyen]{-}(1.2,0){0.2}{0}{180}
\end{pspicture}
}

\def\arcsuptwo{
\psset{unit=0.4cm}
\begin{pspicture}[shift=0.015cm](0,0)(1.6,0.4)
\psline[linewidth=0.5pt]{-}(0,0)(1.6,0)
\psarc[linecolor=blue,linewidth=\moyen]{-}(0.8,0){0.2}{0}{180}
\psbezier[linecolor=blue,linewidth=\moyen]{-}(0.2,0)(0.2,0.6)(1.4,0.6)(1.4,0)
\end{pspicture}
}

\def\arcsupthree{
\psset{unit=0.4cm}
\begin{pspicture}[shift=0.015cm](-0.02,0)(1.62,0.4)
\psline[linewidth=0.5pt]{-}(0,0)(1.6,0)
\psbezier[linecolor=blue,linewidth=\moyen]{-}(0.6,0)(0.6,0.65)(1.5,0.6)(1.62,0.33)
\psbezier[linecolor=blue,linewidth=\moyen]{-}(-0.02,0.40)(-0.0,0.37)(0.2,0.3)(0.2,0)
\psarc[linecolor=blue,linewidth=\moyen]{-}(1.2,0){0.2}{0}{180}
\psframe[fillstyle=solid,linecolor=white,linewidth=0pt](1.6,-0.02)(1.66,0.5)
\psframe[fillstyle=solid,linecolor=white,linewidth=0pt](0.0,-0.02)(-0.06,0.5)
\end{pspicture}
}

\def\arcsupfour{
\psset{unit=0.4cm}
\begin{pspicture}[shift=0.015cm](0,0)(1.6,0.4)
\psline[linewidth=0.5pt]{-}(0,0)(1.6,0)
\psarc[linecolor=blue,linewidth=\moyen]{-}(0.8,0){0.2}{0}{180}
\psarc[linecolor=blue,linewidth=\moyen]{-}(0,0){0.2}{0}{90}
\psarc[linecolor=blue,linewidth=\moyen]{-}(1.6,0){0.2}{90}{180}
\end{pspicture}
}

\def\arcsupfive{
\psset{unit=0.4cm}
\begin{pspicture}[shift=0.015cm](-0.02,0)(1.62,0.4)
\psline[linewidth=0.5pt]{-}(0,0)(1.6,0)
\psbezier[linecolor=blue,linewidth=\moyen]{-}(1,0)(1,0.65)(0.1,0.6)(-0.02,0.33)
\psbezier[linecolor=blue,linewidth=\moyen]{-}(1.62,0.40)(1.6,0.37)(1.4,0.3)(1.4,0)
\psarc[linecolor=blue,linewidth=\moyen]{-}(0.4,0){0.2}{0}{180}
\psframe[fillstyle=solid,linecolor=white,linewidth=0pt](1.6,-0.02)(1.66,0.5)
\psframe[fillstyle=solid,linecolor=white,linewidth=0pt](0.0,-0.02)(-0.06,0.5)
\end{pspicture}
}

\def\arcsupsix{
\psset{unit=0.4cm}
\begin{pspicture}[shift=0.015cm](0,0)(1.6,0.4)
\psline[linewidth=0.5pt]{-}(0,0)(1.6,0)
\psarc[linecolor=blue,linewidth=\moyen]{-}(0,0){0.2}{0}{90}
\psarc[linecolor=blue,linewidth=\moyen]{-}(1.6,0){0.2}{90}{180}
\psbezier[linecolor=blue,linewidth=\moyen]{-}(0.6,0)(0.6,0.5)(0,0.55)(-0.02,0.55)
\psbezier[linecolor=blue,linewidth=\moyen]{-}(1,0)(1,0.5)(1.6,0.55)(1.62,0.55)
\psframe[fillstyle=solid,linecolor=white,linewidth=0pt](1.6,-0.02)(1.66,0.6)
\psframe[fillstyle=solid,linecolor=white,linewidth=0pt](0.0,-0.02)(-0.06,0.6)
\end{pspicture}
}

\def\arcsdownone{
\psset{unit=0.4cm}
\begin{pspicture}[shift=-0.09cm](0,-0.5)(1.6,0)
\psline[linewidth=0.5pt]{-}(0,0)(1.6,0)
\psarc[linecolor=blue,linewidth=\moyen]{-}(0.4,0){0.2}{180}{0}
\psarc[linecolor=blue,linewidth=\moyen]{-}(1.2,0){0.2}{180}{0}
\end{pspicture}
}

\def\arcsdowntwo{
\psset{unit=0.4cm}
\begin{pspicture}[shift=-0.09cm](0,-0.5)(1.6,0)
\psline[linewidth=0.5pt]{-}(0,0)(1.6,0)
\psarc[linecolor=blue,linewidth=\moyen]{-}(0.8,0){0.2}{180}{0}
\psbezier[linecolor=blue,linewidth=\moyen]{-}(0.2,0)(0.2,-0.6)(1.4,-0.6)(1.4,0)
\end{pspicture}
}

\def\arcsdownthree{
\psset{unit=0.4cm}
\begin{pspicture}[shift=-0.09cm](0,-0.5)(1.6,0)
\psline[linewidth=0.5pt]{-}(0,0)(1.6,0)
\psbezier[linecolor=blue,linewidth=\moyen]{-}(0.6,0)(0.6,-0.65)(1.5,-0.6)(1.62,-0.33)
\psbezier[linecolor=blue,linewidth=\moyen]{-}(-0.02,-0.40)(-0.0,-0.37)(0.2,-0.3)(0.2,0)
\psarc[linecolor=blue,linewidth=\moyen]{-}(1.2,0){0.2}{180}{0}
\psframe[fillstyle=solid,linecolor=white,linewidth=0pt](1.6,-0.6)(1.66,0.02)
\psframe[fillstyle=solid,linecolor=white,linewidth=0pt](0.0,-0.5)(-0.06,0.02)
\end{pspicture}
}

\def\arcsdownfour{
\psset{unit=0.4cm}
\begin{pspicture}[shift=-0.09cm](0,-0.5)(1.6,0)
\psline[linewidth=0.5pt]{-}(0,0)(1.6,0)
\psarc[linecolor=blue,linewidth=\moyen]{-}(0.8,0){0.2}{180}{0}
\psarc[linecolor=blue,linewidth=\moyen]{-}(0,0){0.2}{-90}{0}
\psarc[linecolor=blue,linewidth=\moyen]{-}(1.6,0){0.2}{180}{-90}
\end{pspicture}
}

\def\arcsdownfive{
\psset{unit=0.4cm}
\begin{pspicture}[shift=-0.09cm](0,-0.5)(1.6,0)
\psline[linewidth=0.5pt]{-}(0,0)(1.6,0)
\psbezier[linecolor=blue,linewidth=\moyen]{-}(1,0)(1,-0.65)(0.1,-0.6)(-0.02,-0.33)
\psbezier[linecolor=blue,linewidth=\moyen]{-}(1.62,-0.40)(1.6,-0.37)(1.4,-0.3)(1.4,0)
\psarc[linecolor=blue,linewidth=\moyen]{-}(0.4,0){0.2}{180}{0}
\psframe[fillstyle=solid,linecolor=white,linewidth=0pt](1.6,-0.5)(1.66,0.02)
\psframe[fillstyle=solid,linecolor=white,linewidth=0pt](0.0,-0.5)(-0.06,0.02)
\end{pspicture}
}

\def\arcsdownsix{
\psset{unit=0.4cm}
\begin{pspicture}[shift=-0.09cm](0,-0.5)(1.6,0)
\psline[linewidth=0.5pt]{-}(0,0)(1.6,0)
\psarc[linecolor=blue,linewidth=\moyen]{-}(0,0){0.2}{-90}{0}
\psarc[linecolor=blue,linewidth=\moyen]{-}(1.6,0){0.2}{180}{-90}
\psbezier[linecolor=blue,linewidth=\moyen]{-}(0.6,0)(0.6,-0.5)(0,-0.55)(-0.02,-0.55)
\psbezier[linecolor=blue,linewidth=\moyen]{-}(1,0)(1,-0.5)(1.6,-0.55)(1.62,-0.55)
\psframe[fillstyle=solid,linecolor=white,linewidth=0pt](1.6,-0.6)(1.66,0.02)
\psframe[fillstyle=solid,linecolor=white,linewidth=0pt](0.0,-0.6)(-0.06,0.02)
\end{pspicture}
}

\def\arcdownout{
\psset{unit=0.4cm}
\begin{pspicture}[shift=-0.11cm](0,-0.5)(0.8,0)
\psline[linewidth=0.5pt]{-}(0,0)(0.8,0)
\psarc[linecolor=blue,linewidth=\moyen]{-}(0,0){0.2}{-90}{0}
\psarc[linecolor=blue,linewidth=\moyen]{-}(0.8,0){0.2}{180}{270}
\end{pspicture}
}

\def\arcdownin{
\psset{unit=0.4cm}
\begin{pspicture}[shift=-0.11cm](0,-0.5)(0.8,0)
\psline[linewidth=0.5pt]{-}(0,0)(0.8,0)
\psarc[linecolor=blue,linewidth=\moyen]{-}(0.4,0){0.2}{180}{0}
\end{pspicture}
}

\def\halfarcright{
\psset{unit=0.5cm}
\begin{pspicture}[shift=-0.04cm](0,-0.2)(0.4,0)
\psline[linewidth=0.5pt]{-}(0,0)(0.4,0)
\psarc[linecolor=blue,linewidth=\moyen]{-}(0.4,0){0.2}{180}{270}
\end{pspicture}
}

\def\halfarcleft{
\psset{unit=0.5cm}
\begin{pspicture}[shift=-0.04cm](0,-0.2)(0.4,0)
\psline[linewidth=0.5pt]{-}(0,0)(0.4,0)
\psarc[linecolor=blue,linewidth=\moyen]{-}(0,0){0.2}{-90}{0}
\end{pspicture}
}

\def\twohalfarcright{
\psset{unit=0.5cm}
\begin{pspicture}[shift=-0.14cm](0,-0.4)(0.8,0)
\psline[linewidth=0.5pt]{-}(0,0)(0.8,0)
\psarc[linecolor=blue,linewidth=\moyen]{-}(0.8,0){0.2}{180}{270}
\psbezier[linecolor=blue,linewidth=\moyen]{-}(0.2,0)(0.2,-0.3)(0.7,-0.4)(0.8,-0.4)
\end{pspicture}
}

\def\twohalfarcleft{
\psset{unit=0.5cm}
\begin{pspicture}[shift=-0.14cm](0,-0.4)(0.8,0)
\psline[linewidth=0.5pt]{-}(0,0)(0.8,0)
\psarc[linecolor=blue,linewidth=\moyen]{-}(0,0){0.2}{270}{0}
\psbezier[linecolor=blue,linewidth=\moyen]{-}(0.6,0)(0.6,-0.3)(0.1,-0.4)(0,-0.4)
\end{pspicture}
}

\def \||{|}

\renewcommand{\le}{\leqslant}

\begin{document}

\topmargin -15mm
\oddsidemargin 05mm

%
%

\title{\mbox{}\vspace{-.2in}
\bf 
\huge Logarithmic correlation functions for \\[0.5cm] critical dense polymers on the cylinder
}

\date{}
\maketitle 
\vspace{-1.5cm}

\begin{center}
{\Large Alexi Morin-Duchesne$^{1}$, \quad Jesper Lykke Jacobsen$^{2}$}
\end{center}\vspace{0cm}

\begin{itemize}[leftmargin=0.5in]
\item[1\phantom{$^a$}]{\em Institut de Recherche en Math\'ematique et Physique, Universit\'e catholique de Louvain,\\ Louvain-la-Neuve, B-1348, Belgium}
\item[2$^a$]{\em Laboratoire de Physique de l'Ecole Normale Sup\'erieure, ENS, Universit\'e PSL, CNRS,
Sorbonne Universit\'e, Universit\'e de Paris, Paris, France}
\item[2$^b$] {\em Sorbonne Universit\'e, \'Ecole Normale Sup\'erieure, CNRS, Laboratoire de Physique (LPENS), 75005 Paris, France}
\item[2$^c$] {\em Institut de Physique Th\'eorique, Universit\'e Paris Saclay, CEA, CNRS, F-91191 Gif-sur-Yvette, France}
\end{itemize}

\begin{center}
{\tt alexi.morin-duchesne\,@\,uclouvain.be}
\qquad
{\tt jesper.jacobsen\,@\,ens.fr} 
\end{center}
\medskip

%
%
 
\begin{abstract}
We compute lattice correlation functions for the model of critical dense polymers on a semi-infinite cylinder of perimeter $n$. In the lattice loop model, contractible loops have a vanishing fugacity whereas non-contractible loops have a fugacity $\alpha \in (0,\infty)$. These correlators are defined as ratios $Z(x)/Z_0$ of partition functions, where $Z_0$ is a reference partition function wherein only simple half-arcs are attached to the boundary of the cylinder. For $Z(x)$, the boundary of the cylinder is also decorated with simple half-arcs, but it also has two special positions $1$ and $x$ where the boundary condition is different. We investigate two such kinds of boundary conditions: (i) there is a single node at each of these points where a long arc is attached, and (ii) there are pairs of adjacent nodes at these points where two long arcs are attached.

We find explicit expressions for these correlators for finite $n$ using the representation of the enlarged periodic Temperley-Lieb algebra in the XX spin chain. The resulting asymptotics as $n\to \infty$ are expressed as simple integrals that depend on the scaling parameter $\tau = \frac {x-1} n \in (0,1)$. For small $\tau$, the leading behaviours are proportional to $\tau^{1/4}$, $\tau^{1/4}\log \tau$, $\log \tau$ and $\log^2 \tau$.

We interpret the lattice results in terms of ratios of conformal correlation functions. We assume that the corresponding boundary changing fields are highest weight states in irreducible, Kac or staggered Virasoro modules, with central charge $c=-2$ and conformal dimensions $\Delta = -\frac18$ or $\Delta = 0$. With these assumptions, we obtain differential equations of order two and three satisfied by the conformal correlation functions, solve these equations in terms of hypergeometric functions, and find a perfect agreement with the lattice results. We use the lattice results to compute structure constants and ratios thereof which appear in the operator product expansions of the boundary condition changing fields. The fusion of these fields is found to be non-abelian.

\end{abstract}
Keywords: Loop models, correlation functions, logarithmic conformal field theory

%
%

\newpage

\tableofcontents
\clearpage

\section{Introduction}

A synthetic presentation of the study of critical phenomena, taking into account some of the main lessons learned during the first half-century
following Onsager's celebrated exact solution of the two-dimensional Ising model \cite{O44}, would run roughly as follows: A physical system standing
at a second-order phase transition possesses a continuum limit which is scale invariant \cite{WK74} and usually also conformally invariant \cite{BPZ84,dFMS97}. It is characterised by a
set of critical exponents and universal amplitude ratios that define a given universality class, characteristic of all systems with a given set of symmetries.
The exponents are related to the eigenvalues of the dilation operator, and they characterise the algebraic decay of the correlation functions.
This operator can be defined either in lattice models (as the transfer matrix or Hamiltonian) or in
the quantum field theory describing the continuum limit, so a paramount goal of the theoretician is to diagonalise it. In two dimensions,
this task can be efficiently accomplished by identifying suitably tractable lattice models whose spin chain descriptions can be solved exactly by
integrability techniques \cite{FST79,B82,KBI93}, or by studing the properties of suitable highest-weight representations of infinite-dimensional conformal algebras \cite{BPZ84,FQS84,KRR88,dFMS97,IK11}.
The compatibility of the conclusions obtained from these two approaches has been witnessed in thousands of model studies.

As appealing as this outline may appear, a significant proviso was brought forward in the early 1990's \cite{RS92,G93}: What
happens if the dilation operator is not diagonalisable after all? After a lingering start, the importance of this question has become increasingly
clear in the last two decades \cite{C99,GL02} and has lead directly to the detailed study of non semi-simple representations of certain algebras, both in the
lattice model \cite{DJS10,VJS11,RSA14} and conformal field theory (CFT) \cite{R96,KG96,MR07,MR08,KR09} contexts. The latter define what has become known as logarithmic conformal field theory (LCFT) \cite{GRR13}, since the correlation
functions exhibit both power-law and logarithmic factors. Quite remarkably, the interplay between the lattice models and their continuum limit
has turned out to be as tight as ever before, with essentially the same indecomposable structures appearing in either \cite{GJSV13}.

A necessary condition for this logarithmic behaviour is provided by non-unitarity. While there are various ways of providing this ingredient, a very
natural and physically relevant context arises in the study of models formulated in terms of extended degrees of freedom, such as loops and clusters.
In this paper, we shall focus on so-called critical dense polymers \cite{PR07}, of which the basic manifestation is a single completely-packed closed curve that fills up
the whole lattice. It is the first member, $\mathcal {LM}(1,2)$, of the family of logarithmic minimal models \cite{PRZ06}, with its central charge and conformal weights given by
\be
c = -2, \qquad \Delta_{r,s} = \frac{(2r-s)^2-1}{8}.
\ee
Theories with $c=-2$ are the most well understood logarithmic conformal field theories \cite{K95,GK99,K00,KW01,BF02}.
Other cognate models described by logarithmic CFTs include self-avoiding walks and critical percolation. 
A common feature of such lattice models is the appearance
of cellular algebras \cite{GL96} of the Temperley-Lieb type.

Logarithms in correlation functions were previously found in various lattice models, including the abelian sandpile model \cite{PR04,JPR06,R13}, critical dense polymers \cite{I99,IH05,MDJ18}, critical percolation \cite{VJS12,TCDJ19} and the $Q$-state Potts model \cite{VJ14,CJV17}. In many cases, the results were obtained using conformal arguments and verified numerically on a computer. Our goal is to provide new examples of such logarithmic behaviour where the correlators are computed both rigorously from the lattice using the toolbox of integrability, and using the arguments of conformal invariance extended to the logarithmic theories \cite{F02}.

In a previous paper \cite{MDJ18}, we have defined several types of two-point boundary correlation functions of critical dense polymers. We established
their exact finite-size expressions on a semi-infinite strip of width~$n$ and compared the corresponding asymptotic expansions with the field-theoretical
predictions. These correlators describe boundaries with defect points, allowing them to be connected by the loops in various ways.
Some of the correlators turned out to exhibit logarithmic features, while others did not. The main goal of the present paper is to extend parts of this
study to periodic boundary conditions, where now the correlation functions are defined on a semi-infinite cylinder of circumference~$n$. An interesting
by-product of this modification is that we can now allow for non-contractible loops with a generic weight $\alpha$, while 
contractible loops are forbidden (they have the dense polymer weight $\beta = 0$).
This is also interesting from the CFT perspective, as the relevant conformal correlation functions involve boundary fields, 
as in our previous work, but also a bulk field $\psi_\alpha(z,z^*)$, which is responsible for assigning the weight $\alpha$ to the non-contractible loops.

In the usual CFT setting, there is a close connection between conformally invariant boundary conditions and highest-weight representations in the
bulk theory \cite{C89,C04}. Moreover, a number of strong results (see, e.g., \cite{DJS09,GJP16} in the loop model context) have been obtained from the principle of modular invariance, that is, by comparing
the equivalent results of a closed evolution operator (i.e.~subject to periodic boundary conditions) acting between two given boundary states with those of an
open evolution operator (with given boundary conditions) acting in periodic imaginary time. Our setup similarly replaces the open evolution operator
of \cite{MDJ18} with a closed one, but we keep the defect points at the boundary. The correct interpretation of our results thus remains within
the boundary version of LCFT. The relation between bulk and boundary theories is much more involved in the LCFT setting \cite{GRS13a,GRS13b}, and in particular it is known that in bulk LCFT primary operators can mix into Jordan cells of rank higher than two \cite{MDSA13,GRSV15}.

The outline of the paper is as follows. We start out, in \cref{sec:denseloopmodels}, by recalling the definition of the model of critical dense polymers and its connection with the Temperley-Lieb algebra and the XX spin chain. Due to our
setup, we shall need the enlarged periodic version of the Temperley-Lieb algebra, and shall consider the spin chain with periodic boundary conditions. In \cref{sec:two.defs,sec:four.entry.points},
we define two types of lattice correlation functions, where each insertion point on the boundary involves respectively a single node or a pair of nearest-neighbour
nodes. We find exact expressions for each of these correlators, compute the leading large-$n$ asymptotics in the form of integral formulae, and extract the limiting cases of small and large distances. 
The same correlation functions are then discussed in the context of LCFT in \cref{sec:CFT.two.defects,sec:CFT.four.defects}. Each lattice correlator is understood as a ratio of conformal correlation functions. 
We use conformal invariance to obtain differential equations for these conformal correlators and find that the solutions to these equations precisely reproduce the exact results of \cref{sec:two.defs,sec:four.entry.points}.
In \cref{sec:OPE}, we study the operator product expansions of the boundary conformal fields and use the lattice results to compute some of the structure constants. The fusion of the corresponding boundary conformal fields distinguishes between operators that mark the start and end of arcs attached to the boundary, and is found to be non-abelian. The paper ends in \cref{sec:conclusion} with a discussion of our results and with concluding comments.

\section{Dense polymers with periodic boundary conditions}\label{sec:denseloopmodels}

\subsection{Dense polymers and two-point functions}\label{sec:loopmodel}

We study the model of critical dense polymers on the $m\times n$ cylinder. We draw it in the plane, as in Figure 1, and choose $n$ to be even. A configuration of the model is the choice of a tile\,
$\psset{unit=0.3cm}
\begin{pspicture}[shift=-0.16](0,0)(1,1)
\pspolygon[fillstyle=solid,fillcolor=lightlightblue,linewidth=\mince](0,0)(0,1)(1,1)(1,0)
\psarc[linewidth=\moyen,linecolor=blue](1,0){.5}{90}{180}
\psarc[linewidth=\moyen,linecolor=blue](0,1){.5}{-90}{0}
\end{pspicture}$\, 
 or 
\,$\psset{unit=0.3cm}
\begin{pspicture}[shift=-0.16](0,0)(1,1)
\pspolygon[fillstyle=solid,fillcolor=lightlightblue,linewidth=\mince](0,0)(0,1)(1,1)(1,0)
\psarc[linewidth=\moyen,linecolor=blue](0,0){.5}{0}{90}
\psarc[linewidth=\moyen,linecolor=blue](1,1){.5}{180}{-90}
\end{pspicture}$\, 
for each face of the lattice. The boundary conditions are periodic in the horizontal direction, meaning that the left and right ends of the rectangle are identified in the planar representation. The top of the cylinder is decorated exclusively with simple arcs. Labeling the nodes from $1$ to $n$, these arcs connect the points in the pairs $(1,2)$, $(3,4)$, \dots, $(n-1,n)$. In contrast, the bottom of the cylinder is decorated with a collection of defects and arcs. Their organisation depends on the correlation function that we are studying and is detailed below. A loop configuration $\sigma$ has the weight $w_\sigma = \alpha^{n_\alpha} \delta_{n_\beta,0}$ where $n_\alpha$ and $n_\beta$ are the numbers of non-contractible and contractible loops in the configuration. A collection of arcs that connects two defects of the boundary has weight one. The partition function is defined as
\be
Z = \sum_\sigma w_\sigma.
\ee

We denote by $Z_0$ the partition function for the model where the lower segment is identical to the top segment of the cylinder and is therefore exclusively decorated with simple arcs. We consider $\alpha \in (0,\infty)$ for which $Z_0 \neq 0$. Likewise, we denote by $Z_{\twodefdown}(x)$ the partition function wherein the bottom of the cylinder has simple arcs, and two defects inserted in positions $1$ and $x$. Finally, we denote by $Z_{\fourdefdown}(x)$ the partition function where the bottom of the cylinder has two pairs of adjacent defects in the positions $(1,2)$ and $(x,x+1)$, and simple arcs occupying the other nodes. Examples of configurations for the three partition functions are given in Figure 1. The corresponding two-point correlation functions are then defined as follows:
\be
C_{\twodefdown}(x) = \lim_{m \to \infty} \frac{Z_{\twodefdown}(x)}{Z_0}, \qquad C_{\fourdefdown}(x) = \lim_{m \to \infty} \frac{Z_{\fourdefdown}(x)}{Z_0}.
\ee

\begin{figure}
\begin{center}
\psset{unit=0.35cm}
\begin{pspicture}[shift=-2.9](-0.5,-1.5)(10.5,8.5)
\facegrid{(0,0)}{(10,8)}
\rput(0,7){\loopa}\rput(1,7){\loopa}\rput(2,7){\loopb}\rput(3,7){\loopb}\rput(4,7){\loopa}\rput(5,7){\loopb}\rput(6,7){\loopb}\rput(7,7){\loopb}\rput(8,7){\loopb}\rput(9,7){\loopb}
\rput(0,6){\loopb}\rput(1,6){\loopa}\rput(2,6){\loopb}\rput(3,6){\loopa}\rput(4,6){\loopa}\rput(5,6){\loopa}\rput(6,6){\loopa}\rput(7,6){\loopb}\rput(8,6){\loopa}\rput(9,6){\loopb}
\rput(0,5){\loopa}\rput(1,5){\loopb}\rput(2,5){\loopb}\rput(3,5){\loopb}\rput(4,5){\loopb}\rput(5,5){\loopb}\rput(6,5){\loopa}\rput(7,5){\loopa}\rput(8,5){\loopa}\rput(9,5){\loopa}
\rput(0,4){\loopa}\rput(1,4){\loopb}\rput(2,4){\loopa}\rput(3,4){\loopb}\rput(4,4){\loopa}\rput(5,4){\loopb}\rput(6,4){\loopb}\rput(7,4){\loopa}\rput(8,4){\loopb}\rput(9,4){\loopa}
\rput(0,3){\loopa}\rput(1,3){\loopa}\rput(2,3){\loopb}\rput(3,3){\loopb}\rput(4,3){\loopa}\rput(5,3){\loopb}\rput(6,3){\loopb}\rput(7,3){\loopa}\rput(8,3){\loopb}\rput(9,3){\loopa}
\rput(0,2){\loopb}\rput(1,2){\loopb}\rput(2,2){\loopb}\rput(3,2){\loopa}\rput(4,2){\loopa}\rput(5,2){\loopa}\rput(6,2){\loopa}\rput(7,2){\loopa}\rput(8,2){\loopa}\rput(9,2){\loopb}
\rput(0,1){\loopa}\rput(1,1){\loopa}\rput(2,1){\loopa}\rput(3,1){\loopa}\rput(4,1){\loopb}\rput(5,1){\loopb}\rput(6,1){\loopa}\rput(7,1){\loopb}\rput(8,1){\loopa}\rput(9,1){\loopb}
\rput(0,0){\loopb}\rput(1,0){\loopa}\rput(2,0){\loopb}\rput(3,0){\loopb}\rput(4,0){\loopb}\rput(5,0){\loopa}\rput(6,0){\loopb}\rput(7,0){\loopb}\rput(8,0){\loopa}\rput(9,0){\loopa}
\psarc[linewidth=\elegant,linecolor=blue](1,8){0.5}{0}{180}
\psarc[linewidth=\elegant,linecolor=blue](3,8){0.5}{0}{180}
\psarc[linewidth=\elegant,linecolor=blue](5,8){0.5}{0}{180}
\psarc[linewidth=\elegant,linecolor=blue](7,8){0.5}{0}{180}
\psarc[linewidth=\elegant,linecolor=blue](9,8){0.5}{0}{180}
\psarc[linewidth=\elegant,linecolor=blue](1,0){0.5}{180}{360}
\psarc[linewidth=\elegant,linecolor=blue](3,0){0.5}{180}{360}
\psarc[linewidth=\elegant,linecolor=blue](5,0){0.5}{180}{360}
\psarc[linewidth=\elegant,linecolor=blue](7,0){0.5}{180}{360}
\psarc[linewidth=\elegant,linecolor=blue](9,0){0.5}{180}{360}
\end{pspicture}
\qquad
\begin{pspicture}[shift=-2.9](-0.5,-1.5)(10.5,8.5)
\facegrid{(0,0)}{(10,8)}
\rput(0,7){\loopa}\rput(1,7){\loopb}\rput(2,7){\loopb}\rput(3,7){\loopb}\rput(4,7){\loopa}\rput(5,7){\loopa}\rput(6,7){\loopb}\rput(7,7){\loopb}\rput(8,7){\loopa}\rput(9,7){\loopb}
\rput(0,6){\loopa}\rput(1,6){\loopb}\rput(2,6){\loopb}\rput(3,6){\loopa}\rput(4,6){\loopb}\rput(5,6){\loopb}\rput(6,6){\loopb}\rput(7,6){\loopb}\rput(8,6){\loopb}\rput(9,6){\loopb}
\rput(0,5){\loopb}\rput(1,5){\loopb}\rput(2,5){\loopa}\rput(3,5){\loopa}\rput(4,5){\loopb}\rput(5,5){\loopa}\rput(6,5){\loopa}\rput(7,5){\loopa}\rput(8,5){\loopa}\rput(9,5){\loopb}
\rput(0,4){\loopb}\rput(1,4){\loopa}\rput(2,4){\loopb}\rput(3,4){\loopa}\rput(4,4){\loopa}\rput(5,4){\loopb}\rput(6,4){\loopb}\rput(7,4){\loopb}\rput(8,4){\loopb}\rput(9,4){\loopb}
\rput(0,3){\loopb}\rput(1,3){\loopb}\rput(2,3){\loopb}\rput(3,3){\loopb}\rput(4,3){\loopb}\rput(5,3){\loopb}\rput(6,3){\loopa}\rput(7,3){\loopb}\rput(8,3){\loopa}\rput(9,3){\loopa}
\rput(0,2){\loopb}\rput(1,2){\loopb}\rput(2,2){\loopa}\rput(3,2){\loopa}\rput(4,2){\loopb}\rput(5,2){\loopa}\rput(6,2){\loopa}\rput(7,2){\loopa}\rput(8,2){\loopa}\rput(9,2){\loopa}
\rput(0,1){\loopb}\rput(1,1){\loopb}\rput(2,1){\loopa}\rput(3,1){\loopa}\rput(4,1){\loopb}\rput(5,1){\loopb}\rput(6,1){\loopb}\rput(7,1){\loopb}\rput(8,1){\loopb}\rput(9,1){\loopb}
\rput(0,0){\loopa}\rput(1,0){\loopb}\rput(2,0){\loopb}\rput(3,0){\loopa}\rput(4,0){\loopa}\rput(5,0){\loopb}\rput(6,0){\loopa}\rput(7,0){\loopa}\rput(8,0){\loopb}\rput(9,0){\loopb}
\psarc[linewidth=\elegant,linecolor=blue](1,8){0.5}{0}{180}
\psarc[linewidth=\elegant,linecolor=blue](3,8){0.5}{0}{180}
\psarc[linewidth=\elegant,linecolor=blue](5,8){0.5}{0}{180}
\psarc[linewidth=\elegant,linecolor=blue](7,8){0.5}{0}{180}
\psarc[linewidth=\elegant,linecolor=blue](9,8){0.5}{0}{180}
\psarc[linewidth=\elegant,linecolor=blue](2,0){0.5}{180}{360}
\psarc[linewidth=\elegant,linecolor=blue](4,0){0.5}{180}{360}
\psarc[linewidth=\elegant,linecolor=blue](7,0){0.5}{180}{360}
\psarc[linewidth=\elegant,linecolor=blue](9,0){0.5}{180}{360}
\psline[linewidth=\elegant,linecolor=blue](0.5,0)(0.5,-1)\rput(0.5,-1.6){$_1$}
\psline[linewidth=\elegant,linecolor=blue](5.5,0)(5.5,-1)\rput(5.5,-1.6){$_{x=6}$}
\end{pspicture}
\qquad
\begin{pspicture}[shift=-2.9](-0.5,-1.5)(10.5,8.5)
\facegrid{(0,0)}{(10,8)}
\rput(0,7){\loopa}\rput(1,7){\loopa}\rput(2,7){\loopb}\rput(3,7){\loopb}\rput(4,7){\loopa}\rput(5,7){\loopa}\rput(6,7){\loopb}\rput(7,7){\loopb}\rput(8,7){\loopa}\rput(9,7){\loopb}
\rput(0,6){\loopb}\rput(1,6){\loopb}\rput(2,6){\loopb}\rput(3,6){\loopa}\rput(4,6){\loopa}\rput(5,6){\loopa}\rput(6,6){\loopa}\rput(7,6){\loopb}\rput(8,6){\loopa}\rput(9,6){\loopa}
\rput(0,5){\loopa}\rput(1,5){\loopa}\rput(2,5){\loopa}\rput(3,5){\loopa}\rput(4,5){\loopb}\rput(5,5){\loopb}\rput(6,5){\loopb}\rput(7,5){\loopb}\rput(8,5){\loopa}\rput(9,5){\loopa}
\rput(0,4){\loopb}\rput(1,4){\loopa}\rput(2,4){\loopb}\rput(3,4){\loopa}\rput(4,4){\loopb}\rput(5,4){\loopa}\rput(6,4){\loopb}\rput(7,4){\loopa}\rput(8,4){\loopa}\rput(9,4){\loopa}
\rput(0,3){\loopa}\rput(1,3){\loopa}\rput(2,3){\loopa}\rput(3,3){\loopb}\rput(4,3){\loopb}\rput(5,3){\loopa}\rput(6,3){\loopb}\rput(7,3){\loopa}\rput(8,3){\loopa}\rput(9,3){\loopb}
\rput(0,2){\loopa}\rput(1,2){\loopa}\rput(2,2){\loopa}\rput(3,2){\loopb}\rput(4,2){\loopb}\rput(5,2){\loopa}\rput(6,2){\loopb}\rput(7,2){\loopb}\rput(8,2){\loopa}\rput(9,2){\loopb}
\rput(0,1){\loopb}\rput(1,1){\loopb}\rput(2,1){\loopa}\rput(3,1){\loopb}\rput(4,1){\loopb}\rput(5,1){\loopa}\rput(6,1){\loopa}\rput(7,1){\loopb}\rput(8,1){\loopb}\rput(9,1){\loopb}
\rput(0,0){\loopa}\rput(1,0){\loopa}\rput(2,0){\loopa}\rput(3,0){\loopa}\rput(4,0){\loopb}\rput(5,0){\loopb}\rput(6,0){\loopa}\rput(7,0){\loopa}\rput(8,0){\loopb}\rput(9,0){\loopb}
\psarc[linewidth=\elegant,linecolor=blue](1,8){0.5}{0}{180}
\psarc[linewidth=\elegant,linecolor=blue](3,8){0.5}{0}{180}
\psarc[linewidth=\elegant,linecolor=blue](5,8){0.5}{0}{180}
\psarc[linewidth=\elegant,linecolor=blue](7,8){0.5}{0}{180}
\psarc[linewidth=\elegant,linecolor=blue](9,8){0.5}{0}{180}
\psarc[linewidth=\elegant,linecolor=blue](3,0){0.5}{180}{360}
\psarc[linewidth=\elegant,linecolor=blue](5,0){0.5}{180}{360}
\psarc[linewidth=\elegant,linecolor=blue](9,0){0.5}{180}{360}
\psline[linewidth=\elegant,linecolor=blue](0.5,0)(0.5,-1)\rput(0.5,-1.6){$_1$}
\psline[linewidth=\elegant,linecolor=blue](1.5,0)(1.5,-1)
\psline[linewidth=\elegant,linecolor=blue](6.5,0)(6.5,-1)\rput(6.5,-1.6){$_{x=7}$}
\psline[linewidth=\elegant,linecolor=blue](7.5,0)(7.5,-1)
\end{pspicture}
\end{center}
Figure $1$: Loop configurations on the $10\times 8$ cylinder, with the boundary conditions corresponding to $Z_0$, $Z_{\twodefdown}(x=6)$ and $Z_{\fourdefdown}(x=7)$.
\end{figure}

\subsection{The enlarged periodic Temperley-Lieb algebra}

\paragraph{Definition of the algebra.}

This periodic Temperley-Lieb algebra \cite{L91,MS93,GL98,G98,EG99} is a unital associative algebra that is used to describe many classes of statistical models on periodic geometries. The terminology and conventions vary, and here we work with the {\it enlarged} periodic Temperley-Lieb algebra $\eptl_n(\alpha, \beta)$ defined in \cite{PRV10}:
\be
\eptl_n(\alpha, \beta) = \big\langle I, \Omega, \Omega^{-1},\,e_j;\,j=1,\ldots,n\big\rangle.\qquad  
\ee
Each element of the algebra is associated to a connectivity diagram drawn inside a rectangle that has periodic boundary conditions in the horizontal direction. For the generators, this identification is
\begin{subequations}
\be
I=\,
\begin{pspicture}[shift=-0.55](0.0,-0.65)(2.0,0.45)
\pspolygon[fillstyle=solid,fillcolor=lightlightblue,linewidth=0pt](0,-0.35)(2.0,-0.35)(2.0,0.35)(0,0.35)
\rput(1.4,0.0){\small$...$}
\psline[linecolor=blue,linewidth=1.5pt]{-}(0.2,0.35)(0.2,-0.35)\rput(0.2,-0.55){$_1$}
\psline[linecolor=blue,linewidth=1.5pt]{-}(0.6,0.35)(0.6,-0.35)\rput(0.6,-0.55){$_2$}
\psline[linecolor=blue,linewidth=1.5pt]{-}(1.0,0.35)(1.0,-0.35)\rput(1.0,-0.55){$_3$}
\psline[linecolor=blue,linewidth=1.5pt]{-}(1.8,0.35)(1.8,-0.35)\rput(1.8,-0.55){$_n$}
\psframe[fillstyle=solid,linecolor=white,linewidth=0pt](-0.1,-0.4)(0,0.4)
\psframe[fillstyle=solid,linecolor=white,linewidth=0pt](2.0,-0.4)(2.1,0.4)
\end{pspicture} 
\ ,\qquad
 e_j=\,
\begin{pspicture}[shift=-0.55](0.0,-0.65)(3.2,0.45)
\pspolygon[fillstyle=solid,fillcolor=lightlightblue,linewidth=0pt](0,-0.35)(3.2,-0.35)(3.2,0.35)(0,0.35)
\rput(0.6,0.0){\small$...$}
\rput(2.6,0.0){\small$...$}
\psline[linecolor=blue,linewidth=1.5pt]{-}(0.2,0.35)(0.2,-0.35)\rput(0.2,-0.55){$_1$}
\psline[linecolor=blue,linewidth=1.5pt]{-}(1.0,0.35)(1.0,-0.35)
\psline[linecolor=blue,linewidth=1.5pt]{-}(2.2,0.35)(2.2,-0.35)
\psline[linecolor=blue,linewidth=1.5pt]{-}(3.0,0.35)(3.0,-0.35)\rput(3.0,-0.55){$_{n}$}
\psarc[linecolor=blue,linewidth=1.5pt]{-}(1.6,0.35){0.2}{180}{0}\rput(1.35,-0.55){$_j$}
\psarc[linecolor=blue,linewidth=1.5pt]{-}(1.6,-0.35){0.2}{0}{180}\rput(1.85,-0.55){$_{j+1}$}
\psframe[fillstyle=solid,linecolor=white,linewidth=0pt](-0.1,-0.4)(0,0.4)
\psframe[fillstyle=solid,linecolor=white,linewidth=0pt](3.2,-0.4)(3.3,0.4)
\end{pspicture} \ ,
\qquad \ \ 
e_n= \
\begin{pspicture}[shift=-0.45](0,-0.55)(2.4,0.35)
\pspolygon[fillstyle=solid,fillcolor=lightlightblue,linewidth=0pt](0,-0.35)(2.4,-0.35)(2.4,0.35)(0,0.35)
\rput(0.2,-0.55){$_1$}\rput(0.6,-0.55){$_2$}\rput(1.0,-0.55){$_3$}\rput(2.2,-0.55){$_n$}
\rput(1.4,0.0){\small$...$}
\psarc[linecolor=blue,linewidth=1.5pt]{-}(0.0,0.35){0.2}{-90}{0}
\psarc[linecolor=blue,linewidth=1.5pt]{-}(0.0,-0.35){0.2}{0}{90}
\psline[linecolor=blue,linewidth=1.5pt]{-}(0.6,0.35)(0.6,-0.35)
\psline[linecolor=blue,linewidth=1.5pt]{-}(1.0,0.35)(1.0,-0.35)
\psline[linecolor=blue,linewidth=1.5pt]{-}(1.8,0.35)(1.8,-0.35)
\psarc[linecolor=blue,linewidth=1.5pt]{-}(2.4,-0.35){0.2}{90}{180}
\psarc[linecolor=blue,linewidth=1.5pt]{-}(2.4,0.35){0.2}{180}{-90}
\psframe[fillstyle=solid,linecolor=white,linewidth=0pt](-0.1,-0.4)(0,0.4)
\psframe[fillstyle=solid,linecolor=white,linewidth=0pt](2.4,-0.4)(2.5,0.4)
\end{pspicture} \ ,
\ee

\be
\begin{pspicture}[shift=-0.55](-0.7,-0.65)(2.0,0.35)
\rput(0.2,-0.55){$_1$}\rput(0.6,-0.55){$_2$}\rput(1.0,-0.55){$_3$}\rput(1.4,-0.55){\small$...$}\rput(1.8,-0.55){$_n$}
\pspolygon[fillstyle=solid,fillcolor=lightlightblue,linewidth=0pt](0,-0.35)(2.0,-0.35)(2.0,0.35)(0,0.35)
\multiput(0,0)(0.4,0){6}{\psbezier[linecolor=blue,linewidth=1.5pt]{-}(-0.2,-0.35)(-0.2,-0.0)(0.2,0.0)(0.2,0.35)}
\psframe[fillstyle=solid,linecolor=white,linewidth=0pt](-0.3,-0.4)(0,0.4)
\psframe[fillstyle=solid,linecolor=white,linewidth=0pt](2.0,-0.4)(2.4,0.4)
\rput(-0.55,0.042){$\Omega=$}
\end{pspicture} \ ,
\qquad \ \ 
\begin{pspicture}[shift=-0.55](-1.1,-0.65)(2.0,0.35)
\rput(0.2,-0.55){$_1$}\rput(0.6,-0.55){$_2$}\rput(1.0,-0.55){$_3$}\rput(1.4,-0.55){\small$...$}\rput(1.8,-0.55){$_n$}
\pspolygon[fillstyle=solid,fillcolor=lightlightblue,linewidth=0pt](0,-0.35)(2.0,-0.35)(2.0,0.35)(0,0.35)
\multiput(0,0)(0.4,0){6}{\psbezier[linecolor=blue,linewidth=1.5pt]{-}(-0.2,0.35)(-0.2,-0.0)(0.2,0.0)(0.2,-0.35)}
\psframe[fillstyle=solid,linecolor=white,linewidth=0pt](-0.3,-0.4)(0,0.4)
\psframe[fillstyle=solid,linecolor=white,linewidth=0pt](2.0,-0.4)(2.4,0.4)
\rput(-0.75,0.07){$\Omega^{-1}=$}
\end{pspicture} \ .
\ee
\end{subequations}
The defining relations of the algebra are
\begin{subequations}
\label{eq:EPTL.def}
\begin{alignat}{4}
e_j^2=\beta e_j,& \qquad 
e_j e_{j\pm1} e_j = e_j, \qquad 
&&e_i e_j = e_j e_i \qquad &&(|i-j|>1),
\label{eq:TLdef}
\\[0.3cm]
\Omega \Omega^{-1}= \Omega^{-1} \Omega = I,& \qquad
\Omega e_i \Omega^{-1} = e_{i-1}, \qquad
&&\Omega^{n} e_n = e_n \Omega^{n}, \qquad &&
(\Omega^{\pm 1} e_n)^{n-1} = \Omega^{\pm n} (\Omega ^{\pm 1} e_n),
\end{alignat}
\end{subequations}
where the indices are in the set $\{1, \dots, n\}$ and taken modulo $n$.
For $n$ even, there are extra relations which remove each non-contractible loop and replace it by a weight $\alpha$:
\be
\label{eq:EPTL.alpha}
E \Omega^{\pm 1} E = \alpha E \qquad \textrm {where} \qquad E= e_2e_4\ldots e_{n-2}e_n.
\ee

Henceforth, we set $\beta = 0$ corresponding to the model of critical dense polymers for which contractible loops have a vanishing fugacity.

\paragraph{Transfer tangle.}
The transfer tangle for the model of polymers with periodic boundary conditions is an element of $\eptl_n(\alpha, 0)$ defined as
\be
\Tb (u)= \ 
\psset{unit=0.8}
\begin{pspicture}[shift=-1.1](-0.2,-0.7)(5.2,1.0)
\facegrid{(0,0)}{(5,1)}
\psarc[linewidth=0.025]{-}(0,0){0.16}{0}{90}
\psarc[linewidth=0.025]{-}(1,0){0.16}{0}{90}
\psarc[linewidth=0.025]{-}(4,0){0.16}{0}{90}
\psline[linewidth=1.5pt,linecolor=blue]{-}(0,0.5)(-0.2,0.5)
\psline[linewidth=1.5pt,linecolor=blue]{-}(5,0.5)(5.2,0.5)
\rput(2.5,0.5){$\ldots$}
\rput(3.5,0.5){$\ldots$}
\rput(0.5,.5){$u$}
\rput(1.5,.5){$u$}
\rput(4.5,.5){$u$}
\rput(2.5,-0.5){$\underbrace{\ \hspace{3.8cm} \ }_n$}
\end{pspicture}\ \ ,
\qquad 
 \begin{pspicture}[shift=-.40](1,1)
\facegrid{(0,0)}{(1,1)}
\psarc[linewidth=0.025]{-}(0,0){0.16}{0}{90}
\rput(.5,.5){$u$}
\end{pspicture}
\ = \cos u\ \
\begin{pspicture}[shift=-.40](1,1)
\facegrid{(0,0)}{(1,1)}
\rput[bl](0,0){\loopa}
\end{pspicture}
\ + \sin u \ \
\begin{pspicture}[shift=-.40](1,1)
\facegrid{(0,0)}{(1,1)}
\rput[bl](0,0){\loopb}
\end{pspicture}\ \ , 
\label{eq:Tu}
\ee
where $u$ is the so-called spectral parameter.
The isotropic value is $u = \frac \pi 4$, and we use the short-hand notation $\Tb(\frac\pi4) = \Tb$. The transfer tangle commutes at different values of $u$, namely $[\Tb(u),\Tb(v)] = 0$, and satisfies
\be
\Tb(u) = \Omega \big(I - u\, \mathcal H + \mathcal O(u^2)\big),\qquad \mathcal H = - \sum_{j=1}^n e_j,
\ee
where $\mathcal H$ is the Hamiltonian. It also commutes with $\Tb(u)$.

\paragraph{The standard module $\boldsymbol{\mathsf W_{n,0}}$.}
The algebra $\eptl_n(\alpha,0)$ has a family of standard modules $\mathsf W_{n,d}$ labeled by an integer number $d$ of defects. Our calculations below only require the standard module with no defects, $\mathsf W_{n,0}$. This module is defined on the vector space generated by link states with no defects. These are diagrams drawn over a segment where $n$ marked nodes are connected pairwise by non-intersecting loop segments. The boundary conditions are cylindric, namely periodic in the horizontal direction so that the loop segments may connect via the back of the cylinder. Here are the link states for $n=2$ and $n=4$:
\be
\label{eq:link.states.2.4}
\mathsf W_{2,0}: \ \ 
\psset{unit=0.8cm}
\begin{pspicture}[shift=-0.0](-0.0,0)(0.8,0.5)
\psline[linewidth=0.75pt]{-}(0,0)(0.8,0)
\psarc[linecolor=blue,linewidth=1.5pt]{-}(0.4,0){0.2}{0}{180}
\end{pspicture}\ , \ \ 
\begin{pspicture}[shift=-0.0](-0.0,0)(0.8,0.5)
\psline[linewidth=0.75pt]{-}(0,0)(0.8,0)
\psarc[linecolor=blue,linewidth=1.5pt]{-}(0.0,0){0.2}{0}{90}
\psarc[linecolor=blue,linewidth=1.5pt]{-}(0.8,0){0.2}{90}{180}
\end{pspicture}\ ,
\qquad
\mathsf W_{4,0}: \ \ 
\psset{unit=0.8cm}
\begin{pspicture}[shift=-0.0](-0.0,0)(1.6,0.5)
\psline[linewidth=0.75pt]{-}(0,0)(1.6,0)
\psarc[linecolor=blue,linewidth=1.5pt]{-}(0.4,0){0.2}{0}{180}
\psarc[linecolor=blue,linewidth=1.5pt]{-}(1.2,0){0.2}{0}{180}
\end{pspicture}\ , \ \ 
\begin{pspicture}[shift=-0.0](-0.0,0)(1.6,0.5)
\psline[linewidth=0.75pt]{-}(0,0)(1.6,0)
\psbezier[linecolor=blue,linewidth=1.5pt]{-}(0.2,0)(0.2,0.7)(1.4,0.7)(1.4,0)
\psarc[linecolor=blue,linewidth=1.5pt]{-}(0.8,0){0.2}{0}{180}
\end{pspicture}\ , \ \ 
\begin{pspicture}[shift=-0.0](-0.0,0)(1.6,0.5)
\psline[linewidth=0.75pt]{-}(0,0)(1.6,0)
\psbezier[linecolor=blue,linewidth=1.5pt]{-}(0.6,0)(0.6,0.7)(1.8,0.7)(1.8,0)
\psbezier[linecolor=blue,linewidth=1.5pt]{-}(-0.2,0.54)(-0.12,0.43)(0.2,0.4)(0.2,0)
\psarc[linecolor=blue,linewidth=1.5pt]{-}(1.2,0){0.2}{0}{180}
\psframe[fillstyle=solid,linecolor=white,linewidth=0pt](1.6,-0.1)(2.0,0.9)
\psframe[fillstyle=solid,linecolor=white,linewidth=0pt](0.0,-0.1)(-0.4,0.9)
\end{pspicture}\ , \ \ 
\begin{pspicture}[shift=-0.0](-0.0,0)(1.6,0.5)
\psline[linewidth=0.75pt]{-}(0,0)(1.6,0)
\psarc[linecolor=blue,linewidth=1.5pt]{-}(0,0){0.2}{0}{90}
\psarc[linecolor=blue,linewidth=1.5pt]{-}(0.8,0){0.2}{0}{180}
\psarc[linecolor=blue,linewidth=1.5pt]{-}(1.6,0){0.2}{90}{180}
\end{pspicture}\ , \ \ 
\begin{pspicture}[shift=-0.0](-0.0,0)(1.6,0.5)
\psline[linewidth=0.75pt]{-}(0,0)(1.6,0)
\psbezier[linecolor=blue,linewidth=1.5pt]{-}(1.8,0.54)(1.72,0.43)(1.4,0.4)(1.4,0)
\psbezier[linecolor=blue,linewidth=1.5pt]{-}(-0.2,0)(-0.2,0.7)(1.0,0.7)(1.0,0)
\psarc[linecolor=blue,linewidth=1.5pt]{-}(0.4,0){0.2}{0}{180}
\psframe[fillstyle=solid,linecolor=white,linewidth=0pt](1.6,-0.1)(2.0,0.9)
\psframe[fillstyle=solid,linecolor=white,linewidth=0pt](0.0,-0.1)(-0.4,0.9)
\end{pspicture}\ , \ \ 
\begin{pspicture}[shift=-0.0](-0.0,0)(1.6,0.5)
\psline[linewidth=0.75pt]{-}(0,0)(1.6,0)
\psbezier[linecolor=blue,linewidth=1.5pt]{-}(-0.12,0.53)(-0.07,0.545)(0.6,0.545)(0.6,0)
\psbezier[linecolor=blue,linewidth=1.5pt]{-}(1.72,0.53)(1.67,0.545)(1.0,0.545)(1.0,0)
\psarc[linecolor=blue,linewidth=1.5pt]{-}(0.0,0){0.2}{0}{90}
\psarc[linecolor=blue,linewidth=1.5pt]{-}(1.6,0){0.2}{90}{180}
\psframe[fillstyle=solid,linecolor=white,linewidth=0pt](1.6,-0.1)(2.0,0.9)
\psframe[fillstyle=solid,linecolor=white,linewidth=0pt](0.0,-0.1)(-0.4,0.9)
\end{pspicture}\ .
\ee

We define the action of the elements of $\eptl_n(\alpha,0)$ on the link states via the action of the algebra's generators. To compute $a v$ with $a = e_j$ or $a = \Omega^{\pm 1}$, we draw the connectivity diagram corresponding to $a$ below $v$. The new link state is read from the connectivity of the bottom segment. If one (or more) contractible loop is formed, the result is set to zero. If one (or more) non-contractible loop is formed, each such loop is erased and replaced by a multiplicative factor of $\alpha$. Here are examples for $n=4$:
\be
\begin{pspicture}[shift=-0.45](0.0,-0.45)(1.6,0.95)
\pspolygon[fillstyle=solid,fillcolor=lightlightblue,linewidth=0pt](0,-0.35)(1.6,-0.35)(1.6,0.35)(0,0.35)
\psline[linecolor=blue,linewidth=1.5pt]{-}(1.0,0.35)(1.0,-0.35)
\psline[linecolor=blue,linewidth=1.5pt]{-}(1.4,0.35)(1.4,-0.35)
\psarc[linecolor=blue,linewidth=1.5pt]{-}(0.4,0.35){0.2}{180}{0}
\psarc[linecolor=blue,linewidth=1.5pt]{-}(0.4,-0.35){0.2}{0}{180}
\rput(0,0.35){
\psbezier[linecolor=blue,linewidth=1.5pt]{-}(0.6,0)(0.6,0.6)(1.4,0.65)(1.66,0.35)
\psbezier[linecolor=blue,linewidth=1.5pt]{-}(-0.0675,0.47)(-0.05,0.43)(0.2,0.33)(0.2,0)
\psarc[linecolor=blue,linewidth=1.5pt]{-}(1.2,0){0.2}{0}{180}
\psframe[fillstyle=solid,linecolor=white,linewidth=0pt](1.6,0.1)(1.7,0.9)
\psframe[fillstyle=solid,linecolor=white,linewidth=0pt](0.0,0.1)(-0.1,0.9)
}
\end{pspicture} \ \ = \alpha \ \ 
\begin{pspicture}[shift=0](0,0)(1.6,0.6)
\psarc[linecolor=blue,linewidth=1.5pt]{-}(0.4,0){0.2}{0}{180}
\psarc[linecolor=blue,linewidth=1.5pt]{-}(1.2,0){0.2}{0}{180}
\psframe[fillstyle=solid,linecolor=white,linewidth=0pt](1.6,-0.1)(1.7,0.9)
\psframe[fillstyle=solid,linecolor=white,linewidth=0pt](0.0,-0.1)(-0.1,0.9)
\psline[linewidth=0.75pt]{-}(0,0)(1.6,0)
\end{pspicture}\ \ ,
\qquad \qquad
\begin{pspicture}[shift=-0.45](0.0,-0.45)(1.6,0.95)
\pspolygon[fillstyle=solid,fillcolor=lightlightblue,linewidth=0pt](0,-0.35)(1.6,-0.35)(1.6,0.35)(0,0.35)
\psline[linecolor=blue,linewidth=1.5pt]{-}(0.2,0.35)(0.2,-0.35)
\psline[linecolor=blue,linewidth=1.5pt]{-}(0.6,0.35)(0.6,-0.35)
\psarc[linecolor=blue,linewidth=1.5pt]{-}(1.2,0.35){0.2}{180}{0}
\psarc[linecolor=blue,linewidth=1.5pt]{-}(1.2,-0.35){0.2}{0}{180}
\rput(0,0.35){
\psbezier[linecolor=blue,linewidth=1.5pt]{-}(0.6,0)(0.6,0.6)(1.4,0.65)(1.66,0.35)
\psbezier[linecolor=blue,linewidth=1.5pt]{-}(-0.0675,0.47)(-0.05,0.43)(0.2,0.33)(0.2,0)
\psarc[linecolor=blue,linewidth=1.5pt]{-}(1.2,0){0.2}{0}{180}
\psframe[fillstyle=solid,linecolor=white,linewidth=0pt](1.6,0.1)(1.7,0.9)
\psframe[fillstyle=solid,linecolor=white,linewidth=0pt](0.0,0.1)(-0.1,0.9)
}
\end{pspicture} \ \ = 0\ ,
\qquad \qquad
\begin{pspicture}[shift=-0.45](0.0,-0.45)(1.6,0.95)
\pspolygon[fillstyle=solid,fillcolor=lightlightblue,linewidth=0pt](0,-0.35)(1.6,-0.35)(1.6,0.35)(0,0.35)
\multiput(0,0)(0.4,0){5}{\psbezier[linecolor=blue,linewidth=1.5pt]{-}(-0.2,0.35)(-0.2,-0.0)(0.2,0.0)(0.2,-0.35)}
\psframe[fillstyle=solid,linecolor=white,linewidth=0pt](-0.3,-0.4)(0,0.4)
\psframe[fillstyle=solid,linecolor=white,linewidth=0pt](1.6,-0.4)(2.0,0.4)
\rput(0,0.35){
\psbezier[linecolor=blue,linewidth=1.5pt]{-}(0.6,0)(0.6,0.6)(1.4,0.65)(1.66,0.35)
\psbezier[linecolor=blue,linewidth=1.5pt]{-}(-0.0675,0.47)(-0.05,0.43)(0.2,0.33)(0.2,0)
\psarc[linecolor=blue,linewidth=1.5pt]{-}(1.2,0){0.2}{0}{180}
\psframe[fillstyle=solid,linecolor=white,linewidth=0pt](1.6,0.1)(1.7,0.9)
\psframe[fillstyle=solid,linecolor=white,linewidth=0pt](0.0,0.1)(-0.1,0.9)
}
\end{pspicture} \ \ =  \ \ 
\begin{pspicture}[shift=0](0,0)(1.6,0.6)
\psline[linewidth=0.75pt]{-}(0,0)(1.6,0)
\psbezier[linecolor=blue,linewidth=1.5pt]{-}(-0.12,0.53)(-0.07,0.545)(0.6,0.545)(0.6,0)
\psbezier[linecolor=blue,linewidth=1.5pt]{-}(1.72,0.53)(1.67,0.545)(1.0,0.545)(1.0,0)
\psarc[linecolor=blue,linewidth=1.5pt]{-}(0.0,0){0.2}{0}{90}
\psarc[linecolor=blue,linewidth=1.5pt]{-}(1.6,0){0.2}{90}{180}
\psframe[fillstyle=solid,linecolor=white,linewidth=0pt](1.6,-0.1)(2.0,0.9)
\psframe[fillstyle=solid,linecolor=white,linewidth=0pt](0.0,-0.1)(-0.4,0.9)
\end{pspicture}\ \ .
\ee

\paragraph{The bilinear Gram form.}
Let $v,w$ be two link states in $\mathsf W_{n,0}$. Their Gram overlap, denoted $v\cdot w$, is obtained by flipping $v$ vertically and attaching its nodes to those of $w$. The resulting diagram then has $n_\alpha$ non-contractible loops and $n_\beta$ contractible loops. The overlap is then defined as $v\cdot w = \alpha^{n_\alpha} \delta_{n_\beta,0}$. It is then bilinearly extended to all states in $\mathsf W_{n,0}$. The values of the overlaps in the link state basis are encoded in the Gram matrix. To illustrate, in the bases \eqref{eq:link.states.2.4}, the Gram matrices for $n=2$ and $n=4$ are
\be
\label{eq:Gram24}
\begin{pmatrix}
0 & \alpha \\ \alpha & 0
\end{pmatrix} 
\qquad \textrm{and} \qquad
\begin{pmatrix}
0 & 0 & 0 & \alpha  & 0 & 0 \\
 0 & 0 & \alpha  & 0 & \alpha  & \alpha ^2 \\
 0 & \alpha  & 0 & 0 & \alpha ^2 & \alpha  \\
 \alpha  & 0 & 0 & 0 & 0 & 0 \\
 0 & \alpha  & \alpha ^2 & 0 & 0 & \alpha  \\
 0 & \alpha ^2 & \alpha  & 0 & \alpha  & 0
\end{pmatrix}.
\ee

\paragraph{The spin-chain representation $\boldsymbol{\mathsf X_n}$.}
The representation $\mathsf X_n$ of $\eptl_n(\alpha,0)$ is defined on the vector space $(\mathbb C^2)^{\otimes n}$. The generators $e_1, \dots, e_{n-1} $ are represented by the matrices
\be
\mathsf X_n(e_j) = \mathbb I_{j-1} \otimes
\begin{pmatrix}
0 & 0 & 0 & 0 \\
0 & \ir & 1 & 0 \\
0 & 1 & \ir^{-1} & 0 \\
0 & 0 & 0 & 0
\end{pmatrix} 
\otimes \mathbb I_{n-j-1}, \qquad j = 1, \dots, n-1,
\ee
where $\mathbb I_j$ is the identity matrix of size $2^j$. The representants for $e_n$ and $\Omega$ depend on a {\it twist angle} $\phi$,
\be
\mathsf X_n(e_n) = t \left[ 
\begin{pmatrix}
0 & 0 & 0 & 0 \\
0 & \ir & \eE^{\ir \phi} & 0 \\
0 & \eE^{-\ir \phi} & \ir^{-1} & 0 \\
0 & 0 & 0 & 0
\end{pmatrix} 
 \otimes \mathbb I_{n-2}\right] t^{-1}, 
 \qquad
 \mathsf X_n(\Omega) = t\, \eE^{\ir \phi\sigma^z_1/2},
\ee
where $t$ is the translation operator:
\be
t |v_1\rangle \otimes |v_2\rangle \otimes \cdots \otimes |v_n\rangle = |v_2\rangle  \otimes \cdots \otimes |v_n\rangle \otimes |v_1\rangle.
\ee
One can check that the defining relations \eqref{eq:EPTL.def} and \eqref{eq:EPTL.alpha} are satisfied, with the weight $\alpha$ of the non-contractible loops related to the twist angle via the relation
\be
\label{eq:alpha.phi}
\alpha = 2 \cos \Big(\frac \phi 2\Big).
\ee
For $\alpha \in (0,2]$, $\phi$ is real and in the range $[0,\pi)$, whereas for $\alpha \in (2,\infty)$, $\phi$ is purely imaginary in the range $(\ir 0, \ir\infty)$.

\paragraph{Homomorphism and overlaps.}
There exists a homomorphism between the standard module $\mathsf W_{n,0}$ and the spin-chain module $\mathsf X_n$ of $\eptl_n(\alpha,0)$. It is defined via the following local maps:
\be\label{eq:localmaps}
|\,
\psset{unit=0.74}
\begin{pspicture}[shift=-0.08](0.0,0)(0.8,0.5)
\psline[linewidth=\mince](0,0)(0.8,0)
\psarc[linecolor=blue,linewidth=\elegant]{-}(0.4,0){0.2}{0}{180}
\end{pspicture}
\,  \rangle = \omega\, |{\uparrow \downarrow}\rangle + \omega^{-1}\, |{\downarrow \uparrow}\rangle, \qquad
|\,
\begin{pspicture}[shift=-0.08](0.0,0)(0.8,0.5)
\psline[linewidth=\mince](0,0)(0.8,0)
\psarc[linecolor=blue,linewidth=\elegant]{-}(0,0){0.2}{0}{90}
\psarc[linecolor=blue,linewidth=\elegant]{-}(0.8,0){0.2}{90}{180}
\end{pspicture}
\, \rangle = \omega\,\eE^{\ir \phi/2}|{\downarrow\uparrow}\rangle + \omega^{-1}\eE^{-\ir \phi/2} |{\uparrow \downarrow}\rangle, \qquad \omega = \eE^{\ir \pi/4}.
\ee
For link states with more than one arc, the local map is applied multiplicatively to each arc. For instance: 
\be
|\,
\psset{unit=0.74}
\begin{pspicture}[shift=-0.0](-0.0,0)(1.6,0.5)
\psline{-}(0,0)(1.6,0)
\psbezier[linecolor=blue,linewidth=1.5pt]{-}(0.6,0)(0.6,0.6)(1.4,0.65)(1.66,0.35)
\psbezier[linecolor=blue,linewidth=1.5pt]{-}(-0.0675,0.47)(-0.05,0.43)(0.2,0.33)(0.2,0)
\psarc[linecolor=blue,linewidth=1.5pt]{-}(1.2,0){0.2}{0}{180}
\psframe[fillstyle=solid,linecolor=white,linewidth=0pt](1.6,-0.1)(1.7,0.9)
\psframe[fillstyle=solid,linecolor=white,linewidth=0pt](0.0,-0.1)(-0.1,0.9)
\end{pspicture}
\,  \rangle = \eE^{-\ir \phi/2}|{\uparrow\downarrow\uparrow\downarrow}\rangle + \omega^{-2}\eE^{-\ir \phi/2} |{\uparrow \downarrow \downarrow \uparrow}\rangle +  \omega^2\,\eE^{\ir \phi/2} |{\downarrow \uparrow \uparrow \downarrow}\rangle + \eE^{\ir \phi/2}|{\downarrow \uparrow \downarrow \uparrow}\rangle.
\ee
It is well known that this is indeed a homomorphism, namely for each $v$ in $\mathsf W_{n,0}$ we have
\be
\mathsf X_n(e_j) |v\rangle = |e_jv\rangle, \qquad \mathsf X_n(\Omega^{\pm 1}) |v\rangle = |\Omega^{\pm 1}v\rangle,
\ee
where $\alpha$ and $\phi$ are related as in \eqref{eq:alpha.phi}. The dual states are then defined as $\langle v | = | v \rangle^\Tt\big|_{\phi \to -\phi}$. The spin-chain overlap between $v$ and $w$ then equals the Gram overlap between $v$ and $w$:
\be
\langle v | w\rangle = v \cdot w, \qquad v,w \in \mathsf W_{n,0}.
\ee

\subsection{XX Hamiltonian}

The periodic XX Hamiltonian with the twist $\phi$ is defined as 
\be
H = \mathsf X_n(\mathcal H) =  - \Big(\sum_{j=1}^{n-1} \sigma^+_j \sigma^-_{j+1} + \sigma^-_j \sigma^+_{j+1} \Big) - \eE^{\ir \phi} \sigma^+_n \sigma^-_1 - \eE^{-\ir \phi} \sigma^-_n \sigma^+_1.
\ee
It is hermitian and therefore has real eigenvalues.
In terms of the fermions
\be
c_j = (-1)^{j-1}\bigg(\prod_{k=1}^{j-1} \sigma^z_k\bigg) \sigma^-_j, \qquad c_j^\dagger = (-1)^{j-1}\bigg(\prod_{k=1}^{j-1} \sigma^z_k\bigg) \sigma^+_j,
\ee
it is expressed as
\be
H = - \Big(\sum_{j=1}^{n-1} c^\dagger_j c_{j+1} + c^\dagger_{j+1} c_j \Big) - \eE^{\ir (\frac \pi 2(n+2S^z + 2) + \phi)} c_n^\dagger c_1 - \eE^{-\ir (\frac \pi 2(n+2S^z + 2) + \phi)} c_1^\dagger c_n \,,
\ee
where $S^z = \frac12\sum_{j=1}^n \sigma^z_j$ is the magnetisation.
Applying a Fourier transform allows us to put $H$ in diagonal form:
\be
H = - \sum_{k=1}^n2 \cos( \theta_k) \eta_k^\dagger \eta_k \,,
\ee
where
\be
\eta_k = \frac1{\sqrt n} \sum_{j=1}^n \eE^{\ir j \theta_k} c_j, \qquad \eta^\dagger_k = \frac1{\sqrt n} \sum_{j=1}^n \eE^{-\ir j \theta_k} c_j^\dagger, \qquad 
\theta_k =
\left\{\begin{array}{cl}
\frac{2 \pi k - \phi}n & \frac n2 + S^z \textrm{ odd},\\[0.15cm]
\frac{2 \pi (k-\frac12) - \phi}n & \frac n2 + S^z \textrm{ even}.
\end{array}\right.
\ee
A full set of fermionic operators is obtained by taking $k$ in the set $\{1, \dots, n\}$. In what follows, it is however convenient to extend the definition of $\eta_k$ to integer values of $k$ that are negative, using the periodicity properties $\eta_{k+n} = \eta_k$ and $\eta^\dagger_{k+n} = \eta^\dagger_k$. Then, for $\phi \in [0,\pi)$ and $\phi \in (\ir 0, \ir\infty)$, the groundstate of $H$ lies in the magnetisation sector $S^z = 0$ and is given by
\be
|w_0\rangle = \left\{\begin{array}{cl}
\eta^\dagger_{(2-n)/4}\cdots\eta^\dagger_{(n-2)/4} |0\rangle & \frac n2 \textrm{ odd,}\\[0.15cm]
\eta^\dagger_{(4-n)/4}\cdots\eta^\dagger_{n/4} |0\rangle & \frac n2 \textrm{ even,} 
\end{array}\right.
\qquad |0\rangle = |{\downarrow\cdots\downarrow}\rangle.
\ee
This state is also the groundstate of the transfer matrix $T(u)$ for $0 \le u \le \frac \pi 2$. The corresponding eigenvalue $\Lambda_0$ of $T = T(\frac \pi 4)$ is \cite{PRV10,MDPR13}
\be
\Lambda_0 = \frac{\cos (\frac\phi2)}{2^{n-1}} \prod_{j=1}^{n/2} (1+\tan x_j) \prod_{j=n/2+1}^{n} (1-\tan x_j), \qquad 
x_j = \frac {\pi(j-\frac12)-\frac\phi2} n.
\ee

\section{Lattice correlators for single entry points}\label{sec:two.defs}

\subsection{Refined partition functions}

For each loop configuration that contributes to $Z_{\twodefdown}(x)$, the two defects can connect either via the back or the front of the cylinder. We denote these two possible connections as 
$
\psset{unit=0.5cm}
\begin{pspicture}[shift=0.015cm](0,0)(0.8,0.4)
\psline[linewidth=0.5pt]{-}(0,0)(0.8,0)
\psarc[linecolor=blue,linewidth=\moyen]{-}(0,0){0.2}{0}{90}
\psarc[linecolor=blue,linewidth=\moyen]{-}(0.8,0){0.2}{90}{180}
\end{pspicture}
$
and
$
\psset{unit=0.5cm}
\begin{pspicture}[shift=0.015cm](0,0)(0.8,0.4)
\psline[linewidth=0.5pt]{-}(0,0)(0.8,0)
\psarc[linecolor=blue,linewidth=\moyen]{-}(0.4,0){0.2}{0}{180}
\end{pspicture}
$
and define the refined partition functions $Z_{\arcupout}(x)$ and $Z_{\arcupin}(x)$. The partition function $Z_{\twodefdown}(x)$ decomposes as
\be
Z_{\twodefdown}(x) = Z_{\arcupout}(x) + Z_{\arcupin}(x).
\ee
Let us also define two more partition functions: $Z_{\arcdownin}(x)$ and $Z_{\arcdownout}(x)$. They are defined in a similar way to $Z_{\twodefdown}(x)$, but with the two defects on the lower boundary attached together to become a long arc, in the two possible ways. This means that, in the middle panel of Figure 1, the boundary condition at the lower end of the cylinder is replaced with the following states flipped vertically:
\begin{equation}
\psset{unit=0.8}
v^{\rm a}_x = \ \ 
\begin{pspicture}[shift=-0.48](0,-0.3)(7.2,1)
\psline[linewidth=\mince](0,0)(7.2,0)
\psarc[linecolor=blue,linewidth=\notelegant]{-}(0.8,0){0.2}{0}{180}
\psarc[linecolor=blue,linewidth=\notelegant]{-}(1.6,0){0.2}{0}{180}
\rput(2.4,0.15){...}
\rput(0.2,-0.25){\scriptsize$1$}
\rput(3.8,-0.25){\scriptsize$x$}
\psbezier[linecolor=blue,linewidth=\notelegant](0.2,0)(0.2,1.1)(3.8,1.1)(3.8,0)
\psarc[linecolor=blue,linewidth=\notelegant]{-}(3.2,0){0.2}{0}{180}
\psarc[linecolor=blue,linewidth=\elegant]{-}(4.4,0){0.2}{0}{180}
\psarc[linecolor=blue,linewidth=\elegant]{-}(5.2,0){0.2}{0}{180}
\psarc[linecolor=blue,linewidth=\elegant]{-}(6.8,0){0.2}{0}{180}
\psline[linewidth=0.5pt,linestyle= dashed, dash = 2pt 2pt]{-}(0,0)(0,0.9)
\psline[linewidth=0.5pt,linestyle= dashed, dash = 2pt 2pt]{-}(7.2,0)(7.2,0.9)
\rput(6.0,0.15){...}
\end{pspicture}\ \ ,
\qquad\quad
v^{\rm b}_x = \ \ 
\begin{pspicture}[shift=-0.48](0,-0.3)(7.2,1)
\psline[linewidth=\mince](0,0)(7.2,0)
\psarc[linecolor=blue,linewidth=\notelegant]{-}(0.8,0){0.2}{0}{180}
\psarc[linecolor=blue,linewidth=\notelegant]{-}(1.6,0){0.2}{0}{180}
\rput(2.4,0.15){...}
\rput(0.2,-0.25){\scriptsize$1$}
\rput(3.8,-0.25){\scriptsize$x$}
\psbezier[linecolor=blue,linewidth=\notelegant](7.6,0)(7.6,1.1)(3.8,1.1)(3.8,0)
\psbezier[linecolor=blue,linewidth=\notelegant](0.2,0)(0.2,0.2)(0.1,0.34)(-0.1,0.48)
\psarc[linecolor=blue,linewidth=\notelegant]{-}(3.2,0){0.2}{0}{180}
\psarc[linecolor=blue,linewidth=\elegant]{-}(4.4,0){0.2}{0}{180}
\psarc[linecolor=blue,linewidth=\elegant]{-}(5.2,0){0.2}{0}{180}
\psarc[linecolor=blue,linewidth=\elegant]{-}(6.8,0){0.2}{0}{180}
\rput(6.0,0.15){...}
\psframe[fillstyle=solid,fillcolor=white,linecolor=white,linewidth=0pt](-0.2,1.2)(0,0)
\psframe[fillstyle=solid,fillcolor=white,linecolor=white,linewidth=0pt](7.2,1.2)(7.7,0)
\psline[linewidth=0.5pt,linestyle= dashed, dash = 2pt 2pt]{-}(0,0)(0,0.9)
\psline[linewidth=0.5pt,linestyle= dashed, dash = 2pt 2pt]{-}(7.2,0)(7.2,0.9)
\end{pspicture}
\ \ .
\end{equation}
Clearly, we have
\be
\begin{pmatrix}
Z_{\arcdownin}(x)\\
Z_{\arcdownout}(x)
\end{pmatrix} = 
\begin{pmatrix}
0 & \alpha \\ \alpha & 0
\end{pmatrix}
\begin{pmatrix}
Z_{\arcupin}(x)\\
Z_{\arcupout}(x)
\end{pmatrix}.
\ee
The matrix on the right side is the Gram matrix for $\mathsf W_{2,0}$ given in \eqref{eq:Gram24}. We also note that the state attached to the top of the cylinder in Figure 1 is none other than $v^{\rm a}_2$. Its translation by one node is $v^{\rm b}_n$.

We define the refined correlation functions corresponding to
$
\psset{unit=0.5cm}
\begin{pspicture}[shift=-0.11cm](0,-0.5)(0.8,0)
\psline[linewidth=0.5pt]{-}(0,0)(0.8,0)
\psarc[linecolor=blue,linewidth=\moyen]{-}(0,0){0.2}{-90}{0}
\psarc[linecolor=blue,linewidth=\moyen]{-}(0.8,0){0.2}{180}{270}
\end{pspicture}
$
and
$
\psset{unit=0.5cm}
\begin{pspicture}[shift=-0.11cm](0,-0.5)(0.8,0)
\psline[linewidth=0.5pt]{-}(0,0)(0.8,0)
\psarc[linecolor=blue,linewidth=\moyen]{-}(0.4,0){0.2}{180}{0}
\end{pspicture}
$:
\be
C_{\arcdownin}(x) = \lim_{m \to \infty} \frac{Z_{\arcdownin}(x)}{Z_0}, \qquad C_{\arcdownout}(x) = \lim_{m \to \infty} \frac{Z_{\arcdownout}(x)}{Z_0}. 
\ee
In constrast to $C_{\twodefdown}(x)$, these two functions have a well-defined $\alpha \to 0$ limit. Indeed, the numerators and denominators are polynomials in $\alpha$ with the lowest-degree term proportional to $\alpha$. In the limit $\alpha\to 0$, these correlation functions give ratios of partition functions where the configurations with non-zero weights have a single non-contractible loop, and no contractible loops.  

The partition functions $Z_{\arcdownin}(x)$ and $Z_{\arcdownout}(x)$ are expressed in terms of Gram overlaps as 
\be
Z_{\arcdownin}(x) = 2^{mn/2} (v^{\rm a}_x\cdot\Tb^m v^{\rm a}_2), \qquad Z_{\arcdownout}(x) = 2^{mn/2} (v^{\rm b}_x\cdot\Tb^m v^{\rm a}_2),
\ee
where the factor of $2^{mn/2}$ ensures that the tiles have weight $1$ and not $\frac1{\sqrt 2}$, as they do in \eqref{eq:Tu} for $u = \frac \pi 4$.
Because of the invariance under translation, we can equivalently write 
\be
Z_{\arcdownout}(x) = 2^{mn/2} (v^{\rm a}_{n+2-x}\cdot\Tb^m v^{\rm b}_n).
\ee
These partition functions are then expressed as overlaps in the spin chain:
\be
Z_{\arcdownin}(x) = 2^{mn/2} \langle v^{\rm a}_x|T^m | v^{\rm a}_2\rangle, \qquad Z_{\arcdownout}(x) = 2^{mn/2} \langle v^{\rm a}_{n+2-x}|T^m |v^{\rm b}_n\rangle,
\ee
where $T = \mathsf X_n(\Tb)$ is the transfer matrix.

\subsection[Closed-form expressions in the limit $m \to \infty$]{Closed-form expressions in the limit $\boldsymbol{m \to \infty}$}

In the limit $m \to \infty$, the leading contribution to the overlaps comes from the groundstate,
\be
Z_{\arcdownin}(x) \simeq 2^{mn/2} \Lambda_0^m \langle v^{\rm a}_x|w_0\rangle\langle w_0 | v^{\rm a}_2\rangle,
\qquad
Z_{\arcdownout}(x) \simeq 2^{mn/2} \Lambda_0^m \langle  v^{\rm a}_{n+2-x}|w_0\rangle\langle w_0|v^{\rm b}_n\rangle, 
\ee
where $\simeq$ indicates an equality up to terms that are exponentially small in $m$ compared to the leading term. Because $|w_0\rangle$ is invariant under translation, we know that $\langle w_0| v^{\rm a}_2\rangle = \langle w_0 |v^{\rm b}_n\rangle$. The partition function $Z_0$ is equal to $Z_{\arcdownin}(2)$, and as a result we have
\be
\label{eq:2Cs}
C_{\arcdownin}(x) =  \frac{\langle  v^{\rm a}_x|w_0\rangle}{\langle  v^{\rm a}_2|w_0\rangle}, \qquad
C_{\arcdownout}(x) =  \frac{\langle  v^{\rm a}_{n+2-x}|w_0\rangle}{\langle  v^{\rm a}_2|w_0\rangle} = C_{\arcdownin}(n+2-x).
\ee

We therefore focus on computing $C_{\arcdownin}(x)$. Following \eqref{eq:localmaps}, the state $v^{\rm a}_x$ is represented in the spin chain by
\be
\label{eq:vax}
\langle v^{\rm a}_x| = \langle 0 | a_{n-1} a_{n-3} \cdots a_{x+1} a_{x-2} a_{x-4} \cdots a_{2} \big(\omega\, c_1 + (-1)^{\frac{x-2}2}\omega^{-1} c_x\big)
\ee
where 
\be
a_j = \omega c_j + \omega^{-1} c_{j+1}, \qquad \omega = \eE^{\ir \pi/4}.
\ee
The sign $(-1)^{\frac{x-2}2}$ of the last factor in \eqref{eq:vax} comes from anticommuting $c_x$ with the factors $a_2, a_4, \dots, a_{x-2}$.
We have the following identity,
\be
\big(\omega\, c_1 + (-1)^{\frac{x-2}2}\omega^{-1} c_x\big) = \sum_{\ell=1}^{x-1}\ir^{\ell-1} a_\ell,
\ee
and the following commutation relation:
\be
\label{eq:a.eta.acomm}
\{a_j, \eta_k^\dagger\}  = 2n^{-1/2}\cos\big(\tfrac{\theta_k}2+\tfrac \pi 4\big)\eE^{-\ir (j+1/2)\theta_k}.
\ee
Using Wick's theorem, we find
\begin{subequations}
\be
\langle v^{\rm a}_x|w_0\rangle = \sum_{\ell=1}^{x-1} \ir^{\ell-1}\det M^{(\ell)} \,,
\ee
where
\be
M^{(\ell)}_{jk} = 
\left\{
\begin{array}{cl}
\{a_\ell, \eta^\dagger_k\}& j = 1,\\[0.15cm]
\{a_{2j-2}, \eta^\dagger_k\}& j = 2, \dots, \tfrac{x}2,\\[0.15cm]
\{a_{2j-1}, \eta^\dagger_k\}& j = \tfrac{x+2}2, \dots, \frac n2,
\end{array}
\right.\quad 
k \in K = 
\left\{
\begin{array}{cl}
\{\frac{4-n}4, \dots \frac n4 \} & \frac n2 \textrm{ even,}\\[0.15cm]
\{\frac{2-n}4, \dots \frac {n-2}4 \} & \frac n2 \textrm{ odd.} 
\end{array}
\right.
\ee
\end{subequations}
For $\ell$ even, the determinant is zero, as the first row is identical to the row with label $j = \frac {\ell+2}2$. The factors appearing in \eqref{eq:a.eta.acomm} that are constant or depend only on $k$ can be extracted from the denominator, and we find
\begin{subequations}
\begin{alignat}{2}
&C_{\arcdownin}(x) = \frac{\langle v^{\rm a}_x|w_0\rangle}{\langle v^{\rm a}_2|w_0\rangle} = \sum_{\ell = 1, 3, \dots, x-1} (-1)^{(\ell - 1)/2} \frac{\det \tilde M^{(\ell)}}{\det \tilde M^{(1)}},
\\[0.3cm]
&\tilde M^{(\ell)}_{jk} = 
\left\{
\begin{array}{cl}
\eE^{-\ir (\ell + 1/2) \theta_k}& j = 1,\\[0.15cm]
\eE^{-\ir (2j - 3/2) \theta_k}& j = 2, \dots, \tfrac{x}2,\\[0.15cm]
\eE^{-\ir (2j - 1/2) \theta_k}& j = \tfrac{x+2}2, \dots, \frac n2,
\end{array}
\right.
\qquad k \in K.
\end{alignat}
\end{subequations}
The inverse of the matrix $\tilde M^{(1)}$ has the entries
\be
\label{eq:invM1}
(\tilde M^{(1)})^{-1}_{kj} = \frac 2n \eE^{\ir \theta_k (2j-1/2)}, \qquad j = 1, \dots, \tfrac n2, \qquad k \in K.
\ee
The next step is to multiply $\tilde M^{(\ell)}$ by $(\tilde M^{(1)})^{-1}$ from the right: 
\be
\big[\tilde M^{(\ell)}(\tilde M^{(1)})^{-1}\big]_{jk} = 
\left\{\begin{array}{ccc}
\delta_{i, (\ell+1)/2}& j = 1,
\\[0.3cm]
\displaystyle - \frac{2 (-1)^{j+k} \eE^{\ir \phi(2j-2k-1)/n}}{n \sin(\frac \pi n (2j-2k-1))}\quad& 2 \le j \le \frac x 2,
\\[0.5cm]
\delta_{j,k} & \frac{x+2}2 \le j \le \frac n2,
\end{array}\quad k = 1, \dots, \tfrac n2.
\right.
\ee
This holds for both parities of $\frac n2$. The presence of the Kronecker-$\delta$ functions stems from the fact that the rows of $\tilde M^{(\ell)}$ and $\tilde M^{(1)}$ with labels $j=1$ and $j = \frac{x+2}2, \dots, \frac n2$ are identical. The determinant of $\tilde M^{(\ell)}(\tilde M^{(1)})^{-1}$ thus reduces to the determinant of a matrix of size $\frac{x-2}2$. Except for the $\sin(\frac \pi n (2j-2k-1))$ in the denominator, all the other factors can be extracted from the determinant. After simplification, we find
\begin{subequations}
\be
C_{\arcdownin}(x) = \bigg(\frac{-4 \ir}n \bigg)^{\frac{x-2}2} \eE^{-\frac{\ir \phi(x+2)}{2n}} \eE^{\frac{\ir \pi (x-1)(x+2)}{2n}} \sum_{\ell = 1}^{x/2} (-1)^{\ell - 1} \eE^{\frac{2 \ir \ell (\phi-\pi)}{n}}\det N^{(\ell)}
\ee 
\be
N^{(\ell)}_{jk} = 
\left\{\begin{array}{cl}
(w_j - z_k)^{-1} \quad & 1 \le k \le \ell-1,\\[0.15cm]
(w_j - z_{k+1})^{-1} \quad & \ell \le k \le \frac{x-2}2,
\end{array}\right. 
\qquad j = 1, \dots, \tfrac{x-2}2,
\ee
\end{subequations}
where $w_j = \eE^{4 \ir \pi (j-\frac12)/n}$ and $z_k = \eE^{4 \ir \pi k/n}$. The determinants are evaluated using Cauchy's identity:
\be
\label{eq:Cauchy}
\det_{j,k =1}^a \Big(\frac{1}{w_j-z_k}\Big) = \frac{\prod_{1\le j<k\le a} (w_k-w_j)(z_j-z_k)}{\prod_{j,k=1}^a (w_j-z_k)}.
\ee
This gives us a closed-form expression for $C_{\arcdownin}(x)$ in terms of a complicated sum, wherein the summand is written as a ratio of trigonometric functions. Many factors are independent of $\ell$ and can be factored from the sum. After simplification, we find
\begin{alignat}{2}
C_{\arcdownin}(x) &= \bigg(\frac{-2}n\bigg)^{\frac{x-2}2} \frac{
\displaystyle\prod_{1\le j<k\le (x-2)/2}\hspace{-0.3cm}
\sin\big(\tfrac{2 \pi}n(k- j)\big)\prod_{1\le j<k\le x/2}
\sin\big(\tfrac{2 \pi}n(j - k)\big)
}{\displaystyle \prod_{j = 1}^{(x-2)/2}\prod_{k = 1}^{x/2}\sin\big(\tfrac{2 \pi}n(j-k+\tfrac12)\big)}
\label{eq:C.exact}\\[0.3cm]
&\times \sum_{\ell = 1}^{x/2} \cos\big(\tfrac{\phi}n(2\ell - \tfrac{x+2}2)\big) 
\frac{\displaystyle\prod_{j=1}^{(x-2)/2} \sin\big(\tfrac{2\pi}n(j-\ell+\tfrac12)\big)}
{\displaystyle\prod_{\substack{j=1\\{j \neq \ell}}}^{x/2} \sin\big(\tfrac{2\pi}n(\ell - j)\big)}\ .
\nonumber
\end{alignat}

\subsection{Asymptotic behaviour}\label{sec:asy1}

We compute the asymptotic behaviour of \eqref{eq:C.exact} as $n \to \infty$ with the ratio of $x$ and $n$ fixed. We achieve this by setting $\tau = \frac{x-1}n$ and sending $n$ to infinity with $\tau$ fixed in the range $(0,1)$. 
As a first step, we write all the sine functions in terms of the cardinal sine function $s[x] = \frac{\sin(x)}x$. The products of integers that appear are then rewritten in terms of the Gamma and Barnes functions:
\be
\label{eq:Gamma}
\Gamma(z+1)=z\, \Gamma(z), \qquad G(z+1) = \Gamma(z) G(z).
\ee
The result simplifies to
\begin{alignat}{2}
C_{\arcdownin}(x) &=\frac{G(\frac x2)G(\frac{x+2}2)G^2(\frac12)}{G^2(\frac{x+1}2)} 
\prod_{j=0}^{(x-2)/2} \frac{s[\frac{2\pi}n(j+\frac12)]}{s[\frac{2\pi}nj]}
\prod_{j=0}^{(x-2)/2}\prod_{k=0}^{(x-2)/2}\frac{s[\frac{2\pi}n(j-k)]}{s[\frac{2\pi}n(j-k-\frac12)]}
\nonumber
\\[0.3cm]
\label{eq:for.asy}
&\times \sum_{\ell = 0}^{(x-2)/2} 
\frac{\cos\big(\tfrac{\phi}n(2\ell - \tfrac{x-2}2)\big)}{s[\frac{2\pi}n(\ell+\frac12)]}
\frac{\Gamma(\ell+\frac12)\Gamma(\frac{x-1}2-\ell)}{\Gamma(\ell+1)\Gamma(\frac x2-\ell)}
\prod_{j = 0}^{(x-2)/2} \frac{s[\frac{2\pi}n(j-\ell-\frac12)]}{s[\frac{2\pi}n(j-\ell)]},
\end{alignat}
where we use the convention $s[0] = 1$. The large-$z$ asymptotics of $G(z)$ is 
\begin{equation}
\log G(z) = \left(\frac{(z-1)^2}{2} -\frac{1}{12}\right) \log (z-1)-\frac{3(z-1)^2}{4} +\frac{z-1}{2} \log (2 \pi )+\frac{1}{12}-\log A+ \mathcal{O}\left( z^{-1}\right)
 \end{equation}
where $A$ is the Glaisher-Kinkelin constant.
This yields
\be
\frac{G(\frac{n\tau+1}2)G(\frac{n\tau+3}2)}{G^2(\frac{n\tau+2}2)} = \Big(\frac{n \tau}2\Big)^{1/4}+ \mathcal O(n^{-3/4}).
\ee
To compute the asymptotics of the products in \eqref{eq:for.asy}, we use the $1/n$ expansions
\begin{subequations}
\label{eq:X.asy}
\begin{alignat}{2}
&\hspace{-0.3cm}\sum_{j=0}^{(n \tau-1)/2}\hspace{-0.2cm} \log s[\tfrac{2\pi(j-a)}n- b] = n \int_0^{\tau/2}\! \dd x \log s[2\pi x-b] + \big(a+\tfrac12)(\log s[b]- \log s[\pi\tau-b]\big) + \mathcal O(n^{-1}),
\label{eq:X.asy.1}
\\
&\hspace{-0.3cm}\sum_{j,k=0}^{(n \tau-1)/2}\hspace{-0.2cm} \log s[\tfrac{2\pi(j-k-a)}n] = n^2 \int\hspace{-0.25cm} \int_0^{\tau/2}\! \dd x \dd y \log s[2 \pi(x-y)] + (a^2-\tfrac16) \log s[\pi \tau] + \mathcal O(n^{-1}),
\end{alignat}
\end{subequations}
which are obtained from the Euler-Maclaurin formula. This yields
\be
\prod_{j=0}^{(x-2)/2} \frac{s[\frac{2\pi}n(j+\frac12)]}{s[\frac{2\pi}nj]} = s[\pi \tau]^{1/2} + \mathcal O(n^{-1}),
\quad
\prod_{j,k=0}^{(x-2)/2}\frac{s[\frac{2\pi}n(j-k)]}{s[\frac{2\pi}n(j-k-\frac12)]} = s[\pi \tau]^{-1/4} + \mathcal O(n^{-1}).
\ee
To evaluate the asymptotic behaviour of the sum in \eqref{eq:for.asy}, we first change the summation index to $z = \frac \ell n$, so that the sum goes from $0$ to $\tau$ with increments of $\frac 1n$. We have the following asymptotics
\begin{subequations}
\begin{alignat}{2}
&\prod_{j = 0}^{(x-2)/2} \frac{s[\frac{2\pi}n(j-\frac12)-2 \pi z]}{s[\frac{2\pi j}n-2 \pi z]} = \frac{s[2 \pi z]^{1/2}}{s[\pi (\tau -2 z)]^{1/2}} + \mathcal O(n^{-1}).
\\
&\frac{\Gamma(nz+\frac12)\Gamma(\frac{n\tau}2-n z)}{\Gamma(nz+1)\Gamma(\frac{n\tau+1}2-n z)} = \frac1n \frac1{z^{1/2}(\frac\tau2-z)^{1/2}} + \mathcal O(n^{-2}).
\end{alignat}
\end{subequations}
The first is obtained from \eqref{eq:X.asy.1} and the second from the known asymptotic expansion of $\log \Gamma(z)$.
As $n \to \infty$, the sum has a well-defined limit in terms of an integral. Putting all the results together yields
\be
\label{eq:C.asy}
C_{\arcdownin}(x) \Big|_{x = n \tau + 1} = n^{1/4}  \sin(\pi \tau)^{1/4} (2 \pi)^{3/4}G^2(\tfrac12)  \int_0^{\tau/2} \dd z   \frac{\cos\big(\phi(2z-\frac{\tau}2)\big)}{\sin(2 \pi z)^{1/2}\sin\big(\pi (\tau - 2z)\big)^{1/2}} + \mathcal O(n^{-3/4}).
\ee
This is our final formula for the asymptotics for generic values of $\tau$ in $(0,1)$. The results are displayed in Figure 2.

We analyse its behaviour in the limiting cases $\tau \to 0^+$ and $1^-$. The former is easily evaluated to
\be
\label{eq:C2tau0}
C_{\arcdownin}(x)\Big|_{x = n \tau +1} \xrightarrow{n \gg 1,\, \tau \to 0^+} (\tfrac{n \tau}2)^{1/4}\pi\, G^2(\tfrac12).
\ee
The latter requires closer scrutiny, as the integral in \eqref{eq:C.asy} diverges for $\tau = 1$. Let us denote this integral by~$\mathcal J$. To obtain its asymptotics for $\tau \to 1^-$, we first use the reflection symmetry of the integrand under $z \to \frac\tau2 - z$:
\be
\label{eq:mathcalJ}
\mathcal J = \int_0^{\tau/2} \dd z   \frac{\cos\big(\phi(2z-\frac\tau2)\big)}{\sin(2 \pi z)^{1/2}\sin\big(\pi (\tau-2z)\big)^{1/2}} = 2 \int_0^{\tau/4} \dd z   \frac{\cos\big(\phi(2z-\frac\tau2)\big)}{\sin(2 \pi z)^{1/2}\sin\big(2 \pi (z + \frac{1-\tau}2)\big)^{1/2}}.
\ee
We substract and add a counter-term in the form of an integral wherein the numerator and denominator are simpler:
\be
\mathcal J = 2 \int_0^{\tau/4} \dd z \bigg[\frac{\cos\big(\phi(2z-\frac\tau2)\big)}{\sin(2 \pi z)^{1/2}\sin\big(2 \pi (z+\frac{1-\tau}2)\big)^{1/2}} - \frac{\cos \frac \phi 2}{2 \pi z^{1/2}(z+\frac{1-\tau}2)^{1/2}} \bigg]
+  \int_0^{\tau/4}\dd z\frac{\cos \frac\phi 2}{\pi z^{1/2}(z+\frac{1-\tau}2)^{1/2}}.
\ee
The first integral has a well-defined limit for $\tau \to 1$, whereas the second one is evaluated explicitly:
\be
\mathcal J \simeq 2 \int_0^{1/4} \dd z \bigg[\frac{\cos\big(\phi(2z-\frac12)\big)}{\sin(2 \pi z)} - \frac{\cos \frac \phi 2}{2 \pi z} \bigg]
+ \frac{2\cos(\frac{\phi}2)} \pi \log \Big(\frac{\sqrt{1-\frac\tau2}+ \sqrt \frac\tau2}{\sqrt{1-\tau}}\Big)\label{eq:J}
\ee
where $\simeq$ indicates an equality up to terms of order $(1-\tau)^1$. 
The logarithmic divergence as $\tau \to 1^-$ is explicit in the last term. This yields
\be
\label{eq:C2tau12}
C_{\arcdownin}(x)\Big|_{x = n \tau + 1} \xrightarrow{n \gg 1,\, \tau \to 1^-} n^{1/4}(1-\tau)^{1/4}2^{-1/4}G^2(\tfrac12) \Big(K-\alpha\log(\tfrac{1-\tau}2)\Big) \,,
\ee
where 
\be
\label{eq:K}
K=2 \pi \int_{0}^{1/2}\dd y \bigg(\frac{\cos\big(\phi (y-\tfrac12)\big)}{\sin \pi y} - \frac{\cos(\frac \phi 2)}{\pi y}\bigg).
\ee

Recalling from \eqref{eq:2Cs} that $C_{\arcdownout}(x) = C_{\arcdownin}(n+2-x)$, the final results for the asymptotics of these correlation functions for $1 \ll x \ll n$ are
\be\label{eq:final.asy.1}
C_{\arcdownin}(x) \xrightarrow{1 \ll x \ll n} (\tfrac x2)^{1/4}\pi\, G^2(\tfrac12),
\qquad
C_{\arcdownout}(x) \xrightarrow{1 \ll x \ll n} (\tfrac x2)^{1/4} G^2(\tfrac12)\big(\alpha \log n + K - \alpha \log (\tfrac x2)\big).
\ee
The first has a pure power-law behaviour, namely it is proportional to $x^{-2\Delta}$ with $\Delta = -\frac 18$. 
The second has the same power-law exponent as the first, but with an added logarithmic dependence upon the position $x$. The conformal interpretation of these results is discussed in \cref{sec:CFT.two.defects}.

\begin{figure}
\begin{center}
\begin{pspicture}(-3,-2)(3,2.5)
\rput(0,0){\includegraphics[height=4cm]{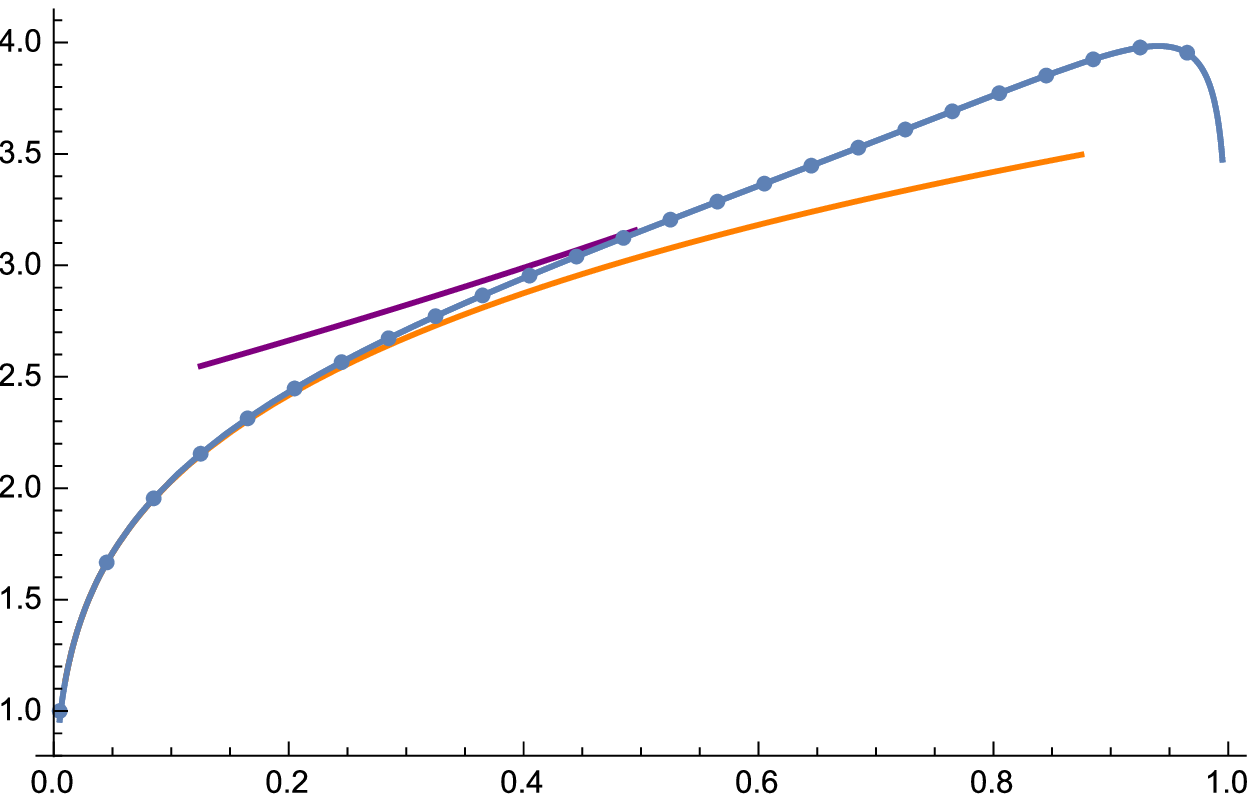}}
\rput(-2.8,2.3){$C_{\arcdownin}(x)$}
\rput(3.35,-1.3){$\tau = \frac{x-1}{n}$}
\end{pspicture}
\end{center}
Figure 2: The correlation function $C_{\arcdownin}(x)$ for $\phi = 1$. The points are the exact values for $n = 200$. The blue, yellow and purple solid curves are respectively drawn from \eqref{eq:C.asy}, \eqref{eq:C2tau0} and \eqref{eq:C2tau12}.
\end{figure}

\section{Lattice correlators for double entry points}\label{sec:four.entry.points}

\subsection{Refined partition functions}

For each loop configuration that contributes to $Z_{\fourdefdown}(x)$, the four defects can connect in six possible ways, according to the six link states of $\mathsf W_{4,0}$, see \eqref{eq:link.states.2.4}. We thus define six refined partition functions, one for each connectivity of the defects:
\be
Z_{\fourdefdown}(x) = Z_{\arcsupone}(x) + Z_{\arcsuptwo}(x) + Z_{\,\arcsupthree}(x) + Z_{\,\arcsupfour}(x) + Z_{\,\arcsupfive}(x) + Z_{\,\arcsupsix}(x).
\ee
To compute these refined partition functions, we define six more partition functions. These are defined similarly to $Z_{\fourdefdown}(x)$, but with the following link states flipped vertically and attached to the lower boundary of the rectangle:
\begin{subequations}
\begin{alignat}{2}
&v_x^{\rm c} = 
\psset{unit=0.7}
\begin{pspicture}[shift=-0.4](-0.2,-0.3)(8.0,1)
\psline[linewidth=\mince](0,0)(8.0,0)
\psarc[linecolor=blue,linewidth=\notelegant]{-}(0.4,0){0.2}{0}{180}
\psarc[linecolor=blue,linewidth=\notelegant]{-}(1.2,0){0.2}{0}{180}
\psarc[linecolor=blue,linewidth=\notelegant]{-}(2.0,0){0.2}{0}{180}
\rput(2.8,0.15){...}
\rput(0.2,-0.25){\scriptsize$1$}
\rput(4.2,-0.25){\scriptsize$x$}
\psarc[linecolor=blue,linewidth=\notelegant]{-}(3.6,0){0.2}{0}{180}
\psarc[linecolor=blue,linewidth=\notelegant]{-}(4.4,0){0.2}{0}{180}
\psarc[linecolor=blue,linewidth=\elegant]{-}(5.2,0){0.2}{0}{180}
\psarc[linecolor=blue,linewidth=\elegant]{-}(6.0,0){0.2}{0}{180}
\psarc[linecolor=blue,linewidth=\elegant]{-}(7.6,0){0.2}{0}{180}
\psline[linewidth=0.5pt,linestyle= dashed, dash = 2pt 2pt]{-}(0,0)(0,0.9)
\psline[linewidth=0.5pt,linestyle= dashed, dash = 2pt 2pt]{-}(8,0)(8,0.9)
\rput(6.8,0.15){...}
\end{pspicture}\ \ ,
\qquad
&&v_x^{\rm d} =
\psset{unit=0.7} 
\begin{pspicture}[shift=-0.4](-0.2,-0.3)(8.0,1)
\psline[linewidth=\mince](0,0)(8.0,0)
\psarc[linecolor=blue,linewidth=\notelegant]{-}(1.2,0){0.2}{0}{180}
\psarc[linecolor=blue,linewidth=\notelegant]{-}(2.0,0){0.2}{0}{180}
\rput(2.8,0.15){...}
\rput(0.2,-0.25){\scriptsize$1$}
\rput(4.2,-0.25){\scriptsize$x$}
\psbezier[linecolor=blue,linewidth=\notelegant](0.2,0)(0.2,1.5)(4.6,1.5)(4.6,0)
\psbezier[linecolor=blue,linewidth=\notelegant](0.6,0)(0.6,1.1)(4.2,1.1)(4.2,0)
\psarc[linecolor=blue,linewidth=\notelegant]{-}(3.6,0){0.2}{0}{180}
\psarc[linecolor=blue,linewidth=\notelegant]{-}(5.2,0){0.2}{0}{180}
\psarc[linecolor=blue,linewidth=\notelegant]{-}(6.0,0){0.2}{0}{180}
\psarc[linecolor=blue,linewidth=\notelegant]{-}(7.6,0){0.2}{0}{180}
\psline[linewidth=0.5pt,linestyle= dashed, dash = 2pt 2pt]{-}(0,0)(0,0.9)
\psline[linewidth=0.5pt,linestyle= dashed, dash = 2pt 2pt]{-}(8,0)(8,0.9)
\rput(6.8,0.15){...}
\end{pspicture}\ \ ,
\\[0.3cm]
&v_x^{\rm e} = 
\psset{unit=0.7}
\begin{pspicture}[shift=-0.4](-0.2,-0.3)(8.0,1)
\psline[linewidth=\mince](0,0)(8.0,0)
\psarc[linecolor=blue,linewidth=\notelegant]{-}(1.2,0){0.2}{0}{180}
\psarc[linecolor=blue,linewidth=\notelegant]{-}(2.0,0){0.2}{0}{180}
\rput(2.8,0.15){...}
\rput(0.2,-0.25){\scriptsize$1$}
\rput(4.2,-0.25){\scriptsize$x$}
\psbezier[linecolor=blue,linewidth=\notelegant](0.6,0)(0.6,1.5)(7.5,1.5)(8.03,0.35)
\psbezier[linecolor=blue,linewidth=\notelegant](0.2,0)(0.2,0.25)(-0.01,0.4)(-0.02,0.4)
\psarc[linecolor=blue,linewidth=\notelegant]{-}(3.6,0){0.2}{0}{180}
\psarc[linecolor=blue,linewidth=\notelegant]{-}(4.4,0){0.2}{0}{180}
\psarc[linecolor=blue,linewidth=\notelegant]{-}(5.2,0){0.2}{0}{180}
\psarc[linecolor=blue,linewidth=\notelegant]{-}(6.0,0){0.2}{0}{180}
\psarc[linecolor=blue,linewidth=\notelegant]{-}(7.6,0){0.2}{0}{180}
\psframe[fillstyle=solid,linecolor=white,linewidth=0pt](8,-0.02)(8.06,0.5)
\psframe[fillstyle=solid,linecolor=white,linewidth=0pt](0,-0.02)(-0.06,0.5)
\psline[linewidth=0.5pt,linestyle= dashed, dash = 2pt 2pt]{-}(0,0)(0,0.9)
\psline[linewidth=0.5pt,linestyle= dashed, dash = 2pt 2pt]{-}(8,0)(8,0.9)
\rput(6.8,0.15){...}
\end{pspicture}\ \ ,
\qquad &&
\psset{unit=0.7}
v_x^{\rm f} = 
\begin{pspicture}[shift=-0.4](-0.2,-0.3)(8.0,1)
\psline[linewidth=\mince](0,0)(8.0,0)
\psarc[linecolor=blue,linewidth=\notelegant]{-}(1.2,0){0.2}{0}{180}
\psarc[linecolor=blue,linewidth=\notelegant]{-}(2.0,0){0.2}{0}{180}
\rput(2.8,0.15){...}
\rput(0.2,-0.25){\scriptsize$1$}
\rput(4.2,-0.25){\scriptsize$x$}
\psbezier[linecolor=blue,linewidth=\notelegant](0.2,0)(0.2,0.25)(-0.01,0.4)(-0.02,0.4)
\psbezier[linecolor=blue,linewidth=\notelegant](0.6,0)(0.6,1.3)(4.2,1.3)(4.2,0)
\psarc[linecolor=blue,linewidth=\notelegant]{-}(3.6,0){0.2}{0}{180}
\psbezier[linecolor=blue,linewidth=\notelegant](4.6,0)(4.6,1.3)(8.2,1.3)(8.2,0)
\psarc[linecolor=blue,linewidth=\notelegant]{-}(5.2,0){0.2}{0}{180}
\psarc[linecolor=blue,linewidth=\notelegant]{-}(6.0,0){0.2}{0}{180}
\psarc[linecolor=blue,linewidth=\notelegant]{-}(7.6,0){0.2}{0}{180}
\psframe[fillstyle=solid,linecolor=white,linewidth=0pt](8,-0.02)(8.26,0.5)
\psframe[fillstyle=solid,linecolor=white,linewidth=0pt](0,-0.02)(-0.06,0.5)
\psline[linewidth=0.5pt,linestyle= dashed, dash = 2pt 2pt]{-}(0,0)(0,0.9)
\psline[linewidth=0.5pt,linestyle= dashed, dash = 2pt 2pt]{-}(8,0)(8,0.9)
\rput(6.8,0.15){...}
\end{pspicture}\ \ ,
\\[0.3cm]
&v_x^{\rm g} = 
\psset{unit=0.7}
\begin{pspicture}[shift=-0.4](-0.2,-0.3)(8.0,1)
\psline[linewidth=\mince](0,0)(8.0,0)
\psarc[linecolor=blue,linewidth=\notelegant]{-}(0.4,0){0.2}{0}{180}
\psarc[linecolor=blue,linewidth=\notelegant]{-}(1.2,0){0.2}{0}{180}
\psarc[linecolor=blue,linewidth=\notelegant]{-}(2.0,0){0.2}{0}{180}
\rput(2.8,0.15){...}
\rput(0.2,-0.25){\scriptsize$1$}
\rput(4.2,-0.25){\scriptsize$x$}
\psarc[linecolor=blue,linewidth=\notelegant]{-}(3.6,0){0.2}{0}{180}
\psbezier[linecolor=blue,linewidth=\notelegant](4.6,0)(4.6,1.0)(8.0,1.1)(8.06,1.1)
\psbezier[linecolor=blue,linewidth=\notelegant](-0.06,1.15)(0,1.15)(4.2,1.15)(4.2,0)
\psarc[linecolor=blue,linewidth=\notelegant]{-}(5.2,0){0.2}{0}{180}
\psarc[linecolor=blue,linewidth=\notelegant]{-}(6.0,0){0.2}{0}{180}
\psarc[linecolor=blue,linewidth=\notelegant]{-}(7.6,0){0.2}{0}{180}
\psframe[fillstyle=solid,linecolor=white,linewidth=0pt](8,-0.02)(8.06,1.2)
\psframe[fillstyle=solid,linecolor=white,linewidth=0pt](0,-0.02)(-0.06,1.2)
\psline[linewidth=0.5pt,linestyle= dashed, dash = 2pt 2pt]{-}(0,0)(0,0.9)
\psline[linewidth=0.5pt,linestyle= dashed, dash = 2pt 2pt]{-}(8,0)(8,0.9)
\rput(6.8,0.15){...}
\end{pspicture}\ \ ,
\qquad
&&v_x^{\rm h} = 
\psset{unit=0.7}
\begin{pspicture}[shift=-0.4](-0.2,-0.3)(8.0,1)
\psline[linewidth=\mince](0,0)(8.0,0)
\psarc[linecolor=blue,linewidth=\notelegant]{-}(1.2,0){0.2}{0}{180}
\psarc[linecolor=blue,linewidth=\notelegant]{-}(2.0,0){0.2}{0}{180}
\rput(2.8,0.15){...}
\rput(0.2,-0.25){\scriptsize$1$}
\rput(4.2,-0.25){\scriptsize$x$}
\psarc[linecolor=blue,linewidth=\notelegant]{-}(3.6,0){0.2}{0}{180}
\psbezier[linecolor=blue,linewidth=\notelegant](4.2,0)(4.2,1.5)(8.6,1.5)(8.6,0)
\psbezier[linecolor=blue,linewidth=\notelegant](4.6,0)(4.6,1.1)(8.2,1.1)(8.2,0)
\psbezier[linecolor=blue,linewidth=\notelegant](-0.02,0.83)(0,0.77)(0.6,0.6)(0.6,0)
\psbezier[linecolor=blue,linewidth=\notelegant](0.2,0)(0.2,0.25)(-0.01,0.4)(-0.02,0.4)
\psarc[linecolor=blue,linewidth=\notelegant]{-}(5.2,0){0.2}{0}{180}
\psarc[linecolor=blue,linewidth=\notelegant]{-}(6.0,0){0.2}{0}{180}
\psarc[linecolor=blue,linewidth=\notelegant]{-}(7.6,0){0.2}{0}{180}
\psframe[fillstyle=solid,linecolor=white,linewidth=0pt](8,-0.02)(9,1.2)
\psframe[fillstyle=solid,linecolor=white,linewidth=0pt](0,-0.02)(-0.06,1.2)
\psline[linewidth=0.5pt,linestyle= dashed, dash = 2pt 2pt]{-}(0,0)(0,0.9)
\psline[linewidth=0.5pt,linestyle= dashed, dash = 2pt 2pt]{-}(8,0)(8,0.9)
\rput(6.8,0.15){...}
\end{pspicture}\ \ .
\end{alignat}
\end{subequations}
We denote them $Z_{\arcsdownone}(x)$, $Z_{\arcsdowntwo}(x)$, $Z_{\arcsdownthree}(x)$, $Z_{\arcsdownfour}(x)$, $Z_{\arcsdownfive}(x)$ and $Z_{\arcsdownsix}(x)$. We have the following decompositions:
\be
\begin{pmatrix}
Z_{\arcsdownone}(x) \\ Z_{\arcsdowntwo}(x) \\ Z_{\arcsdownthree}(x) \\ Z_{\arcsdownfour}(x) \\ Z_{\arcsdownfive}(x) \\ Z_{\arcsdownsix}(x)
\end{pmatrix} = 
\begin{pmatrix}
0 & 0 & 0 & \alpha  & 0 & 0 \\
 0 & 0 & \alpha  & 0 & \alpha  & \alpha ^2 \\
 0 & \alpha  & 0 & 0 & \alpha ^2 & \alpha  \\
 \alpha  & 0 & 0 & 0 & 0 & 0 \\
 0 & \alpha  & \alpha ^2 & 0 & 0 & \alpha  \\
 0 & \alpha ^2 & \alpha  & 0 & \alpha  & 0
\end{pmatrix}
\begin{pmatrix}
Z_{\arcsupone}(x) \\ Z_{\arcsuptwo}(x) \\ Z_{\arcsupthree}(x) \\ Z_{\arcsupfour}(x) \\ Z_{\arcsupfive}(x) \\ Z_{\arcsupsix}(x)
\end{pmatrix}.
\ee
The matrix appearing on the right side is the Gram matrix for $\mathsf W_{4,0}$ given in \eqref{eq:Gram24}. Clearly, we have 
\begin{subequations}
\label{eq:Z4.relations}
\begin{alignat}{2}
&Z_{\arcsdownone}(x) = Z_0, \qquad && Z_{\arcsdownthree}(x) = Z_{\arcsdownfive}(x) =  Z_{\arcdownout}(n), \\[0.15cm]
&Z_{\arcsdownsix}(x) = Z_{\arcsdowntwo}(n+2-x),\qquad &&Z_{\arcsdownfour}(x) = Z_{\arcsdownfour}(n+2-x).
\label{eq:Z4.syms}
\end{alignat}
\end{subequations}
Indeed, for the link states $\arcsdownone$\,, $\arcsdownthree$ and $\arcsdownfive$\,, either the nodes $1$ and $2$ are tied together or the nodes $x$ and $x+1$ are tied together, or both. The corresponding partition functions reduce to partition functions with fewer entry points. The resulting correlation functions, obtained by dividing by $Z_0$ and taking the limit $m \to \infty$, were computed in \cref{sec:two.defs} and are independent of $x$.
There are therefore only two new independent quantities to compute: 
\be
\label{eq:C24}
C_{\arcsdowntwo}(x) =\lim_{m \to \infty} \frac{Z_{\arcsdowntwo}(x)}{Z_0}, \qquad  
C_{\arcsdownfour}(x) =\lim_{m \to \infty} \frac{Z_{\arcsdownfour}(x)}{Z_0}.
\ee
The partition functions in the numerators in \eqref{eq:C24} are expressed in terms of spin-chain overlaps as
\be
Z_{\arcsdowntwo}(x) = 2^{mn/2} \langle v^{\rm d}_x|T^m | v^{\rm a}_2\rangle, \qquad Z_{\arcsdownfour}(x) = 2^{mn/2} \langle v^{\rm f}_{x}|T^m |v^{\rm a}_2\rangle.
\ee

\subsection[Closed-form expressions in the limit $m \to \infty$]{Closed-form expressions in the limit $\boldsymbol{m \to \infty}$}

In the limit $m \to \infty$, the leading contribution for the overlaps is
\be
Z_{\arcsdowntwo}(x) \simeq 2^{mn/2} \Lambda_0^m \langle v^{\rm d}_x|w_0\rangle\langle w_0 | v^{\rm a}_2\rangle,
\qquad
Z_{\arcsdownfour}(x) \simeq 2^{mn/2} \Lambda_0^m \langle v^{\rm f}_x|w_0\rangle\langle w_0 | v^{\rm a}_2\rangle,
\ee
and therefore
\be
\label{eq:2Cs4}
C_{\arcsdowntwo}(x) =  \frac{\langle  v^{\rm d}_x|w_0\rangle}{\langle  v^{\rm a}_2|w_0\rangle}, \qquad
C_{\arcsdownfour}(x) =  \frac{\langle  v^{\rm f}_x|w_0\rangle}{\langle  v^{\rm a}_2|w_0\rangle}.
\ee
The states $v^{\rm d}_x$ and $v^{\rm f}_x$ are represented in the spin chain by
\begin{subequations}
\begin{alignat}{2}
\langle v^{\rm d}_x| &= \langle 0 | a_{n-1} a_{n-3} \cdots a_{x+2} a_{x-2} a_{x-4} \cdots a_3 \big(\omega\, c_2 + (-1)^{\frac{x-3}2}\omega^{-1} c_x\big)\big(\omega\, c_1 + (-1)^{\frac{x-1}2}\omega^{-1} c_{x+1}\big),\\[0.15cm]
\langle v^{\rm f}_x| &= \langle 0 | a_{n-1} a_{n-3} \cdots a_{x+2} a_{x-2} a_{x-4} \cdots a_3 \big(\omega\, c_2 + (-1)^{\frac{x-3}2}\omega^{-1} c_x\big)\nonumber\\&\hspace{5.90cm}\times\big(\omega^{-1}\eE^{\ir \phi/2} c_1 + (-1)^{\frac{x-1}2}\omega\, \eE^{-\ir \phi/2} c_{x+1}\big).
\end{alignat}
\end{subequations}
As a result, we have
\begin{subequations}
\begin{alignat}{2}
C_{\arcsdowntwo}(x) &= \omega^2 f_{1,2} + (-1)^{\frac{x-3}2} f_{1,x} + (-1)^{\frac{x-3}2} f_{2,x+1} + \omega^{-2} f_{x,x+1},\\[0.15cm]
C_{\arcsdownfour}(x) &= \eE^{\ir \phi/2} f_{1,2} + \eE^{\ir \phi/2}\omega^{-2}(-1)^{\frac{x-3}2} f_{1,x} + \eE^{-\ir \phi/2}\omega^{2}(-1)^{\frac{x-3}2} f_{2,x+1} + \eE^{-\ir \phi/2} f_{x,x+1},
\end{alignat}
\end{subequations}
where
\be
f_{a,b} = \frac{\det P^{(a,b)}}{\det M^{(1)}}, \qquad 
P^{(a,b)}_{ij} =\left\{ \begin{array}{ll}
\{c_a, \eta^\dagger_k\}& j = 1, \\[0.15cm]
\{c_b, \eta^\dagger_k\}& j = 2,\\[0.15cm]
\{a_{2j-3}, \eta^\dagger_k\}& j = 3, \dots, \frac{x+1}2, \\[0.15cm]
\{a_{2j-1}, \eta^\dagger_k\}& j = \frac{x+3}2, \dots, \frac n 2,
\end{array}\right. \quad k \in K.
\ee
We use \eqref{eq:a.eta.acomm}, remove from the determinant some of the factors that depend on $k$, and find
\be
f_{a,b} = \frac{\det \tilde P^{(a,b)}}{\det \tilde M^{(1)}}, \qquad 
\tilde P^{(a,b)}_{jk} =\left\{ \begin{array}{ll}
\frac{\eE^{-\ir a \theta_k }}{2 \cos(\frac{\theta_k}2 + \frac \pi 4)}& j = 1, \\[0.45cm]
\frac{\eE^{-\ir b \theta_k }}{2 \cos(\frac{\theta_k}2 + \frac \pi 4)}& j = 2,\\[0.45cm]
\eE^{-\ir\theta_k(2j-5/2)}& j = 3, \dots, \frac{x+1}2, \\[0.15cm]
\eE^{-\ir\theta_k(2j-1/2)}& j = \frac{x+3}2, \dots, \frac n 2,
\end{array}\right. \quad k \in K.
\ee
All the rows of $P^{(a,b)}$ except for the first two appear in $M^{(1)}$. Using \eqref{eq:invM1}, we find
\be
Q^{(a,b)}_{j\ell} = \big[\tilde P^{(a,b)}(\tilde M^{(1)})^{-1}\big]_{j\ell} = 
\left\{ \begin{array}{cl}
\displaystyle\frac1n \sum_{k \in K} \frac{\eE^{\ir \theta_k(2\ell - a - 1/2)}}{\cos(\frac{\theta_k}2 + \frac \pi 4)}& j = 1, \\[0.55cm]
\displaystyle\frac1n \sum_{k \in K} \frac{\eE^{\ir \theta_k(2\ell - b - 1/2)}}{\cos(\frac{\theta_k}2 + \frac \pi 4)}& j = 2,\\[0.55cm]
\delta_{j,k+1}& j = 3, \dots, \frac{x+1}2, \\[0.15cm]
\delta_{j,k}& j = \frac{x+3}2, \dots, \frac n 2.
\end{array}\right.
\ee
The determinant of $Q^{(a,b)}$ thus reduces to the determinant of a $2\times2$ matrix:
\begin{alignat}{2}
f_{a,b} &= (-1)^{\frac{x-3}2} \det 
\begin{pmatrix}
Q^{(a,b)}_{1,1} & Q^{(a,b)}_{1,(x+1)/2} \\[0.25cm]
Q^{(a,b)}_{1,2} & Q^{(a,b)}_{2,(x+1)/2}
\end{pmatrix} 
= \frac{(-1)^{\frac{x-3}2}}{n^2} \sum_{k,\ell \in K} \eE^{-\ir (a \theta_k + b \theta_\ell)}\frac{
\eE^{\frac32 \ir \theta_k + \ir \theta_\ell(x+\frac12)}-\eE^{\frac32\ir \theta_\ell + \ir \theta_k(x+\frac12)}}{\cos(\frac{\theta_k}2 + \frac \pi 4)\cos(\frac{\theta_\ell}2 + \frac \pi 4)}.
\end{alignat}
After simplifications, this yields
\begin{subequations}
\begin{alignat}{2}
C_{\arcsdowntwo}(x) &= \frac{2}{n^2} \sum_{k,\ell \in K} \bigg( 
\frac{\cos(\frac{\theta_\ell + \theta_k}2) -  \cos\big((x-\frac12)\theta_\ell - (x-\frac32)\theta_k\big)}
{\cos(\frac{\theta_k}2 + \frac \pi 4)\cos(\frac{\theta_\ell}2 + \frac \pi 4)}\nonumber\\[0.15cm]
&\hspace{2.0cm} + (-1)^{\frac{x-1}2} \frac{\sin\big((x-\frac32) \theta_\ell+\frac{\theta_k}2\big)-\sin\big((x-\frac12) \theta_\ell-\frac{\theta_k}2 \big)}
{\cos(\frac{\theta_k}2 + \frac \pi 4)\cos(\frac{\theta_\ell}2 + \frac \pi 4)}\bigg),\\[0.15cm]
C_{\arcsdownfour}(x) &= \frac{2}{n^2} \sum_{k,\ell \in K} \bigg( 
\frac{\sin(\frac{\theta_\ell + \theta_k}2+\frac \phi 2) -  \sin\big((x-\frac12)\theta_\ell - (x-\frac32)\theta_k + \frac \phi 2\big)}
{\cos(\frac{\theta_k}2 + \frac \pi 4)\cos(\frac{\theta_\ell}2 + \frac \pi 4)}\nonumber\\[0.15cm]
&\hspace{2.0cm} + (-1)^{\frac{x-3}2} \frac{\cos\big((x-\frac32) \theta_\ell+\frac{\theta_k}2+ \frac \phi 2\big)-\cos\big((x-\frac12) \theta_\ell-\frac{\theta_k}2 + \frac \phi 2 \big)}
{\cos(\frac{\theta_k}2 + \frac \pi 4)\cos(\frac{\theta_\ell}2 + \frac \pi 4)}\bigg).
\end{alignat}
\end{subequations}
Many of these double sums can be reduced to single sums. For instance,
\begin{alignat}{2}
\sum_{k,\ell \in K} \frac{\cos(\frac{\theta_\ell + \theta_k}2)}{\cos(\frac{\theta_k}2 + \frac \pi 4)\cos(\frac{\theta_\ell}2 + \frac \pi 4)} &= 
\sum_{k,\ell \in K} \frac{\cos(\frac{\theta_\ell}2 + \frac \pi 4)\sin(\frac{\theta_k}2 + \frac \pi 4) + \cos(\frac{\theta_k}2 + \frac \pi 4)\sin(\frac{\theta_\ell}2 + \frac \pi 4)}{\cos(\frac{\theta_k}2 + \frac \pi 4)\cos(\frac{\theta_\ell}2 + \frac \pi 4)}
\nonumber\\[0.3cm]
& = \sum_{k,\ell \in K} \tan(\tfrac{\theta_k}2 + \tfrac \pi 4) + \tan(\tfrac{\theta_\ell}2 + \tfrac \pi 4)= n \sum_{k \in K} \tan(\tfrac{\theta_k}2 + \tfrac \pi 4).
\end{alignat}
The other terms are simplified using similar ideas, and after some algebra we find
\begin{subequations}
\begin{alignat}{2}
C_{\arcsdowntwo}(x) &=\frac 4n \sum_{k = 0}^{(n-2)/2} \cot \big(\tfrac \pi n(k+\tfrac12) + \tfrac \phi {2n}\big)\sin^2 \Big((x-1)\big(\tfrac \pi n(k+\tfrac12) + \tfrac \phi {2n}\big)\Big),\\[0.15cm]
C_{\arcsdownfour}(x) &= \frac {4\sin (\frac \phi 2)}n \sum_{k = 0}^{(n-2)/2} \cot \big(\tfrac \pi n(k+\tfrac12) + \tfrac \phi {2n}\big)\sin^2 \Big((x-1)\big(\tfrac \pi n(k+\tfrac12) + \tfrac \phi {2n}\big)\Big)
\nonumber\\[0.15cm]
&+  \frac{2\cos(\frac \phi 2)}{n} \sum_{k=0}^{(n-2)/2} \cot \big(\tfrac \pi n(k+\tfrac12) + \tfrac \phi {2n}\big) \sin \Big(2(x-1)\big(\tfrac \pi n(k+\tfrac12) + \tfrac \phi {2n}\big)\Big)
\nonumber\\[0.15cm]
&-\frac{2\cos(\frac \phi 2)}{n^2} \bigg[
\sum_{k=0}^{(n-2)/2} \cot \big(\tfrac \pi n(k+\tfrac12) + \tfrac \phi {2n}\big) \cos \Big(2(x-1)\big(\tfrac \pi n(k+\tfrac12) + \tfrac \phi {2n}\big)\Big) \bigg]^2
\\[0.15cm]
&-\frac{2\cos(\frac \phi 2)}{n^2} \bigg[
\sum_{k=0}^{(n-2)/2} \cot \big(\tfrac \pi n(k+\tfrac12) + \tfrac \phi {2n}\big) \sin \Big(2(x-1)\big(\tfrac \pi n(k+\tfrac12) + \tfrac \phi {2n}\big)\Big) \bigg]^2
\nonumber\\[0.15cm]
& + \frac{2\cos(\frac \phi 2)}{n^2} \bigg[\sum_{k=0}^{(n-2)/2} \cot \big(\tfrac \pi n(k+\tfrac12) + \tfrac \phi {2n}\big) \bigg]^2 - \frac{\cos(\frac \phi 2)}2.\nonumber
\end{alignat}
\end{subequations}
In the second expression, some of the double sums could not be simplified to single sums and were instead rewritten as the square of single sums. We simplify these expressions further using the identity
\be
\label{eq:sin.over.cos}
\frac{\sin (\frac{\ell \xi}n)}{\cos(\frac {\xi} n)} = \sum_{t=0}^{\ell-1}(-1)^{t} \sin \big(\tfrac{\xi}n(\ell -1-2t)\big), \qquad \ell \textrm{ even}.
\ee
After some manipulations, we find
\begin{subequations}
\begin{alignat}{2}
C_{\arcsdowntwo}(x) &= \frac 4n \sum_{t = 0}^{(x-3)/2} \frac{\cos \big(\frac \phi n(2t+1)\big)}{\sin \big(\frac \pi n(2t+1)\big)},\label{eq:finalC4two}\\[0.3cm]
C_{\arcsdownfour}(x) &= \bigg(\frac{4 \sin(\frac \phi 2)}n + \frac{8 Y(n) \cos(\frac \phi 2)}{n^2} \bigg) \sum_{t = 0}^{(x-3)/2} \frac{\cos \big(\frac \phi n(2t+1)\big)}{\sin \big(\frac \pi n(2t+1)\big)}
\nonumber\\[0.3cm]
& - \frac{8 \cos(\frac \phi 2)}{n^2} \bigg(\bigg[ \sum_{t = 0}^{(x-3)/2} \frac{\cos \big(\frac \phi n(2t+1)\big)}{\sin \big(\frac \pi n(2t+1)\big)}\bigg]^2+\bigg[ \sum_{t = 0}^{(x-3)/2} \frac{\sin \big(\frac \phi n(2t+1)\big)}{\sin \big(\frac \pi n(2t+1)\big)}\bigg]^2 \bigg),
\end{alignat}
\end{subequations}
where
\be
Y(n) = \sum_{k=0}^{(n-2)/2} \cot\big(\tfrac\pi n(k+\tfrac12) + \tfrac \phi{2n}\big).
\ee

\subsection{Asymptotic behaviour}

We compute the asymptotics for $\tau = \frac{x-1}n$ with $\tau \in (0,1)$ and $n \to \infty$. We find
\begin{subequations}
\begin{alignat}{2}
&\frac 1n \sum_{t = 0}^{(x-3)/2} \frac{\cos \big(\frac \phi n(2t+1)\big)}{\sin \big(\frac \pi n(2t+1)\big)} =  \frac{1}{2\pi} \big(\log (n \tau) + \log 2 + \gamma\big) + \mathcal I_1(\tau) + \mathcal O(n^{-1}),\\[0.2cm]
&\frac 1n \sum_{t = 0}^{(x-3)/2} \frac{\sin \big(\frac \phi n(2t+1)\big)}{\sin \big(\frac \pi n(2t+1)\big)} = \mathcal I_2(\tau) + \mathcal O(n^{-1}),\\[0.2cm]
&\frac {Y(n)}n = \frac 1 \pi \big(\log n  - \log \pi - \psi(\tfrac 12 + \tfrac{\phi}{2\pi})\big)+ \mathcal O(n^{-1}),
\end{alignat}
\end{subequations}
where $\psi(z) = \Gamma'(z)/\Gamma(z)$ is the digamma function, $\gamma$ is the Euler-Mascheroni constant, and 
\be
\mathcal I_1(\tau) = \frac12\int_0^{\tau} \dd t \bigg(\frac{\cos(\phi t)}{\sin(\pi t)}-\frac1{\pi t} \bigg),\qquad
\mathcal I_2(\tau) = \frac12\int_0^{\tau} \dd t \bigg(\frac{\sin(\phi t)}{\sin( \pi t)}\bigg).
\ee
From \eqref{eq:finalC4two}, we obtain
\be
\label{eq:Cd.asy}
C_{\arcsdowntwo}(x) \Big|_{x = n \tau + 1} = \frac{2}\pi \big(\log (n \tau) + \log 2 + \gamma + 2 \pi\, \mathcal I_1(\tau)\big) + \mathcal O(n^{-1}).
\ee
The functions $\mathcal I_1(\tau)$ and $\mathcal I_2(\tau)$ have the following asymptotic behaviour for $\tau \to 0^+$:
\be
\mathcal I_1(\tau) = \frac{\pi \tau^2}8\bigg[\frac13 - \Big(\frac \phi \pi\Big)^2\bigg] + \mathcal O(\tau^4), \qquad 
\mathcal I_2(\tau) = \frac{\phi \tau}{2\pi} + \mathcal O(\tau^3).
\ee
For $\tau \to 1^-$, we have
\begin{alignat}{2}
\mathcal I_1(\tau) = &-\frac{\cos \phi}{2\pi}\log(1-\tau) + \cos^2(\tfrac \phi 2) \int_0^{1/2}\dd t\,\Big(\frac{\cos(\phi t)}{\sin(\pi t)}-\frac1{\pi t}\Big) + \frac{\sin \phi}2 \int_0^{1/2}\dd t \,\frac{\sin(\phi t)}{\sin(\pi t)} 
\nonumber\\[0.15cm]
& - \frac{\cos^2 (\tfrac\phi2)\log 2}{\pi} +  \frac 1 {2\pi}(1-\tau)(1-\phi \sin \phi)  + \mathcal O\big((1-\tau)^2\big).
\end{alignat}
The asymptotic behaviour for $C_{\arcsdowntwo}(x)$ is then given by
\begin{subequations}
\begin{alignat}{2}
&C_{\arcsdowntwo}(x)\Big|_{x = n \tau + 1} \xrightarrow{n \gg 1,\, \tau \to 0^+} \frac 2 \pi \big(\log (n\tau) + \log 2 + \gamma\big), 
\label{eq:Cd.asy.0+}
\\[0.3cm]
&C_{\arcsdowntwo}(x)\Big|_{x = n \tau + 1} \xrightarrow{n \gg 1,\, \tau \to 1^-} \frac 2 \pi \big(\log n - \cos \phi \log(1-\tau)-\cos \phi\log 2 + \gamma\big) + \widehat K, 
\label{eq:Cd.asy.1/2-}
\end{alignat}
\end{subequations}
where
\be
\widehat K = \big(2\cos (\tfrac\phi2)\big)^2 \int_0^{1/2} \dd t \bigg(\frac{\cos (\phi t)}{\sin (\pi t)} - \frac1{\pi t}\bigg) + 2 \sin \phi \int_0^{1/2} \dd t \,\frac{\sin (\phi t)}{\sin(\pi t)}.
\ee

Likewise from \eqref{eq:finalC4two}, we obtain
\begin{alignat}{2}
\label{eq:Cf.asy}
&C_{\arcsdownfour}(x) \Big|_{x = n \tau + 1} = \frac{2}{\pi^2} \Big[\cos(\tfrac \phi 2) (\log n)^2 - \cos(\tfrac \phi 2)\big((2 \pi\, \mathcal I_2(\tau))^2 + (\gamma + 2\pi\, \mathcal I_1(\tau)+\log2+\log \tau)^2 \big)
\nonumber\\[0.3cm]
&\hspace{1.2cm} + \big(\gamma + 2 \pi\, \mathcal I_1(\tau) + \log 2 + \log (n \tau)\big) \big(\pi \sin(\tfrac \phi2) -2 \cos(\tfrac \phi 2)(\log \pi + \psi(\tfrac12+\tfrac{\phi}{2\pi}))\big)\Big] + \mathcal O(n^{-1}).
\end{alignat}
In the limit $\tau \to 0^+$, this yields
\begin{alignat}{2}
\label{eq:Cf.asy.0+}
C_{\arcsdownfour}(x) \Big|_{x = n \tau + 1} \xrightarrow{n \gg 1,\, \tau \to 0^+}& \frac{2}{\pi^2} \Big[\cos(\tfrac \phi 2) (\log n)^2 - \cos(\tfrac \phi 2)\big((\gamma +\log2+\log \tau)^2 \big)
\\[0.3cm]
&+ \big(\gamma + \log 2 + \log (n \tau)\big) \big(\pi \sin(\tfrac \phi2) -2 \cos(\tfrac \phi 2)(\log \pi + \psi(\tfrac12+\tfrac{\phi}{2\pi}))\big)\Big] 
.
\nonumber
\end{alignat}
The results are plotted in the right panel of Figure 2. As expected from \eqref{eq:Z4.syms}, $C_{\arcsdownfour}(x)$ is symmetric under the transformation $\tau \to 1 - \tau$. This is not immediately obvious from \eqref{eq:Cf.asy}, but can be shown using the identities
\begin{subequations}
\begin{alignat}{2}
\mathcal I_1(1-\tau) = &- \frac{\cos \phi}{2 \pi} \Big(\log \tau -\log 2 + 2 \pi\, \mathcal I_1(\tau)\Big) - \mathcal I_2(\tau)\sin \phi - \frac{\log(\frac{1-\tau}2)}{2\pi} 
\nonumber\\[0.15cm]&
- \frac{\cos^2(\frac\phi 2)}{\pi}(\gamma + \log \pi + 2 \log 2 + \psi(\tfrac12 + \tfrac \phi{2\pi})) + \frac{\sin \phi}4,
\\[0.15cm]
\mathcal I_2(1-\tau) = &-\frac{\sin \phi}{2 \pi}\Big(\gamma + \log 2 + \log \pi + \log \tau + \psi(\tfrac12 + \tfrac \phi{2\pi})+2\pi\,\mathcal I_1(\tau)\Big) + \mathcal I_2(\tau)\cos \phi  + \tfrac12 \sin^2(\tfrac \phi 2).
\end{alignat}
\end{subequations}
Recalling from \eqref{eq:Z4.relations} that $C_{\arcsdownsix}(x) = C_{\arcsdowntwo}(n+2-x)$, the final results for the asymptotics for $1 \ll x \ll n$ are
\begin{subequations}
\begin{alignat}{2}
C_{\arcsdowntwo}(x) \xrightarrow{1 \ll x \ll n}&\,\frac 2 \pi \big(\log x + \log 2 + \gamma\big), \label{eq:finalasy2}
\\[0.3cm]
C_{\arcsdownsix}(x) \xrightarrow{1 \ll x \ll n}&\, \frac 2 \pi \big(2 \cos^2(\tfrac\phi 2)\log n - \cos \phi \log x\big)+\widehat K,\label{eq:finalasy6}
\\[0.3cm]
C_{\arcsdownfour}(x) \xrightarrow{1 \ll x \ll n}&\, \frac 2 {\pi^2}(\gamma +  \log 2 + \log x)\Big(\pi \sin(\tfrac \phi 2) - 2 \cos(\tfrac \phi 2)\big(\log n + \log \pi + \psi(\tfrac12+\tfrac{\phi}{2\pi})\big)\Big)\label{eq:finalasy4}\\[0.3cm]
& - \frac {2\cos(\tfrac \phi 2)}{\pi^2} \big(\gamma + \log 2 + \log x\big)^2.\nonumber
\end{alignat}
\end{subequations}
These lattice correlators have a purely logarithmic behavior. The power-law behavior is thus absent, consistent with the conformal weight $\Delta = 0$. We discuss the conformal interpretation of these results in \cref{sec:CFT.four.defects}.

\begin{figure}
\begin{center}
\begin{pspicture}(-4,-2)(4,2.5)
\rput(0,0){\includegraphics[height=4cm]{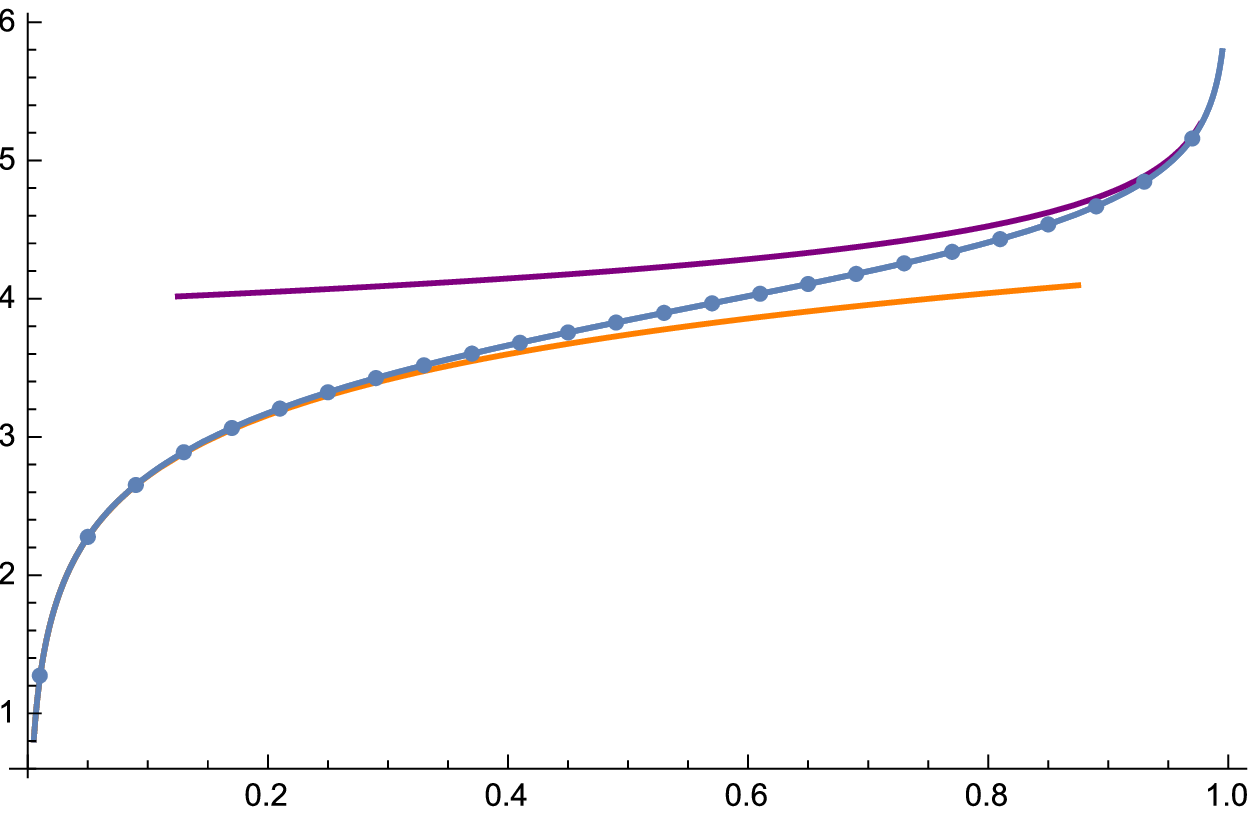}}
\rput(-2.8,2.3){$C_{\arcsdowntwo}(x)$}
\rput(3.05,-1.5){$\tau = \frac{x-1}{n}$}
\end{pspicture}
\qquad
\begin{pspicture}(-4,-2)(4,2.5)
\rput(0,0){\includegraphics[height=4cm]{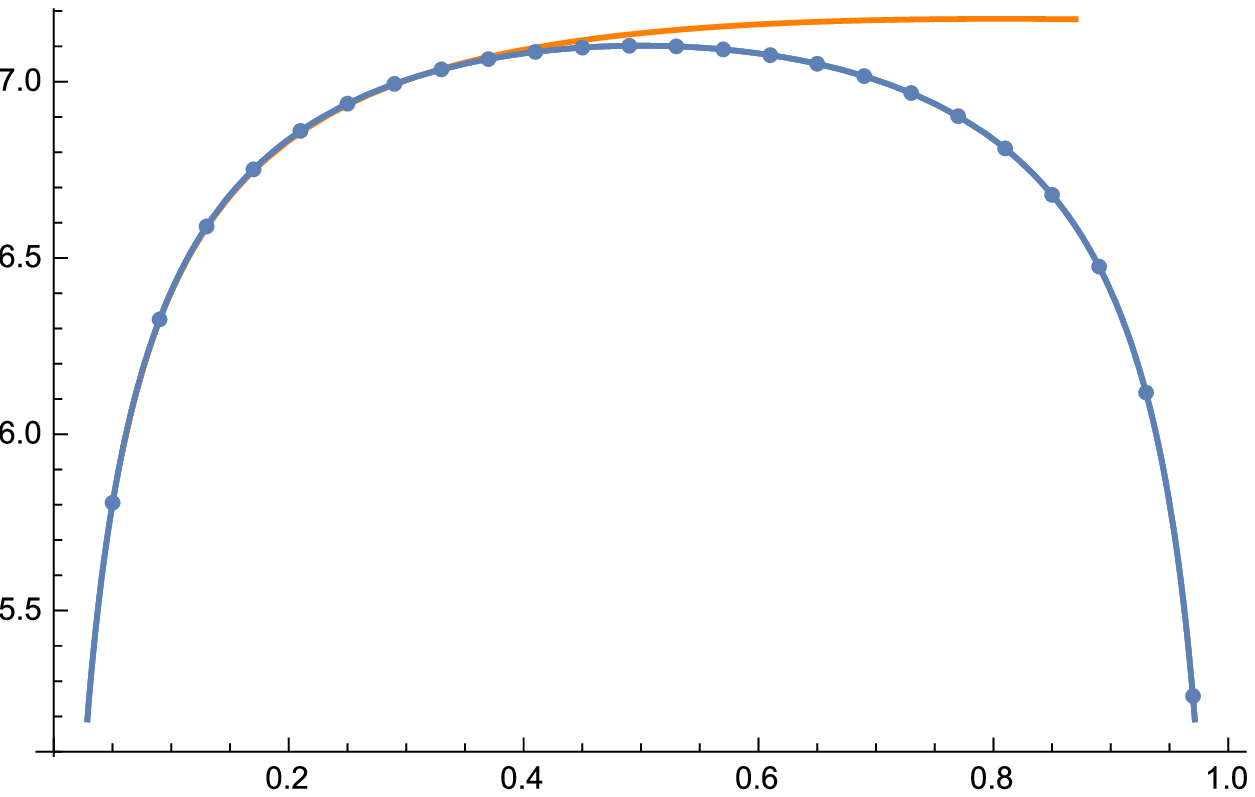}}
\rput(-2.8,2.3){$C_{\arcsdownfour}(x)$}
\rput(2.05,-1.4){$\tau = \frac{x-1}{n}$}
\end{pspicture}
\end{center}
Figure 2: The correlation functions $C_{\arcsdowntwo}(x)$ and $C_{\arcsdownfour}(x)$ for $\phi = 1$. The points are the exact values for $n = 200$. In the left panel, the blue, orange and purple solid curves are drawn from \eqref{eq:Cd.asy}, \eqref{eq:Cd.asy.0+} and \eqref{eq:Cd.asy.1/2-}. In the right panel, the blue and orange curves are drawn from \eqref{eq:Cf.asy} and \eqref{eq:Cf.asy.0+}.
\end{figure}

\section{Conformal correlators for single entry points}\label{sec:CFT.two.defects}

\subsection{Preliminaries}

In this section, we provide an interpretation of the lattice results of \cref{sec:two.defs} in terms of logarithmic conformal field theory. We denote by $\mathbb V$ the semi-infinite cylinder drawn in the plane. We claim that, in the conformal description, the lattice correlation function $C_{\arcdownin}(x)$ is a ratio of correlation functions on $\mathbb V$:
\be
\label{eq:cft.corr}
C_{\arcdownin}(x_{21}) = \lim_{x_3 \to \ir \infty}
\frac{\langle \phi_{\halfarcright}(x_1)\phi_{\halfarcleft}(x_2) \psi_\alpha(x_3,x_3^*)\rangle_{\mathbb V}}{\langle \psi_\alpha(x_3,x_3^*)\rangle_{\mathbb V}},
\ee
where $x_{21} = x_2 - x_1$.

The fields $\phi_{\halfarcright}(x)$ and $\phi_{\halfarcleft}(x)$ are chiral primary fields of conformal weight $-\frac18$ that live on the boundary. As their labels suggest, their insertion on a boundary respectively marks the start or end of a long boundary arc. In our derivation below, we assume that these fields are highest weight states $|\phi \rangle$ in irreducible representations of the Virasoro algebra with $c = -2$ and $\Delta=\Delta_{1,2} = -\frac18$.
By contrast, $\psi_\alpha(z,z^*)$ is a field that lives in the bulk and is therefore not chiral. It changes the weight of the loops encircling the point $z$ from $0$ to $\alpha$. Its action is therefore non-local. The derivation below uses the method of images \cite{C89} and works with the assumption that correlators involving $\psi_\alpha(z,z^*)$ on the upper-half plane are equal to correlators on the full plane with this field replaced by $\psi_\alpha(z)\psi_\alpha(z^*)$, where $\psi_\alpha(z)$ is a primary field. The calculations below strongly support this claim and will allow us to extract the value of the conformal weight $\Delta$ of $\psi_\alpha(z)$ as a function of $\alpha$, see \eqref{eq:Delta.solution}. 
The transfomation laws of the fields under conformal maps are
\be\label{eq:transform}
\phi(z) \mapsto \phi(y) =  \Big(\frac{\dd y}{\dd z}\Big)^{1/8} \phi(z), \qquad \psi_\alpha(z) \mapsto \psi_\alpha(y) =  \Big(\frac{\dd y}{\dd z}\Big)^{-\Delta} \psi_\alpha(z),
\ee
and the two-point functions are
\be
\label{eq:2pt.functions.phiphi}
\langle \phi(z_1)\phi(z_2)\rangle = \tilde\kappa^{\phi\phi}
(z_2-z_1)^{1/4},
\qquad
\langle \psi_\alpha(z_1)\psi_\alpha(z_2)\rangle = \frac{
\kappa^{\psi\psi}
}{(z_2-z_1)^{2\Delta}},
\ee
where $\tilde\kappa^{\phi\phi}$ and $\kappa^{\psi\psi}$ are constants. The fields $\phi_{\halfarcright}(x)$ and $\phi_{\halfarcleft}(x)$ transform identically under conformal transformations. As we shall see in \cref{sec:OPEa}, the distinction between these two fields lies in their fusion rules and in the values of the structure constants. For convenience, their labels are omitted in this section.

The conformal map from $\mathbb V$ to the upper-half plane $\mathbb H$ is
\be
z(x) = \frac{\sin\!\big(\frac \pi n(x-x_1)\big)}{\sin\!\big(\frac \pi n(x_2-x_1)\big)}\frac{\sin(\frac {\pi x_2} n)}{\sin(\frac {\pi x} n)}.
\ee
It is illustrated in \cref{fig:domains}.
The points $x_1$ and $x_2$ are sent to $0$ and $1$, whereas the point at $x = \ir \infty$ is mapped to a finite point in the upper-half plane:
\be
\label{eq:z.infty}
\lim_{x \to \ir \infty} z(x) = \eE^{\ir \pi x_1/n} \frac{\sin(\frac {\pi x_2} n)}{\sin\big(\frac \pi n(x_2-x_1)\big)}.
\ee
The derivative of this transformation is
\be
\frac{\dd z(x)}{\dd x} =  \frac\pi n\frac{\sin\!\big(\frac {\pi x_1} n\big)\sin(\frac {\pi x_2} n)}{\sin\!\big(\frac \pi n(x_2-x_1)\big)\sin(\frac {\pi x} n)^2}.
\ee

\setcounter{figure}{2}
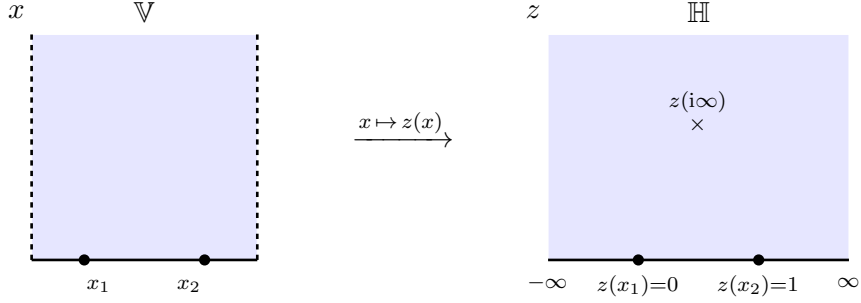
\begin{figure}
\begin{center}
\begin{pspicture}[shift=-2.0](-1.5,-0.5)(1.5,3.5)
\psframe[fillstyle=solid,fillcolor=lightlightblue,linewidth=0pt,linecolor=white](-1.5,0)(1.5,3)
\psline[linewidth=1pt]{-}(-1.5,0)(1.5,0)
\psline[linewidth=1pt,linestyle=dashed,dash=2pt 2pt]{-}(-1.5,0)(-1.5,3)
\psline[linewidth=1pt,linestyle=dashed,dash=2pt 2pt]{-}(1.5,0)(1.5,3)
\psarc[fillstyle=solid,fillcolor=black](-0.8,0){0.06}{0}{360}\rput(-0.6,-0.35){$_{x_1}$}
\psarc[fillstyle=solid,fillcolor=black](0.8,0){0.06}{0}{360}\rput(0.6,-0.35){$_{x_2}$}
\rput(-1.7,3.3){$x$}
\rput(0,3.3){$\mathbb V$}
\end{pspicture}
\hspace{1.0cm}
$\xrightarrow{x \,\mapsto\, z(x)}$
\hspace{1.0cm}
\begin{pspicture}[shift=-2.0](-2,-0.5)(2,3.5)
\psframe[fillstyle=solid,fillcolor=lightlightblue,linewidth=0pt,linecolor=white](-2,0)(2,3)
\psline[linewidth=1pt]{-}(-2,0)(2,0)
\psarc[fillstyle=solid,fillcolor=black](-0.8,0){0.06}{0}{360}\rput(-0.8,-0.35){$_{z(x_1) = 0}$}
\psarc[fillstyle=solid,fillcolor=black](0.8,0){0.06}{0}{360}\rput(0.8,-0.35){$_{z(x_2) = 1}$}
\rput(0,1.8){$_\times$}
\rput(0,2.1){$_{z(\ir \infty)}$}
\rput(-2,-0.3){$_{-\infty}$}\rput(2,-0.3){$_{\infty}$}
\rput(-2.2,3.3){$z$}
\rput(0,3.3){$\mathbb H$}
\end{pspicture}
\end{center}
\caption{The map $z(x)$ from the semi-infinite cylinder $\mathbb V$ to the upper half-plane $\mathbb H$.}
  \label{fig:domains}
  \end{figure}

\subsection{Differential equation for the four-point function}\label{sec:4pt.a}

We consider the correlation functions appearing in \eqref{eq:cft.corr}, but defined on $\mathbb H$. Using the method of images, the correlation functions on $\mathbb H$ are equal to correlation functions in the full complex plane: 
\be
\langle \phi(z_1)\phi(z_2) \psi_\alpha(z,z^*)\rangle_{\mathbb H} = \langle \phi(z_1)\phi(z_2) \psi_\alpha(z)\psi_\alpha(z^*)\rangle_{\mathbb C},
\qquad
\langle \psi_\alpha(z,z^*)\rangle_{\mathbb H} =  \langle \psi_\alpha(z)\psi_\alpha(z^*)\rangle_{\mathbb C}.
\ee
We proceed to compute the four-point function 
\be
\label{eq:G4a}
G = \langle \phi(z_1)\phi(z_2) \psi_\alpha(z_3)\psi_\alpha(z_4)\rangle_{\mathbb C}.
\ee 
We use the known techniques of conformal field theory to compute four-point functions of primary fields in the case where one of them has a null descendant \cite{BPZ84,dFMS97}.
The conformal Ward identities are
\be
\Big(\sum_{i=1}^4 \frac{\partial}{\partial z_i}\Big) G=0, 
\qquad 
\Big(\sum_{i=1}^4 z_i\frac{\partial}{\partial z_i}+ \Delta_i\Big) G=0, 
\qquad
\Big(\sum_{i=1}^4 z_i^2\frac{\partial}{\partial z_i}+ 2 \Delta_iz_i\Big) G=0,
\ee
and imply that $G$ is of the form
\be
\label{eq:G.cross}
G = \frac{(z_2-z_1)^{1/4}}{(z_4-z_3)^{2\Delta}}\tilde G(\eta),\qquad
\eta = \frac{(z_1 - z_2)(z_3 - z_4)}{(z_1-z_3)(z_2-z_4)},
\ee
where $\eta$ is the cross-ratio.
Because $|\phi\rangle$ has the null descendant $(2 L_{-1}^2 - L_{-2})|\phi\rangle$ at level two, $G$ satisfies the second-order partial differential equation
\be
\bigg[2 \frac{\partial^2}{\partial z_1^2} - \sum_{i=2}^4 \Big(\frac{1}{z_1-z_i}\frac{\partial}{\partial z_i} + \frac{\Delta_i}{(z_1-z_i)^2}\Big)\bigg] G = 0.
\ee
This differential equation is rewritten as an ordinary differential equation for $\tilde G(\eta)$:
\be
\label{eq:diffG}
\eta(1-\eta) \tilde G''(\eta) + (1-\tfrac{3\eta}2)\tilde G'(\eta) -\frac{\Delta}2 \frac \eta{1-\eta} \tilde G(\eta) = 0.
\ee
By setting
\be
\label{eq:GF}
\tilde G(\eta) = (1-\eta)^{\rho/2} F(\eta), \qquad \rho = \tfrac12(1-\sqrt{1+8\Delta}),
\ee
we find that $F(\eta)$ satisfies the hypergeometric differential equation
\be
\label{eq:hyper}
z(1-z) F''(z) + \big(c-(1+a+b)z\big)F'(z) - a b\, F(z) =0
\ee
for
\be
a = \rho,\qquad b=\tfrac12,\qquad c = 1.
\ee

For $c$ generic, the two solutions are $_2F_1(a,b;c|z)$ and $z^{1-c}\, {_2F_1}(1+a-c,1+b-c;2-c|z)$. For $c=1$, these two solutions coincide and a second independent solution \cite{OLBC10} is given by 
\be
\label{eq:second.solution}
H(a,b|z) = {_2}F_1(a,b;1|z) \log z + \sum_{k=0}^\infty \frac{(a)_k(b)_k z^k}{(k!)^2} \big(\psi(a+k)+\psi(b+k)-2\psi(1+k)\big)
\ee
where $\psi(x)$ is the digamma function. The general solution is a linear combination of these two solutions.

Before continuing, let us immediately use the transformation laws \eqref{eq:transform} of the fields to obtain the form of the lattice correlator on the cylinder. This computation is achieved by first considering the insertion points of the fields $\psi_\alpha(x_3,x_4)$ on $\mathbb V$ to be in finite positions $x_3$ and $x_4$, instead of $\pm \ir \infty$. This yields
\begin{subequations}
\begin{alignat}{2}
\label{eq:sol1}
\langle \phi(x_1)\phi(x_2) \psi_\alpha(x_3,x_4)\rangle_{\mathbb V} &= \Big(\tfrac n \pi \sin\big(\tfrac \pi n(x_2-x_1)\big)\Big)^{1/4}
 \Big(\tfrac n \pi \sin\big(\tfrac \pi n(x_4-x_3)\big)\Big)^{-2\Delta} (1-\eta)^{\rho/2} F(\eta),
\\[0.15cm]
\langle \psi_\alpha(x_3,x_4)\rangle_{\mathbb V} &= 
 \Big(\tfrac n \pi \sin\big(\tfrac \pi n(x_4-x_3)\big)\Big)^{-2\Delta} \kappa^{\psi \psi}.
\end{alignat}
\end{subequations}
For $x_4 = x_3^*$ and $x_3 \to \ir \infty$, the expression $\sin\big(\tfrac \pi n(x_4-x_3)\big)$ diverges. However, the ratio of these correlation functions, as in \eqref{eq:cft.corr}, has a well-defined limit. In the same $x_3,x_4$ limit, the cross-ratio is given by
\be
\eta = 1-\eE^{2 \pi \ir \tau}, \qquad \tau = \frac{x_2 - x_1}{n}.
\ee
The parameter $\tau$ is the same as the one that was defined in \cref{sec:asy1}. It lies in the range $(0, 1)$. As $\tau$ explores this range, $\eta$ takes complex values and draws a counterclockwise circle of unit radius around the point $\eta = 1$, starting and ending at the origin. We therefore have
\be
C_{\arcdownin}(x_{21}) = \frac{1}{\kappa^{\psi\psi}}\tilde G(1-\eE^{2 \pi \ir \tau}).
\ee

The hypergeometric function $_2F_1(a,b;c|\eta)$ has a branch cut on the real $\eta$-axis for $\eta>1$, which the circle crosses at $\eta = 2$, for $\tau = \frac12$. One way to obtain a smooth function for $C_{\arcdownin}(x_{21})$ is to search for two solutions to the differential equation, one for $\tau \in (0,\frac12)$ and another for $\tau \in (\frac12,1)$, and to impose that the function be smooth at $\tau = \frac12$.

Before proceeding with this plan, we note that, instead of \eqref{eq:GF}, one can choose to set 
\be
\tilde G(\eta) = \tilde F\Big(\frac{\eta}{4(\eta-1)}\Big).
\ee
Then the function $\tilde F(z)$ satisfies \eqref{eq:hyper} with
\be
a = \frac \rho 2, \qquad b = \frac{1-\rho}2, \qquad c = 1.
\ee
The general solution is again given by a linear combination of $_2F_1(a,b;1|z)$ and $H(a,b|z)$ for the appropriate values of $a$ and $b$. This basis of solutions is more convenient because, for $z_1 = 0$, $z_2 = 1$, $z_3 = z(\ir \infty)$ and $z_4 = z(-\ir \infty)$, the expression
\be
\frac{\eta}{4(\eta-1)} = \sin^2(\pi \tau)
\ee
is real for $\tau \in (0,1)$. In this basis, the argument $z$ of the two solutions remains real and bounded in $(0,1]$. It makes contact with the endpoint of the branch cut at $\tau = \frac12$. The derivative of $_2F_1\big(\frac \rho 2,\frac {1-\rho} 2;1|\sin^2(\pi \tau)\big)$ with respect to $\tau$ is discontinuous at $\tau = \frac 12$, and likewise for $H\big(\frac \rho 2,\frac {1-\rho} 2|\sin^2(\pi \tau)\big)$, so we search for separate solutions on the two sub-intervals. We write the general solution as
\be
\label{eq:gen.sol}
C_{\arcdownin}(x_{21}) =\big(n\sin(\pi \tau)\big)^{1/4} \times \left\{
\begin{array}{cc}
A_1\  {_2F_1}\big(\frac \rho 2,\frac {1-\rho} 2;1|\sin^2(\pi \tau)\big) + A_2\, H\big(\frac \rho 2,\frac {1-\rho}2|\sin^2(\pi \tau)\big)
&\quad \tau <\frac12,\\[0.3cm]
B_1\  {_2F_1}\big(\frac \rho 2,\frac {1-\rho} 2;1|\sin^2(\pi \tau)\big) + B_2\, H\big(\frac \rho 2,\frac {1-\rho}2|\sin^2(\pi \tau)\big)
&\quad \tau > \frac12,
\end{array}\right.
\ee
where $A_1, A_2, B_1, B_2$ are constants to be determined.

\subsection{Solving for the unknowns and verifying smoothness}

Near $z = 0$, the functions $_2F_1(a,b;1|z)$ and $H(a,b|z)$ behave as 
\be
_2F_1(a,b;1|z) = 1 + \mathcal O(z), \qquad H(a,b|z) = \log(z) + \big(\psi(a) + \psi(b) + 2 \gamma\big) + \mathcal O(z).
\ee
We fix $A_1, A_2, B_1, B_2$ by imposing that the expression \eqref{eq:gen.sol} for $C_{\arcdownin}(x_{21})$ has the same asymptotic behaviour, for $\tau \to 0^+$ and $\tau \to 1^-$, as the result we obtained from the lattice, see \eqref{eq:C2tau0} and \eqref{eq:C2tau12}. We find
\begin{subequations}
\begin{alignat}{2}
A_1 &= \frac{\pi^{3/4}}{2^{1/4}}G^2(\tfrac12), \qquad A_2 = 0, \qquad B_2 = -\frac \alpha 2 \frac{G^2(\frac12)}{(2\pi)^{1/4}}, \label{eq:A1A2B2}\\[0.2cm]
B_1 &= \frac{G^2(\frac12)}{(2\pi)^{1/4}}\bigg[K + \alpha\Big(\log 2\pi + \gamma + \tfrac12\big(\psi(\tfrac\rho2)+\psi(\tfrac{1-\rho}2)\big)\Big)\bigg].\label{eq:B1}
\end{alignat}
\end{subequations}

The only remaining unknown is the conformal weight $\Delta$. It is fixed by imposing that the solution is continuous at $\tau = \frac12$:
\be
\label{eq:continuous}
(A_1 - B_1) \ _2F_1(\tfrac{\rho}2,\tfrac{1-\rho}2;1|1) = B_2 H(\tfrac{\rho}2,\tfrac{1-\rho}2|1). 
\ee
The left side is evaluated using the relation
\be
\label{eq:2F1at1}
_2F_1(a,b;c|1) = \frac{\Gamma(c)\Gamma(c-a-b)}{\Gamma(c-a)\Gamma(c-b)}, \qquad \textrm{Re}(c-a-b)>0.
\ee
To evaluate the right side, we use the identity
\be
\label{eq:H.diff}
H(a,b|z) = \lim_{c \to 1} \Big(\frac{\partial}{\partial a}+\frac{\partial}{\partial b} + 2 \frac{\partial}{\partial c} + \psi(a) + \psi(b) - 2 \psi(c)\Big)\ _2F_1(a,b;c|z).
\ee
Combining this with \eqref{eq:2F1at1} yields
\be
H(a,b|1) = \frac{\Gamma(1-a-b)}{\Gamma(1-a)\Gamma(1-b)}\big(\psi(a) + \psi(b) - \psi(1-a)-\psi(1-b)\big).
\ee
The equality \eqref{eq:continuous} then simplifies to
\be
\label{eq:Kpsi}
K = \pi - \alpha \Big(\log 2\pi + \gamma + \tfrac12\big(\psi(\tfrac{2-\rho}2)+\psi(\tfrac{1+\rho}2)\big)\Big).
\ee

We compare this with the integral expression \eqref{eq:K} for $K$, which we now rewrite in terms of digamma functions:
\begin{alignat}{2}
K&=2 \pi \int_{0}^{1/2}\dd y \,\frac{\cos\big(\phi (y-\tfrac12)\big)-\cos(\frac \phi 2)}{\sin \pi y} + 2\pi \cos(\tfrac \phi 2) \int_{0}^{1/2}\dd y \bigg(\frac{1}{\sin(\pi y)}- \frac{1}{\pi y} \bigg)\\[0.2cm]
& = \eE^{\ir \phi/2}L(\phi) + \eE^{-\ir \phi/2}L(-\phi) + \alpha (2 \log 2 - \log \pi)\nonumber
\end{alignat}
where
\begin{alignat}{2}
L(\phi) &= 2 \pi \ir\,  \int_{0}^{1/2}\dd y\, \frac{\eE^{-\ir \phi y}-1}{\eE^{\ir \pi y}-\eE^{-\ir \pi y}} = - \int_{\eE^{\ir 0}}^{\eE^{-\ir \pi}}\frac{\dd t}{t^{1/2}}\frac{t^{\frac{\phi}{2\pi}}- 1}{1-t} = \int_{-1}^{1}\frac{\dd t}{t^{1/2}}\frac{t^{\frac{\phi}{2\pi}}- 1}{1-t}\nonumber\\[0.2cm]
& = \int_{0}^{1}\frac{\dd t}{t^{1/2}}\frac{t^{\frac{\phi}{2\pi}}- 1}{1-t} + \int_{0}^{1} \frac{\dd t}{t^{1/2}\eE^{-\ir \pi/2}}\frac{t^{\frac{\phi}{2\pi}}\eE^{-\frac{\ir \phi}2}- 1}{1+t}\nonumber\\[0.2cm]
& = \psi(\tfrac12) - \psi(\tfrac{\phi}{2\pi}+\tfrac12)- \tfrac{\ir \pi}2 + \ir \eE^{-\ir \phi/2} \big(\psi(\tfrac{\phi}{4 \pi}+\tfrac34) - \psi(\tfrac{\phi}{4 \pi}+\tfrac14) \big).
\end{alignat}
At the last step, we used the following integral formula for the digamma function:
\be
\psi(z) = -\gamma + \int_0^1 \dd t\, \frac{1-t^{z-1}}{1-t}, \qquad \textrm{Re}(z) >0.
\ee
The corresponding expression for $K$ is  simplified using properties of $\psi(z)$:
\be
\psi(1-z) - \psi(z) = \pi \cot \pi z, \qquad \psi(2z) = \log 2 + \tfrac12\big(\psi(z) + \psi(z+\tfrac12)\big), \qquad \psi(\tfrac12) = -2 \log 2 - \gamma.
\ee
We find
\be
K = \pi - \alpha \Big(\log 2\pi + \gamma + \tfrac12\big(\psi(\tfrac 34 + \tfrac{\phi}{4 \pi})+\psi(\tfrac 34 - \tfrac{\phi}{4 \pi})\big)\Big).
\ee
Comparing with \eqref{eq:Kpsi}, we see that the equality is satisfied for $\rho = \frac12 \pm \frac{\phi}{2\pi}$, both of which lead to the following value for the conformal weight:
\be
\label{eq:Delta.solution}
\Delta = \frac{(\tfrac{\phi}{\pi})^2-1}8.
\ee
This is precisely the value expected from Coulomb gas arguments. Indeed, the groundstate of the transfer matrix of the six-vertex model with periodic twisted boundary condition in the zero-magnetization sector has precisely this conformal dimension \cite{N84,PS90}. We also note that, with these values of $\rho$, the expression for $B_1$ in \eqref{eq:B1} simplifies to
\be
\label{eq:B1v2}
B_1 = - \frac{\pi^{3/4}}{2^{1/4}}G^2(\tfrac12) = -A_1.
\ee

All the unknowns have been solved for. It remains to check that the expression \eqref{eq:gen.sol} is a smooth function of $\tau$ at $\tau = \frac12$, namely
\be
\label{eq:diffs}
\lim_{\tau \to (1/2)^-}\frac{\dd^k C_{\arcdownin}(x_{21})}{\dd \tau^k} = \lim_{\tau \to (1/2)^+}\frac{\dd^k C_{\arcdownin}(x_{21})}{\dd \tau^k}, \qquad k = 1, 2, \dots \ .
\ee
It in fact suffices to check that the equality holds for $k=1$. Indeed, the function $C_{\arcdownin}(x_{21})$ satisfies the same second-order differential equation in each of the two sub-intervals. The higher-order derivatives at $\tau = \frac 12$ are then uniquely fixed by $C_{\arcdownin}(x_{21})$ and $\frac{\dd}{\dd \tau}C_{\arcdownin}(x_{21})$ at this point. The continuity of $C_{\arcdownin}(x_{21})$ and of its first derivative at $\tau = \frac 12$ therefore implies that all the higher-order derivatives are continuous and that the function is smooth at this point.

The equality \eqref{eq:diffs} for $k=1$ boils down to
\be
\lim_{\tau \to 1/2}\frac{\dd}{\dd \tau} \sum_{k=0}^\infty \frac{(a)_k(b)_k \sin^{2k}(\pi \tau)}{(k!)^2} \big(\psi(a+k)+\psi(b+k)-2\psi(1+k)\big) = 0.
\ee
This identity is readily verified by applying the derivative to each term in the sum, rewriting the result in terms of a differential operator acting on a hypergeometric function similarly to \eqref{eq:H.diff}, and finally evaluating the limit $\tau \to \frac12$ using \eqref{eq:2F1at1}.

We have therefore produced a prediction for $C_{\arcdownin}(x_{21})$ using conformal invariance. By plotting the corresponding curve alongside the exact data for $n=200$, we find that the expression \eqref{eq:gen.sol} with the constants fixed as in \eqref{eq:A1A2B2} and \eqref{eq:B1v2} precisely reproduces the solid curve in Figure 2. It is indeed non-trivial that the solution obtained from conformal invariance, which is defined separately on the two sub-intervals, is equal to the integral expression \eqref{eq:C.asy}, which is defined with a unique expression on the full interval. 

To prove that the two expressions are equal, we check that the function in \eqref{eq:C.asy} satisfies the differential equation predicted by conformal field theory. Rewriting \eqref{eq:diffG} as a second-order differential equation in $\tau$ and setting $\Delta$ to its value \eqref{eq:Delta.solution}, we find that the following identity should hold:
\be
\label{eq:J.diff}
\Big( \frac{\dd^2}{\dd \tau^2} + \pi \cot(\pi \tau) \frac{\dd}{\dd \tau} + 2 \pi^2 \Delta\Big) \mathcal J = 0,
\ee
where $\mathcal J$ is defined in \eqref{eq:mathcalJ}. With a simple rescaling of the integration variable, it is expressed as
\be
\mathcal J =\frac12 \int_0^1 \dd z \, \frac{ \tau \cos (\frac{\phi \tau}2 (1-2z))}{\sin(\pi \tau z)^{1/2} \sin(\pi \tau(1-z))^{1/2}}.
\ee
The differential operator in \eqref{eq:J.diff} can then be applied to the integrand. After a bit of work, we find that \eqref{eq:J.diff} translates to
\begin{alignat}{2}
\frac14\int_{0}^1 \dd z \frac{\dd}{\dd z}\bigg[&\frac{\pi \cos\big(\frac{\phi \tau}2(1-2z)\big)\sin\big(\pi\tau(1-z)\big)^{1/2}\sin(\pi \tau z)^{1/2}}{\sin(\pi \tau)}\bigg(\frac{(1-z)^2}{\sin\big(\pi\tau(1-z)\big)^2}-\frac{z^2}{\sin(\pi\tau z)^2}\bigg)
\nonumber\\[0.3cm]
&- \frac{2 \phi z (1-z)\sin\big(\frac{\phi \tau}2(1-2z)\big)}{\sin\big(\pi\tau(1-z)\big)^{1/2}\sin(\pi\tau z)^{1/2}}
\bigg] = 0.
\end{alignat}
The expression in the bracket vanishes at $z=0$ and $z=1$, so the equality indeed holds.

As a result, the lattice expression for $C_{\arcdownin}(x_{21})$ is a linear combination of the two hypergeometric solutions on each of the sub-intervals, as in \eqref{eq:gen.sol}. It is a continuous function of $\tau$ and, by construction of the conformal solution in terms of the lattice data, the constants $A_1$, $A_2$, $B_1$ and $B_2$ are equal to those given in \eqref{eq:A1A2B2} and \eqref{eq:B1v2}. This concludes the proof that the lattice and conformal expressions for $C_{\arcdownin}(x_{21})$ are equal.

\section{Conformal correlators for double entry points}\label{sec:CFT.four.defects}

\subsection{Preliminaries}

In the conformal picture, the lattice correlation functions $C_{\arcsdowntwo}(x)$ and $C_{\arcsdownfour}(x)$ are ratios of conformal correlation functions on the semi-infinite cylinder:
\begin{subequations}
\label{eq:cft.corr.2}
\begin{alignat}{2}
\label{eq:Ctwo.ratio}
C_{\arcsdowntwo}(x_{21}) &= \lim_{x_3 \to \ir \infty}
\frac{\langle \mu_{\twohalfarcright}(x_1)\mu_{\twohalfarcleft}(x_2) \psi_\alpha(x_3,x_3^*)\rangle_{\mathbb V}}{\langle \psi_\alpha(x_3,x_3^*)\rangle_{\mathbb V}},
\\[0.15cm]
\label{eq:Cfour.ratio}
C_{\arcsdownfour}(x_{21}) &= \lim_{x_3 \to \ir \infty}
\frac{\langle \omega_{\arcdownout}(x_1)\omega_{\arcdownout}(x_2) \psi_\alpha(x_3,x_3^*)\rangle_{\mathbb V}}{\langle \psi_\alpha(x_3,x_3^*)\rangle_{\mathbb V}}.
\end{alignat}
\end{subequations}
The evidence given below will support the following claims for the conformal interpretation:
\begin{itemize}
\item[(a)] The fields $\mu_{\twohalfarcright}(z)$ and $\mu_{\twohalfarcleft}(z)$ form a pair of highest weight states of dimension $\Delta = \Delta_{1,3} =  0$. One of these fields belongs to a Kac module and the other belongs to a staggered module.
\item[(b)] The field $\omega_{\arcdownout}(z)$ is a logarithmic highest weight state in a staggered module, also with $\Delta=  0$.
\end{itemize}
The structure and specifics of these representations are described in \cref{sec:vir}.

The transformation law for $\omega(z)$ under a conformal transformation $z \mapsto y(z)$ involves its primary partner $\varphi(z)$:
\be
\label{eq:trans.laws}
\omega(z) \mapsto \omega(y) = \bigg( \omega(z) -  \lambda \varphi(z)\log \Big ( \frac{\dd y}{\dd z}\Big)\bigg).
\ee
In CFT derivations, the value of $\lambda$ can be adjusted by changing the normalisation of the field $\omega(z)$, so
it is common to fix $\lambda=1$.
In the current context, however, the field $\omega(z)$ arises from the lattice with a natural normalisation, so we leave $\lambda$ free. We will in fact compute its value below, see \eqref{eq:lambda.and.ratio}.
The invariance under the global conformal transformations fixes \cite{G93} the two-point functions to 
\be
\label{eq:2pt.fcts.omega.id}
\langle \omega(z_1)\omega(z_2)\rangle_{\mathbb C} =\kappa^{\omega\omega}-2 \lambda \kappa^\omega \log z_{21},\qquad
\langle \omega(z_1)\varphi(z_2)\rangle_{\mathbb C} =\kappa^\omega,\qquad
\langle \varphi(z_1)\varphi(z_2)\rangle_{\mathbb C}=0.
\ee 
Likewise, the three-point functions involving two fields $\psi_\alpha$ and one field of the pair $(\varphi,\omega)$ are fixed by conformal invariance to
\be
\label{eq:3pt.psi.psi}
\langle \varphi(z_1)\psi_\alpha(z_2)\psi_\alpha(z_3)\rangle_{\mathbb C} = \frac{\kappa^{\psi \psi}}{z_{32}^{2\Delta}}, 
\qquad
\langle \omega(z_1)\psi_\alpha(z_2)\psi_\alpha(z_3)\rangle_{\mathbb C} = \frac{\kappa^{\omega \psi \psi} - \lambda \kappa^{\psi \psi}\log (\frac{z_{21}z_{31}}{z_{32}})}{z_{32}^{2\Delta}}. 
\ee

In the context of loop models, the boundary field $\varphi(z)$ has a simple interpretation: It inserts a simple half-arc $\arcdownin$ at the point $z$ on the boundary. The Dirichlet boundary conditions that we are considering consist of a macroscopic collection of these simple arcs. The insertion of $\varphi(z)$ on such a boundary is thus trivial. This implies that, in any correlator involving other fields, if $\varphi(z)$ appears, it can be simply removed, with the corresponding correlator independent of the position $z$. For instance, from \eqref{eq:2pt.fcts.omega.id}, we have $\langle \omega(z)\rangle_{\mathbb C} =\kappa^\omega$. For generic value of $\beta$, the one-point function $\langle\varphi(z)\rangle_\mathbb V$ then corresponds to the partition function of the loop model on the cylinder, with all loops (contractible and non-contractible) having fugacity $\beta$. This partition function vanishes for $\beta = 0$, implying that the one-point function of $\varphi(z)$ equals zero at this value. 

\subsection[Representations of the Virasoro algebra at $c=-2$]{Representations of the Virasoro algebra at $\boldsymbol{c=-2}$}\label{sec:vir}

This subsection discusses the structure of five modules over the Virasoro algebra with $c=-2$ and $\Delta=0$. The Loewy diagrams of these modules are given in \cref{fig:Loewy}. To start, let us consider the module $\mathsf M$ defined from the free action of the Virasoro generators $L_m$, $m\le 0$, on two highest-weight states $|\varphi\rangle$ and $|\omega\rangle$ satisfying
\be
L_0 |\varphi\rangle = 0, \qquad L_0 |\omega\rangle = \lambda |\varphi\rangle, \qquad L_m |\varphi\rangle =  L_m |\omega\rangle = 0, \qquad m>0.
\ee

The state $|\varphi\rangle$ and its descendants generate a submodule of $\mathsf M$ isomorphic to the Verma module of highest weight $\Delta = 0$, which we denote by $\mathsf V$. This submodule is reducible. In particular, it has the singular vector $L_{-1}|\varphi\rangle$ at level $1$ and the singular vector $(L_{-1}^2-2L_{-2})L_{-1}|\varphi\rangle$ at level $3$. This last state is a descendant of $L_{-1}|\varphi\rangle$, which is consistent with the known structure for this Verma module for $c=-2$ in the form of an imbricated chain \cite{KRR88,IK11}. By quotienting $\mathsf V$ by the singular state at level~$1$, we obtain the irreducible module $\mathsf I$. Likewise, by quotienting $\mathsf V$ by the singular state at level $3$, we obtain the so-called Kac module $\mathsf K$.

We now investigate the descendants of the logarithmic state $|\omega\rangle$ in $\mathsf M$. Clearly, if $\lambda = 0$, $\mathsf M$ splits as the direct sum of two copies of $\mathsf V$. For $\lambda \neq 0$, we observe that the state $L_{-1}|\omega\rangle$ is not singular because $L_1L_{-1}|\omega\rangle = 2\lambda |\varphi\rangle$. It is however {\it sub-singular}: if one quotients $\mathsf M$ by the submodule generated by $|\varphi\rangle$, then the state $L_{-1}|\omega\rangle$ becomes singular. One can check that the state 
\be
\label{eq:null.state.level.3}
(L_{-1}^3-2L_{-2}L_{-1} + L_{-3}L_0)|\omega\rangle = (L_{-1}^2-2L_{-2})L_{-1}|\omega\rangle + \lambda L_{-3}|\varphi\rangle
\ee
is also subsingular: by acting on it with any element of the Virasoro algebra that has a positive multi-index, one obtains a linear combination of states belonging to the tower of states generated by $L_{-1}|\varphi\rangle$. In Rohsiepe's original paper \cite{R96}, these free modules (like $\mathsf M$) were referred to as {\it Jordan-Verma modules}. Their structure is known \cite{R96,KR09}, and for the case we are considering, it is as depicted in the fourth panel of \cref{fig:Loewy}. 

The staggered module $\mathsf S$ is obtained by quotienting $\mathsf M$ by the submodule generated by the singular state at level $3$ given in \eqref{eq:null.state.level.3}. The three composition factors of $\mathsf S$ are irreducible modules of weights $\Delta = 0, 0,1$, and $L_0$ has rank-two Jordan cells tying the states from the two irreducible factors with $\Delta = 0$. We also note that for $\lambda = 0$, the module $\mathsf S$ decomposes as the direct sum $\mathsf I \oplus \mathsf K$. This is used in our calculations below, where we obtain a differential equation for a correlation function involving the field $|\omega\rangle$ in $\mathsf S$, and obtain the corresponding result for $\mathsf K$ by setting $\lambda = 0$.

\begin{figure}
\begin{center}
\begin{pspicture}[shift=-2.0](-0.8,-0.5)(1,4.5)
\psarc[fillstyle=solid,fillcolor=black](0,3){0.06}{0}{360}
\psarc[fillstyle=solid,fillcolor=black](0,2){0.06}{0}{360}
\psarc[fillstyle=solid,fillcolor=black](0,1){0.06}{0}{360}
\psarc[fillstyle=solid,fillcolor=black](0,0){0.06}{0}{360}
\rput(0,-0.5){.}\rput(0,-0.6){.}\rput(0,-0.7){.}
\rput(-0.8,0){$_6$}\rput(-0.8,1){$_3$}\rput(-0.8,2){$_1$}\rput(-0.8,3){$_0$}\rput(-0.8,3.6){$_\Delta$}
\rput(0,3.6){$_{|\varphi\rangle}$}
\psline{->}(0,0.8)(0,0.2)
\psline{->}(0,1.8)(0,1.2)
\psline{->}(0,2.8)(0,2.2)
\rput(-0.5,4.3){$\mathsf V$}
\end{pspicture}
\qquad%
\begin{pspicture}[shift=-2.0](-0.8,-0.5)(1,4.5)
\psarc[fillstyle=solid,fillcolor=black](0,3){0.06}{0}{360}
\psarc[fillstyle=solid,fillcolor=white,linewidth=0.5pt](0,2){0.06}{0}{360}
\psarc[fillstyle=solid,fillcolor=white,linewidth=0.5pt](0,1){0.06}{0}{360}
\psarc[fillstyle=solid,fillcolor=white,linewidth=0.5pt](0,0){0.06}{0}{360}
\rput(0,-0.5){.}\rput(0,-0.6){.}\rput(0,-0.7){.}
\rput(-0.8,0){$_6$}\rput(-0.8,1){$_3$}\rput(-0.8,2){$_1$}\rput(-0.8,3){$_0$}\rput(-0.8,3.6){$_\Delta$}
\rput(0,3.6){$_{|\varphi\rangle}$}
\multiput(0,0)(0,-1){3}{\psline[linewidth=0.5pt]{-}(0.1,2.1)(-0.1,1.9)\psline[linewidth=0.5pt]{-}(-0.1,2.1)(0.1,1.9)}
\rput(-0.5,4.3){$\mathsf I$}
\end{pspicture}
\qquad%
\begin{pspicture}[shift=-2.0](-0.8,-0.5)(1,4.5)
\psarc[fillstyle=solid,fillcolor=black](0,3){0.06}{0}{360}
\psarc[fillstyle=solid,fillcolor=black](0,2){0.06}{0}{360}
\psarc[fillstyle=solid,fillcolor=white,linewidth=0.5pt](0,1){0.06}{0}{360}
\psarc[fillstyle=solid,fillcolor=white,linewidth=0.5pt](0,0){0.06}{0}{360}
\rput(0,-0.5){.}\rput(0,-0.6){.}\rput(0,-0.7){.}
\rput(-0.8,0){$_6$}\rput(-0.8,1){$_3$}\rput(-0.8,2){$_1$}\rput(-0.8,3){$_0$}\rput(-0.8,3.6){$_\Delta$}
\rput(0,3.6){$_{|\varphi\rangle}$}
\psline{->}(0,2.8)(0,2.2)
\multiput(0,-1)(0,-1){2}{\psline[linewidth=0.5pt]{-}(0.1,2.1)(-0.1,1.9)\psline[linewidth=0.5pt]{-}(-0.1,2.1)(0.1,1.9)}
\rput(-0.5,4.3){$\mathsf K$}
\end{pspicture}
\qquad%
\begin{pspicture}[shift=-2.0](-0.8,-0.5)(2,4.5)
\psarc[fillstyle=solid,fillcolor=black](0,3){0.06}{0}{360}\psarc[fillstyle=solid,fillcolor=black](1,3){0.06}{0}{360}
\psarc[fillstyle=solid,fillcolor=black](0,2){0.06}{0}{360}\psarc[fillstyle=solid,fillcolor=black](1,2){0.06}{0}{360}
\psarc[fillstyle=solid,fillcolor=black](0,1){0.06}{0}{360}\psarc[fillstyle=solid,fillcolor=black](1,1){0.06}{0}{360}
\psarc[fillstyle=solid,fillcolor=black](0,0){0.06}{0}{360}\psarc[fillstyle=solid,fillcolor=black](1,0){0.06}{0}{360}
\rput(0,-0.5){.}\rput(0,-0.6){.}\rput(0,-0.7){.}
\rput(1,-0.5){.}\rput(1,-0.6){.}\rput(1,-0.7){.}
\rput(-0.8,0){$_6$}\rput(-0.8,1){$_3$}\rput(-0.8,2){$_1$}\rput(-0.8,3){$_0$}\rput(-0.8,3.6){$_\Delta$}
\rput(0,3.6){$_{|\varphi\rangle}$}\rput(1,3.6){$_{|\omega\rangle}$}
\psline{->}(0.8,0)(0.2,0)\psline{->}(0.8,1)(0.2,1)\psline{->}(0.8,2)(0.2,2)\psline{->}(0.8,3)(0.2,3)
\psline{->}(0,0.8)(0,0.2)\psline{->}(1,0.8)(1,0.2)
\psline{->}(0,1.8)(0,1.2)\psline{->}(1,1.8)(1,1.2)
\psline{->}(0,2.8)(0,2.2)\psline{->}(1,2.8)(1,2.2)
\psline{->}(0.8,0.2)(0.2,0.8)\psline{->}(0.8,1.2)(0.2,1.8)\psline{->}(0.8,2.2)(0.2,2.8)
\rput(0.2,4.3){$\mathsf M$}
\end{pspicture}
\qquad%
\begin{pspicture}[shift=-2.0](-0.8,-0.5)(2,4.5)
\psarc[fillstyle=solid,fillcolor=black](0,3){0.06}{0}{360}\psarc[fillstyle=solid,fillcolor=black](1,3){0.06}{0}{360}
\psarc[fillstyle=solid,fillcolor=white,linewidth=0.5pt](0,2){0.06}{0}{360}\psarc[fillstyle=solid,fillcolor=black](1,2){0.06}{0}{360}
\psarc[fillstyle=solid,fillcolor=white,linewidth=0.5pt](0,1){0.06}{0}{360}\psarc[fillstyle=solid,fillcolor=white,linewidth=0.5pt](1,1){0.06}{0}{360}
\psarc[fillstyle=solid,fillcolor=white,linewidth=0.5pt](0,0){0.06}{0}{360}\psarc[fillstyle=solid,fillcolor=white,linewidth=0.5pt](1,0){0.06}{0}{360}
\rput(0,-0.5){.}\rput(0,-0.6){.}\rput(0,-0.7){.}
\rput(1,-0.5){.}\rput(1,-0.6){.}\rput(1,-0.7){.}
\rput(-0.8,0){$_6$}\rput(-0.8,1){$_3$}\rput(-0.8,2){$_1$}\rput(-0.8,3){$_0$}\rput(-0.8,3.6){$_\Delta$}
\rput(0,3.6){$_{|\varphi\rangle}$}\rput(1,3.6){$_{|\omega\rangle}$}
\multiput(0,0)(0,-1){3}{\psline[linewidth=0.5pt]{-}(0.1,2.1)(-0.1,1.9)\psline[linewidth=0.5pt]{-}(-0.1,2.1)(0.1,1.9)}
\multiput(1,-1)(0,-1){2}{\psline[linewidth=0.5pt]{-}(0.1,2.1)(-0.1,1.9)\psline[linewidth=0.5pt]{-}(-0.1,2.1)(0.1,1.9)}
\psline{->}(0.8,3)(0.2,3)
\psline{->}(1,2.8)(1,2.2)
\psline{->}(0.8,2.2)(0.2,2.8)
\rput(0.2,4.3){$\mathsf S$}
\end{pspicture}
\end{center}
\caption{The Loewy diagrams for the modules $\mathsf V$, $\mathsf I$, $\mathsf K$, $\mathsf M$ and $\mathsf S$.}
  \label{fig:Loewy}
  \end{figure}

\subsection{Differential equation for the four-point function}\label{sec:diff.eqns.stagg}

The correlators on the right sides of \eqref{eq:cft.corr.2} involve one non-chiral field, which we rewrite as the product of two chiral fields using the method of images. We proceed to compute the correlator
\be
G = \langle \omega_1(z_1)\omega_2(z_2) \psi_\alpha(z_3,z_4)\rangle_{\mathbb H} = \langle \omega_1(z_1)\omega_2(z_2) \psi_\alpha(z_3)\psi_\alpha(z_4)\rangle_{\mathbb C}.
\ee
We work under the hypothesis that $\omega_1(z)$ and $\omega_2(z)$ are potentially different fields, and in particular that the corresponding constants $\lambda_1$ and $\lambda_2$ dictating their behaviour under conformal transformations, as in \eqref{eq:trans.laws}, may not be equal. The correlator $G$ satisfies the logarithmic version \cite{F02} of the Ward identities, namely 
\be
\Big(\sum_{i=1}^4 \frac{\partial}{\partial z_i}\Big) G = 0, \qquad
\Big(\sum_{i=1}^4 z_i\frac{\partial}{\partial z_i}+ \Delta_i + \hat\delta_i\Big) G = 0,
\qquad
\Big(\sum_{i=1}^4 z_i^2\frac{\partial}{\partial z_i}+ 2 \Delta_iz_i + 2z_i\hat\delta_i\Big) G = 0,
\ee
with $\Delta_1 = \Delta_2 = 0$ and $\Delta_3 = \Delta_4 = \Delta$ given in \eqref{eq:Delta.solution}. 
In its action on $G$, the operator $\hat\delta_i$ replaces $\omega_i$ by its primary partner $\varphi_i$ and gives zero otherwise.  These identities imply that $G$ has the following form:
\be
G = \frac1{z_{43}^{2\Delta}} \bigg[F(\eta) - \lambda_1 \kappa_2^{\omega \psi \psi} \log\Big(\frac{z_{41}z_{31}}{z_{43}}\Big)- \lambda_2 \kappa_1^{\omega \psi \psi} \log\Big(\frac{z_{42}z_{32}}{z_{43}}\Big)+ \lambda_1\lambda_2 \kappa^{\psi \psi} \log\Big(\frac{z_{41}z_{31}}{z_{43}}\Big)\log\Big(\frac{z_{42}z_{32}}{z_{43}}\Big)\bigg] \,,
\label{G3ptWardForm}
\ee
where $\eta$ is the cross-ratio, defined in \eqref{eq:G.cross}, and $\kappa_1^{\omega \psi \psi}$ and $\kappa_2^{\omega \psi \psi}$ are the constants appearing in the three-point function \eqref{eq:3pt.psi.psi} for $\omega_1(z)$ and $\omega_2(z)$. Mapping this result to the cylinder, we find
\begin{alignat}{2}
\langle \omega_1(x_1)\omega_2(x_2) \psi_\alpha(x_3,x_4)\rangle_{\mathbb V} = \Big(\frac n \pi s_{43}\Big)^{-2\Delta}\bigg[&F(\eta) 
- \lambda_1 \kappa_2^{\omega \psi \psi} \log\Big(\frac n \pi \frac{s_{41}s_{31}}{s_{43}}\Big)
- \lambda_2 \kappa_1^{\omega \psi \psi} \log\Big(\frac n \pi \frac{s_{42}s_{32}}{s_{43}}\Big)
\nonumber\\[0.15cm] &
+ \lambda_1\lambda_2 \kappa^{\psi \psi} \log\Big(\frac n \pi \frac{s_{41}s_{31}}{s_{43}}\Big)\log\Big(\frac n \pi \frac{s_{42}s_{32}}{s_{43}}\Big)\bigg] \,,
\end{alignat}
where we use the compact notation
\be
s_{ij} = \sin(\tfrac {\pi x_{ij}}  n).
\ee
For $x_4 = x_3^*$ and $x_3 \to \ir \infty$, the ratios $s_{41}s_{31}/s_{43}$ and $s_{42}s_{32}/s_{43}$ tend to $\frac{\ir}2$, and as a result we have
\be
\label{eq:ratio.of.log.correlators}
\lim_{x_3 \to \ir \infty}
\frac{\langle \omega_1(x_1)\omega_2(x_2) \psi_\alpha(x_3,x_3^*)\rangle_{\mathbb V}}{\langle \psi_\alpha(x_3,x_3^*)\rangle_{\mathbb V}} = 
\frac1{\kappa^{\psi \psi}}
\bigg[F(\eta) 
- (\lambda_1 \kappa_2^{\omega \psi \psi}+\lambda_2\kappa_1^{\omega \psi \psi})  \log\Big(\frac{\ir n}{2 \pi}\Big)
+ \lambda_1\lambda_2 \kappa^{\psi \psi} \log^2\Big(\frac{\ir n}{2 \pi}\Big)
\bigg].
\ee

In the staggered module, the state $|\omega\rangle$ has the null descendant \eqref{eq:null.state.level.3} at level $3$, which results in the following differential equation for $G$:
\be
\label{eq:diff.LG}
(\mathcal L_{-1}^3-2\mathcal L_{-2}\mathcal L_{-1} + \mathcal L_{-3} \mathcal L_0)G = 0 \,,
\ee
where
\be
\mathcal L_{-n} = \sum_{i=2}^4 \frac{(n-1) (\Delta_i + \hat \delta_i)}{(z_i-z_1)^n} - \frac{1}{(z_i - z_1)^{n-1}} \frac{\partial}{\partial z_i}.
\ee
 The partial differential equation \eqref{eq:diff.LG} simplifies to an ordinary differential equation satisfied by $F(\eta)$:
\be
\label{eq:diff.log}
\eta^3(1-\eta)^2 F'''(\eta) + 2(1-2\eta)(1-\eta)\eta^2F''(\eta) -2 \eta^2(1-(1-\Delta)\eta) F'(\eta) =(2-\eta)\lambda_1\lambda_2 \kappa^{\psi \psi}.
\ee

\subsection{Correlation function for the Kac module}\label{sec:Kac.correlators}

We solve the differential equation \eqref{eq:diff.log} for the case $\lambda_1\lambda_2 = 0$. This means that either $\lambda_1$ or $\lambda_2$ vanishes. The analysis below is independent of which one is set to zero. We choose $\lambda_1 = 0$, for which $\omega_1$ becomes a highest state $\mu$ of a Kac module $\mathsf K$, and we write $\kappa_1^{\omega \psi\psi} = \kappa_1^{\mu \psi\psi}$. Comparing \eqref{eq:Cd.asy} and \eqref{eq:ratio.of.log.correlators}, we see that the terms proportional to $\log n$ are equal for
\be
\label{eq:lambda2.relation}
\lambda_2 \frac{ \kappa_1^{\mu \psi\psi}}{\kappa^{\psi\psi}} = - \frac 2 \pi,
\ee
implying that $\lambda_2 \neq 0$. We note that, for $\tau \to 0$, the term proportional to $\log \tau$ in \eqref{eq:Cd.asy.0+} has the overall constant $2/\pi$, which is identical to the universal value predicted in \cite[equation (76)]{VJ14}. The result obtained here is however more general, as the correlation function $C_{\arcsdowntwo}(x_{21})$ involves the field $\psi_\alpha(z,z^*)$ as well.

We use the notation $F_0(\eta) = F(\eta)\big|_{\lambda_1\lambda_2 = 0}$ for the solution to \eqref{eq:diff.log} in the homogeneous case. The differential equation is in fact a second-order homogeneous differential equation for $F_0'(\eta)$. Setting $\Delta$ to its value given in \eqref{eq:Delta.solution}, the general solution is
\be
\label{eq:F0'}
F_0'(\eta) = A_1\, \frac{(1-\eta)^{-\frac12(1-\frac\phi\pi)}}\eta +  A_2\, \frac{(1-\eta)^{-\frac12(1+\frac\phi\pi)}}\eta \,,
\ee
where $A_1$ and $A_2$ are constants to be determined. According to \eqref{eq:Ctwo.ratio} and \eqref{eq:ratio.of.log.correlators}, we have
\be
C_{\arcsdowntwo}(x_{21}) = \frac{1}{\kappa^{\psi \psi}}F_0(\eta), \qquad \eta = 1 - \eE^{2 \pi \ir \tau},
\ee
and therefore
\be
\label{eq:dCa}
\frac{\partial C_{\arcsdowntwo}(x_{21})}{\partial \tau} =\frac1{\kappa^{\psi \psi}} \frac{\partial F_0}{\partial \eta} \frac{\partial \eta}{\partial \tau} =\frac{\pi}{\kappa^{\psi \psi}} \bigg(\frac{A_1 \eE^{\ir \phi \tau}+A_2 \eE^{-\ir \phi \tau}}{\sin{\pi \tau}}\bigg).
\ee

Comparing this with the lattice result, we see that the derivative of the expression \eqref{eq:Cd.asy} with respect to $\tau$ is easily evaluated because $\tau$ only appears in the upper integration limit:
\be
\label{eq:dCb}
\frac{\partial C_{\arcsdowntwo}(x_{21})}{\partial \tau} = \frac {2 \cos(\phi \tau)}{\sin(\pi \tau)}.
\ee
We find that \eqref{eq:dCa} and \eqref{eq:dCb} coincide for 
\be 
A_1 = A_2 = \frac{\kappa^{\psi \psi}}\pi.
\ee 
This consistency between the lattice and conformal results supports our claim that the fields $\mu_{\twohalfarcright}$ and $\mu_{\twohalfarcleft}$ form a pair of heighest-weight states of dimension $\Delta = 0$, one of which belongs to a Kac module~$\mathsf K$ and the other to a staggered module $\mathsf S$.

\subsection{Correlation function for the staggered module}

For the correlator \eqref{eq:Cfour.ratio}, the four-point correlator contains two copies of the same logarithmic field: $\omega_1 = \omega_2 = \omega_{\arcdownout}$\,. We therefore write $\lambda_1 = \lambda_2 = \lambda$ and $\kappa^{\omega\psi\psi}_1 = \kappa^{\omega\psi\psi}_2 = \kappa^{\omega\psi\psi}$. For $\lambda_1\lambda_2 = \lambda^2 \neq 0$, the differential equation \eqref{eq:diff.log} has an extra inhomogeneous term. The complete solution is equal to the two-parameter solution of the homogeneous equation plus a second function that takes into account the inhomogeneous term:\footnote{This solution to the differential equation \eqref{eq:diff.log} was obtained using Mathematica.}
\begin{alignat}{2}
\label{eq:diff.F.staggered}
F'(\eta) = F'_0(\eta) + \frac{2 \lambda^2 \kappa^{\psi \psi}}{\frac \phi \pi(1-\eta)\eta^2}&\bigg[2 \bigg(\frac{_2F_1(\frac12+\frac\phi{2\pi}, 1, \frac52+\frac\phi{2\pi}\big| \frac1{1-\eta})}{3+\frac \phi \pi} - \frac{_2F_1(\frac12-\frac\phi{2\pi}, 1, \frac52-\frac\phi{2\pi}\big| \frac1{1-\eta})}{3-\frac \phi \pi}\bigg)
\nonumber\\[0.15cm]
&\hspace{-0.2cm}-\eta \bigg(\frac{_2F_1(\frac12+\frac\phi{2\pi}, 1, \frac32+\frac\phi{2\pi}\big| \frac1{1-\eta})}{1+\frac \phi \pi} - \frac{_2F_1(\frac12-\frac\phi{2\pi}, 1, \frac32-\frac\phi{2\pi}\big| \frac1{1-\eta})}{1-\frac \phi \pi}\bigg)\bigg],
\end{alignat}
where the constants $A_1$ and $A_2$ of $F'_0(\eta)$ in \eqref{eq:F0'} remain to be fixed.

We fix the remaining unknowns using the lattice expression. Comparing \eqref{eq:Cf.asy} with \eqref{eq:ratio.of.log.correlators}, we see that the $\log^2(n)$ and $\log(n)$ terms are identical for
\be
\label{eq:lambda.and.ratio}
\lambda = \frac{(2 \cos \frac \phi 2)^{1/2}}{\pi}, \qquad \frac{\kappa^{\omega\psi\psi}}{\kappa^{\psi\psi}} = \frac{1}{\pi(2 \cos \frac \phi 2)^{1/2}}\Big(2\cos (\tfrac \phi 2)\big(\tfrac{\ir \pi}2-\log 2 + \psi(\tfrac12 + \tfrac \phi{2\pi})\big)- \pi  \sin (\tfrac{\phi}2) \Big).
\ee
It may seem odd that $\lambda$ depends on $\phi$, as this parameter dictates the behaviour of the boundary field $\omega_{\arcdownout}(z)$, which a priori one expects not to depend on $\alpha$. This will be discussed further in \cref{sec:OPE}. We also note that the structure constant $\kappa^{\omega\psi\psi}$ originates from the correlator, given in \eqref{eq:3pt.psi.psi}, of a bulk field and two boundary fields. So it is perhaps not surprising that the ratio $\kappa^{\omega\psi\psi}/\kappa^{\psi\psi}$ is complex. The remaining constant factors in \eqref{eq:Cf.asy}, which are independent of $n$ and $\tau$, are accounted for in the conformal solution by the integration constant that appears when one integrates \eqref{eq:diff.F.staggered} to obtain $F(\eta)$.

To fix the constants $A_1$ and $A_2$, we compare the asymptotic behaviour of the lattice and conformal expressions for $F'(\eta)$ in the neighbourhood of $\eta=1$, corresponding to $\tau \to 0^+$. For the lattice expression, \eqref{eq:Cf.asy.0+} yields
\be
\label{eq:dC.asy.1}
\frac{\partial C_{\arcsdownfour}}{\partial \tau} \xrightarrow{n \gg 1, \tau \to 0^+} \frac{2}{\pi^2 \tau} \bigg[\pi \sin(\tfrac \phi 2)-2 \cos(\tfrac \phi 2)\Big(\gamma + \log 2 + \log \pi + \psi(\tfrac 12 + \tfrac \phi{2\pi})+ \log \tau\Big)\bigg].
\ee
For the conformal expression, we extract the behaviour of the hypergeometric functions in \eqref{eq:diff.F.staggered} at $z=1$ by using the relations
\begin{subequations}
\begin{alignat}{2}
_2F_1(a,b,a+b|z) &= - \frac{\Gamma(a+b)}{\Gamma(a)\Gamma(b)}H(a,b|1-z), \\[0.15cm]
_2F_1(a,b,a+b+1|z) &= - \frac{\Gamma(a+b+1)}{\Gamma(a+1)\Gamma(b+1)}(1-z)\frac{\partial H(a,b|1-z)}{\partial z},
\end{alignat}
\end{subequations}
and the series expansion \eqref{eq:second.solution} of $H(a,b|z)$. This yields
\begin{alignat}{2}
\frac{\partial C_{\arcsdownfour}}{\partial \tau} = \frac 1{\kappa^{\psi\psi}} \frac{\partial \eta}{\partial \tau}\frac{\partial F}{\partial \eta}
=& -\frac 1 \tau \Big(A_1 + A_2 + 2 \lambda^2\big(\gamma + \log \pi + \log 2 + \psi(\tfrac12 + \tfrac \phi{2\pi}) + \log \tau + \tfrac {\ir \pi}2 + \tfrac \pi 2 \tan(\tfrac \phi{2\pi})\big)\Big)
\nonumber\\[0.1cm]
& - 2 \ir \phi\,\big(A_1 - A_2 + \pi \lambda^2 \tan (\tfrac \phi 2)\big).
\label{eq:dC.asy.2}
\end{alignat}
With $\lambda$ evaluated to its value in \eqref{eq:lambda.and.ratio}, we find that \eqref{eq:dC.asy.1} and \eqref{eq:dC.asy.2} are equal for
\be
\frac{A_1}{\kappa^{\psi\psi}} = \frac{\eE^{-\ir \phi/2}}{\ir \pi}, \qquad \frac{A_2}{\kappa^{\psi\psi}} = \frac{\eE^{\ir \phi/2}}{\ir \pi}.
\ee
The final conformal expression for $\frac{\partial C_{\arcsdownfour}}{\partial \tau}$ is
\begin{alignat}{2}
\frac{\partial C_{\arcsdownfour}}{\partial \tau} =&  \frac{8 \cos (\tfrac \phi 2)}{ \ir\phi (1-\eE^{2\pi \ir \tau})^2}\bigg[2 \bigg(\frac{_2F_1(\frac12+\frac\phi{2\pi}, 1, \frac52+\frac\phi{2\pi}\big| \eE^{-2 \pi \ir \tau})}{3+\frac \phi \pi} - \frac{_2F_1(\frac12-\frac\phi{2\pi}, 1, \frac52-\frac\phi{2\pi}\big| \eE^{-2 \pi \ir \tau})}{3-\frac \phi \pi}\bigg)
\nonumber\\[0.15cm]
&-(1-\eE^{2 \pi \ir \tau}) \bigg(\frac{_2F_1(\frac12+\frac\phi{2\pi}, 1, \frac32+\frac\phi{2\pi}\big| \eE^{-2 \pi \ir \tau})}{1+\frac \phi \pi} - \frac{_2F_1(\frac12-\frac\phi{2\pi}, 1, \frac32-\frac\phi{2\pi}\big| \eE^{-2 \pi \ir \tau})}{1-\frac \phi \pi}\bigg)\bigg]
\nonumber\\[0.15cm]
&+\frac{2\ir \cos\big(\phi(\tau-\tfrac12)\big)}{\sin(\pi \tau)}.
\end{alignat}

It is in fact not obvious that this function is real for $\tau \in (0,1)$, but indeed it is. Plotting this function alongside the lattice expression for $\frac{\partial C_{\arcsdownfour}}{\partial \tau}$, we find that the two functions precisely coincide, on the full interval. To prove the equality, we show that the lattice expression also satisfies the differential equation \eqref{eq:diff.log}. In terms of $\tau$, this equation reads
\be
\label{eq:tau.diff}
\Big(\frac{\partial^2}{\partial \tau^2} + 2 \pi \cot(\pi \tau) \frac{\partial}{\partial \tau} + 8 \pi^2 \Delta \Big) \frac{\partial C_{\arcsdownfour}}{\partial \tau} = 2 \pi^3 \lambda^2 \frac{\cos(\pi \tau)}{\sin^3( \pi \tau)}.
\ee
To check this, we take the derivative of \eqref{eq:Cf.asy} and simplify it to
\begin{alignat}{2}
\frac{\partial C_{\arcsdownfour}}{\partial \tau} = & -8 \cos(\tfrac \phi 2) \int_0^{\tau} \dd t\, \frac{\sin (\frac{t \phi}2)\sin\!\big((\tau-\frac t2)\phi\big)}{\sin(\pi \tau)\sin(\pi t)} + 2\sin(\tfrac \phi 2)\frac{\cos(\phi \tau)}{\sin(\pi \tau)}
\nonumber\\[0.1cm]
&- \frac{4 \cos (\frac\phi2)}{\pi}\frac{\cos(\phi \tau)}{\sin(\pi \tau)}\big(\gamma + 2 \log 2 + \log \tan(\tfrac{\pi \tau}2) + \psi(\tfrac12+ \tfrac\phi{2\pi})\big).
\label{eq:simplified.dC4}
\end{alignat}
Applying the operator that appears on the left-hand side of \eqref{eq:tau.diff} to the terms proportional to $\frac{\cos(\phi\tau)}{\sin(\pi \tau)}$ gives zero, as this is a solution of the homogeneous differential equation discussed in \cref{sec:Kac.correlators}. For the remaining terms, after simplifications we find
\begin{alignat}{2}
\Big( \frac{\partial^2}{\partial \tau^2} &+ 2 \pi \cot(\pi \tau) \frac{\partial}{\partial \tau} + 8 \pi^2 \Delta \Big)
\bigg[\frac{4 \cos (\frac \phi 2)}\pi \frac{\cos(\phi \tau)}{\sin (\pi \tau)}\log \tan(\tfrac{\pi \tau}2)\bigg]
\nonumber\\[0.15cm]
 & = 8 \pi \cos(\tfrac \phi 2) \bigg(\frac{\cos(\pi \tau)\sin(\frac{\phi \tau}2)^2}{\sin^3(\pi\tau)} - \frac \phi \pi \frac{\sin(\phi \tau)}{\sin^2(\pi\tau)}\bigg) - 2 \pi^3 \lambda^2 \frac{\cos(\pi \tau)}{\sin^3(\pi \tau)}
\end{alignat}
and
\begin{alignat}{2}
&\int_0^1 \dd t\Big( \frac{\partial^2}{\partial \tau^2} + 2 \pi \cot(\pi \tau) \frac{\partial}{\partial \tau} + 8 \pi^2 \Delta \Big)
\bigg[\frac{\tau \sin(\frac{\phi \tau t}2)\sin\big(\phi \tau(1-\frac t2)\big)}{\sin(\pi \tau t)\sin(\pi \tau)}\bigg]
\nonumber\\[0.15cm]
& =\int_0^1 \dd t \frac{\dd}{\dd t} \bigg[\frac{t}{\sin(\pi \tau) \sin(\pi \tau t)}\bigg(\phi \sin(\phi\tau)-(1-\tfrac t 2)\phi \sin\big(\phi \tau(1-t)\big) 
\nonumber\\[0.15cm]
&\hspace{5.5cm}- \pi t \cot(\pi \tau t)\sin(\tfrac{\phi \tau t}2)\sin\big(\phi \tau(1-\tfrac t 2)\big)\bigg)\bigg]
\nonumber\\[0.0cm]& 
= \phi \frac{\sin(\phi \tau)}{\sin^2(\pi \tau)} - \pi \frac{\cos(\pi \tau)\sin^2(\frac{\phi\tau}2)}{\sin^3(\pi \tau)}.
\end{alignat}
Combining these results, we confirm that the lattice expression \eqref{eq:simplified.dC4} indeed satisfies \eqref{eq:tau.diff}. The lattice and conformal solutions therefore satisfy the same differential equation. Recalling that the constants $A_1$ and $A_2$ were fixed so that the lattice and conformal expressions have the same asymptotics, we conclude that the two expressions coincide.

\section{Operator product expansion and structure constants}\label{sec:OPE}

\subsection[Analysis for the fields $\phi$]{Analysis for the fields $\boldsymbol{\phi(z)}$}\label{sec:OPEa}

In this section, we use the lattice expressions for the correlators $C_{\arcdownin}(x)$ and $C_{\arcdownout}(x)$ to obtain conformal structure constants and ratios thereof. We claim that the fields $\phi_{\halfarcright}(z)$ and $\phi_{\halfarcleft}(z)$ satisfy the following operator product expansions (OPEs): 
\begin{subequations}
\label{eq:OPEs.phi}
\begin{alignat}{2}
\phi_{\halfarcright}(z_1)\phi_{\halfarcleft}(z_2) &= z_{21}^{1/4} \kappa^{\phi\phi}_{\arcdownin}\varphi(z_2) + \dots\ ,\\[0.2cm] 
\phi_{\halfarcleft}(z_1)\phi_{\halfarcright}(z_2) &= z_{21}^{1/4} \Big(\kappa^{\phi\phi\omega}_{\arcdownout} \big(\omega_{\arcdownout}(z_2) + \lambda\, \varphi(z_2) \log z_{21}\big)  + \kappa^{\phi\phi}_{\arcdownout}\, \varphi(z_2)\Big) +\dots\ .
\end{alignat}
\end{subequations}
From \eqref{eq:2pt.fcts.omega.id}, this yields the following two-point functions:
\be
\langle \phi_{\halfarcright}(z_1)\phi_{\halfarcleft}(z_2) \rangle_{\mathbb C} = 0, \qquad 
\langle \phi_{\halfarcleft}(z_1)\phi_{\halfarcright}(z_2) \rangle_{\mathbb C} = z_{21}^{1/4} \kappa^{\phi\phi\omega}_{\arcdownout} \kappa^\omega.
\ee
We therefore see that the constant $\tilde \kappa^{\phi\phi}$ appearing in \eqref{eq:2pt.functions.phiphi} vanishes for the first correlator, but equals $\kappa^{\phi\phi\omega}_{\arcdownout}\kappa^\omega$ for the second correlator.

With these OPEs, we can easily obtain the behaviour of the four-point functions studied in \cref{sec:4pt.a} in the regime where $z_1$ approaches $z_2$. Indeed, in this regime, the four-point functions reduce to two- and three-point functions of the form \eqref{eq:3pt.psi.psi}, and we find
\begin{subequations}
\begin{alignat}{2}
 \langle \phi_{\halfarcright}(z_1)\phi_{\halfarcleft}(z_2) \psi_\alpha(z_3)\psi_\alpha(z_4)\rangle_{\mathbb C} &\xrightarrow{z_1 \to z_2} 
 \frac{z_{21}^{1/4}}{z_{43}^{2\Delta}} \kappa^{\phi\phi}_{\arcdownin} \kappa^{\psi\psi},
\\[0.15cm]
 \langle \phi_{\halfarcleft}(z_1)\phi_{\halfarcright}(z_2) \psi_\alpha(z_3)\psi_\alpha(z_4)\rangle_{\mathbb C} &\xrightarrow{z_1 \to z_2}
  \frac{z_{21}^{1/4}}{z_{43}^{2\Delta}} \Big[\kappa^{\phi\phi\omega}_{\arcdownout}\kappa^{\omega\psi\psi} +\kappa^{\phi\phi\omega}_{\arcdownout}\kappa^{\psi\psi} \lambda \log \Big(\frac{z_{21}z_{43}}{z_{32}z_{42}} \Big) + \kappa^{\phi\phi}_{\arcdownout}\kappa^{\psi\psi}\,  \Big].
\end{alignat}
\end{subequations}
The corresponding correlators on the cylinder are obtained by applying the transformation laws \eqref{eq:transform}. The regime $z_1 \to z_2$ then corresponds to $x_2-x_1 \ll n$, or equivalently to $\tau \to 0^+$. In this regime, we also have $s_{21} \simeq \pi \tau$ and $\frac{x_1}n \simeq \frac {x_2}n$, where $\simeq$ indicates an equality up to corrections of order $\tau$. We obtain expressions for the correlators $C_{\arcdownin}(x_{21})$ and $C_{\arcdownout}(x_{21})$ by dividing by $\langle \psi_\alpha(x_3)\psi_\alpha(x_4)\rangle_{\mathbb V}$, setting $x_4 = x_3^*$ and sending $x_3$ to $\pm \ir \infty$. This yields
\be
C_{\arcdownin}(x_{21}) \xrightarrow{\tau \to 0^+} (n \tau)^{1/4} \kappa^{\phi\phi}_{\arcdownin}, \qquad
C_{\arcdownout}(x_{21}) \xrightarrow{\tau \to 0^+} (n \tau)^{1/4}\Big[\kappa^{\phi\phi\omega}_{\arcdownout}\frac{\kappa^{\omega\psi\psi}}{\kappa^{\psi\psi}} +\kappa^{\phi\phi\omega}_{\arcdownout} \lambda \log (-2 \pi \ir \tau) + \kappa^{\phi\phi}_{\arcdownout}\,  \Big].
\ee
We compare these expressions with \eqref{eq:C2tau0} and \eqref{eq:C2tau12}, assuming that the structure constants are independent of $n$ and $\tau$. We find
\be
\label{eq:constants.phi}
\kappa^{\phi\phi}_{\arcdownin} = \frac{\pi\, G^2(\frac12)}{2^{1/4}}, \qquad 
\kappa^{\phi\phi\omega}_{\arcdownout} = -  \frac{\big(2 \cos(\frac \phi 2)\big)^{1/2}\pi\, G^2(\frac12)}{2^{1/4}}, \qquad 
\kappa^{\phi\phi}_{\arcdownout} = \frac{2 \cos(\frac \phi 2) G^2(\frac12)}{2^{1/4}}(- \gamma+\log 2),
\ee
where we used \eqref{eq:lambda.and.ratio} for the ratio $\frac{\kappa^{\omega\psi\psi}}{\kappa^{\psi\psi}}$.

We have therefore computed the structure constants for the fields $\phi_{\halfarcright}(z)$ and $\phi_{\halfarcleft}(z)$ in their OPEs \eqref{eq:OPEs.phi}. We see that the fusion of these fields is non-abelian. We expect that this is a general feature of logarithmic CFT. If the fusion product of two fields involves a logarithmic field, then interchanging the order of the two fields induces a non-trivial change of the value of the correlator. 
The (weak) non-locality of the two fields endows them with a non-trivial monodromy, which can for instance be compared with that of disorder operators in the Ising model or in more general parafermionic field theories \cite{ZF86}. The unusual distinction made here between defect insertions that mark the start and end of boundary arcs goes beyond the introduction of a mere phase.
We note that the same behaviour was observed for the abelian sandpile model in \cite{PR04}, wherein two fields $\mu_{D,N}(z)$ and $\mu_{N,D}(z)$ that mark the transition between dissipative and non-dissipative sites on the boundary satisfy different fusion rules according to the order in which they are fused. 

Finally, we note that, similarly to the value of $\lambda$ in \eqref{eq:lambda.and.ratio}, the structure constants $\kappa^{\phi\phi\omega}_{\arcdownout}$ and $\kappa^{\phi\phi}_{\arcdownout}$ depend on $\alpha$. This is somewhat unexpected, as one might expect the boundary fields $\phi_{\halfarcleft}(z)$, $\phi_{\halfarcright}(z)$ and $\omega_{\arcdownout}(z)$ not to depend on $\alpha$. We believe that the resolution to this conundrum lies in the non-local nature of the field $\psi_\alpha(z)$. Its action on the lattice is indeed non-local, as it modifies the weight of all loops that encircle its insertion point, independently of how distant these are. We believe that, in the conformal interpretation, the non-locality of this field means that it modifies some conformal properties of the other fields of the theory, namely their structure constants and transformation laws, even if these are inserted a large distance away.

\subsection[Analysis for the fields $\mu$]{Analysis for the fields $\boldsymbol{\mu(z)}$}

The results of \cref{sec:Kac.correlators} are derived with the assumption that $\mu_{\twohalfarcright}(z)$ is a field that belongs to a Kac module, whereas $\mu_{\twohalfarcleft}(z)$ belongs to a staggered module. (We recall that these identifications could equivalently have been made the other way around). Here, we claim that the OPEs for these fields are of the form
\begin{subequations}
\label{eq:OPEs.mu}
\begin{alignat}{2}
\mu_{\twohalfarcright}(z_1)\mu_{\twohalfarcleft}(z_2) &=  \kappa^{\mu\mu}_{\arcsdowntwo}\, \varphi(z_2) +\dots\ ,\\[0.2cm] 
\mu_{\twohalfarcleft}(z_1)\mu_{\twohalfarcright}(z_2) &= \kappa^{\mu\mu\tilde\omega}_{\arcsdownsix} \big(\tilde\omega_{\arcsdownsix}(z_2) + \lambda_2\, \varphi(z_2) \log z_{21}\big)  + \kappa^{\mu\mu}_{\arcsdownsix}\, \varphi(z_2) +\dots\ .
\end{alignat}
\end{subequations}
We believe that the field $\tilde\omega_{\arcsdownsix}(z)$ is a logarithmic field with weight $\Delta = 0$ in a rank-two Jordan cell. It is potentially different from the field $\omega_{\arcdownout}(z)$, although the evidence below cannot rule out that they are in fact identical. Similarly to $\mu_{\twohalfarcleft}(z)$, $\tilde\omega_{\arcsdownsix}(z)$ transforms conformally according to \eqref{eq:trans.laws}, with the constant $\lambda $ replaced by $\lambda_2$. Its three-point correlator with two copies of the field $\psi_\alpha(z)$ is
\be
\langle \tilde\omega_{\arcsdownsix}(z_1)\psi_\alpha(z_2)\psi_\alpha(z_3)\rangle_{\mathbb C} = \frac{\kappa^{\tilde\omega \psi \psi} - \lambda_2 \kappa^{\psi \psi}\log (\frac{z_{21}z_{31}}{z_{32}})}{z_{32}^{2\Delta}}
\ee
where $\kappa^{\tilde\omega \psi \psi}$ is a constant.

In the regime $z_1 \to z_2$, we have the following four-point functions:
\begin{subequations}
\begin{alignat}{2}
 \langle \mu_{\twohalfarcright}(z_1)\mu_{\twohalfarcleft}(z_2) \psi_\alpha(z_3)\psi_\alpha(z_4)\rangle_{\mathbb C} &\xrightarrow{z_1 \to z_2}  \frac{1}{z_{43}^{2\Delta}} \kappa^{\mu\mu}_{\arcsdowntwo}\kappa^{\psi\psi},
\\[0.2cm]
 \langle \mu_{\twohalfarcleft}(z_1)\mu_{\twohalfarcright}(z_2) \psi_\alpha(z_3)\psi_\alpha(z_4)\rangle_{\mathbb C} &\xrightarrow{z_1 \to z_2}
  \frac{1}{z_{43}^{2\Delta}} \bigg[\kappa^{\mu\mu\tilde\omega}_{\arcsdownsix}\kappa^{\tilde\omega\psi\psi} +\lambda_2\, \kappa^{\mu\mu\tilde\omega}_{\arcsdownsix}\kappa^{\psi\psi}  \log \Big(\frac{z_{21}z_{43}}{z_{32}z_{42}} \Big) 
+ \kappa^{\mu\mu}_{\arcsdownsix}\kappa^{\psi\psi} 
 \bigg].
\end{alignat}
\end{subequations}
We map these correlators on the cylinder, divide by $\langle \psi_\alpha(x_3)\psi_\alpha(x_4)\rangle_{\mathbb V}$, take the limit $x_3 \to \ir \infty$ with $x_4 = x_3^*$, and obtain
\begin{subequations}
\label{eq:CtwoCsix}
\begin{alignat}{2}
C_{\arcsdowntwo}(x_{21}) &\xrightarrow{\tau \to 0^+} \kappa^{\mu\mu}_{\arcsdownsix} -\lambda_2\frac{\kappa_1^{\mu\psi\psi}}{\kappa^{\psi\psi}}  \log(n \tau),
\\[0.2cm]
C_{\arcsdownsix}(x_{21}) &\xrightarrow{\tau \to 0^+} \kappa^{\mu\mu\tilde\omega}_{\arcsdownsix}\frac{\kappa^{\tilde\omega\psi\psi}}{\kappa^{\psi\psi}} +\lambda_2\, \kappa^{\mu\mu\tilde\omega}_{\arcsdownsix}\,  \log (-2 \pi \ir \tau) + \kappa^{\mu\mu}_{\arcsdownsix}
- \lambda_2\frac{\kappa_1^{\mu\psi\psi}}{\kappa^{\psi\psi}}  \log(n \tau).
\end{alignat}
\end{subequations}
In each of these expressions, the last term originates from the logarithmic transformation law for the field $\mu_{\twohalfarcleft}(z)$, and $\kappa_1^{\mu\psi\psi}$ is the constant that appears in the correlator $\langle \mu_{\twohalfarcright}(z_1) \psi_\alpha(z_2)\psi_\alpha(z_3)\rangle_{\mathbb C}$.
Comparing these expressions with \eqref{eq:finalasy2} and \eqref{eq:finalasy6}, we equate the different terms, assuming that the structure constants are independent of $n$ and $\tau$. We find
\begin{subequations}
\begin{alignat}{2}
&\kappa^{\mu\mu}_{\arcsdowntwo} = \frac{2}\pi(\log 2+ \gamma), \qquad \lambda_2\, \kappa^{\mu\mu \tilde\omega}_{\arcsdownsix} = - \frac{\big(2 \cos(\frac \phi 2)\big)^2}\pi,
\\[0.15cm]
& \kappa^{\mu\mu\tilde\omega}_{\arcsdownsix}\frac{\kappa^{\tilde\omega\psi\psi}}{\kappa^{\psi\psi}} + \kappa^{\mu\mu}_{\arcsdownsix}
 = \frac2\pi\Big(\gamma + \log 2 + 2 \cos^2(\tfrac \phi 2) (\log \pi - \tfrac{\ir \pi}2)\Big) + \widehat K,
\label{eq:some.constraint}
\end{alignat}
\end{subequations}
where we used \eqref{eq:lambda2.relation}.

The lattice derivations of \cref{sec:four.entry.points} are therefore insufficient to fix individually all the structure constants appearing in \eqref{eq:OPEs.mu}. In comparison, all the constants for the OPEs of the fields $\phi_{\halfarcright}(z)$ and $\phi_{\halfarcleft}(z)$ were obtained in \eqref{eq:constants.phi}. This was possible because the four-point correlator involving the field $\omega_{\arcdownout}(z)$ was computed from the lattice in \cref{sec:four.entry.points}. Presumably, by computing
\be
\lim_{x_3 \to \ir \infty}\frac{\langle\tilde\omega_{\arcsdownsix}(x_1)\tilde\omega_{\arcsdownsix}(x_2)\psi_{\alpha}(x_3,x_3^*)\rangle_{\mathbb V}}{\langle\psi_{\alpha}(x_3,x_3^*)\rangle_{\mathbb V}}
\ee 
from the lattice, one would be able to compute $\lambda_2$ and $\kappa^{\tilde\omega\psi\psi}/\kappa^{\psi\psi}$ directly and then solve for the remaining unknowns, namely $\kappa^{\mu\mu \tilde\omega}_{\arcsdownsix}$, $\kappa^{\mu\mu}_{\arcsdownsix}$ and $\kappa_1^{\mu\psi\psi}/\kappa^{\psi\psi}$. \medskip

Thus, in the present state of affairs, these results do not allow us to determine whether $\omega_{\arcdownout}(z)$ and $\tilde\omega_{\arcsdownsix}(z)$ are the same field. If they turn out not to be, one may have to consider the possibility that the LCFT for the model of critical dense polymers involves an infinite number of boundary logarithmic fields of conformal dimension $\Delta = 0$, one for each link state of the form 
$
\psset{unit=0.4cm}
\begin{pspicture}[shift=-0.09cm](0,-0.5)(0.8,0)
\psline[linewidth=0.5pt]{-}(0,0)(0.8,0)
\psarc[linecolor=blue,linewidth=\moyen]{-}(0,0){0.2}{-90}{0}
\psarc[linecolor=blue,linewidth=\moyen]{-}(0.8,0){0.2}{180}{-90}
\end{pspicture}
$
, $\psset{unit=0.4cm}
\begin{pspicture}[shift=-0.09cm](0,-0.5)(1.6,0)
\psline[linewidth=0.5pt]{-}(0,0)(1.6,0)
\psarc[linecolor=blue,linewidth=\moyen]{-}(0,0){0.2}{-90}{0}
\psarc[linecolor=blue,linewidth=\moyen]{-}(1.6,0){0.2}{180}{-90}
\psbezier[linecolor=blue,linewidth=\moyen]{-}(0.6,0)(0.6,-0.5)(0,-0.55)(-0.02,-0.55)
\psbezier[linecolor=blue,linewidth=\moyen]{-}(1,0)(1,-0.5)(1.6,-0.55)(1.62,-0.55)
\psframe[fillstyle=solid,linecolor=white,linewidth=0pt](1.6,-0.6)(1.66,0.02)
\psframe[fillstyle=solid,linecolor=white,linewidth=0pt](0.0,-0.6)(-0.06,0.02)
\end{pspicture}
$\,, 
$
\psset{unit=0.4cm}
\begin{pspicture}[shift=-0.09cm](0,-0.5)(2.4,0)
\psline[linewidth=0.5pt]{-}(0,0)(2.4,0)
\psarc[linecolor=blue,linewidth=\moyen]{-}(0,0){0.2}{-90}{0}
\psarc[linecolor=blue,linewidth=\moyen]{-}(2.4,0){0.2}{180}{-90}
\psbezier[linecolor=blue,linewidth=\moyen]{-}(0.6,0)(0.6,-0.5)(0,-0.55)(-0.02,-0.55)
\psbezier[linecolor=blue,linewidth=\moyen]{-}(1.0,0)(1.0,-0.85)(0,-0.9)(-0.02,-0.9)
\psbezier[linecolor=blue,linewidth=\moyen]{-}(1.4,0)(1.4,-0.85)(2.4,-0.9)(2.42,-0.9)
\psbezier[linecolor=blue,linewidth=\moyen]{-}(1.8,0)(1.8,-0.5)(2.4,-0.55)(2.42,-0.55)
\psframe[fillstyle=solid,linecolor=white,linewidth=0pt](2.4,-0.95)(2.46,0.02)
\psframe[fillstyle=solid,linecolor=white,linewidth=0pt](0.0,-0.95)(-0.06,0.02)
\end{pspicture}
$\,,
etc.

\subsection[Analysis for the field $\omega$]{Analysis for the fields $\boldsymbol{\omega(z)}$}

Following the notation of \cite{PR04}, we write the OPE of the field $\omega_{\arcdownout}(z)$ with itself as
\be
\omega_{\arcdownout}(z_1)\omega_{\arcdownout}(z_2) = (a-2\lambda \log z_{21}) \omega_{\arcdownout}(z_2) + (b + a \lambda \log z_{21} - \lambda^2 \log^2 z_{21}) \varphi(z_2) + \dots\ 
\ee
where $a$ and $b$ are constants. The two-point function is then given by \eqref{eq:2pt.fcts.omega.id} with $\kappa^{\omega\omega} = a \kappa^\omega$. We use this OPE to obtain the four-point function involving two copies of $\omega_{\arcdownout}(z)$ and two copies of $\psi_\alpha(z)$, in the regime $z_1 \to z_2$:
\begin{alignat}{2}
\langle \omega_{\arcdownout}(z_1)\omega_{\arcdownout}(z_2) \psi_\alpha(z_3)\psi_\alpha(z_4)\rangle_{\mathbb C} \xrightarrow{z_1 \to z_2}
  \frac{1}{z_{43}^{2\Delta}} \bigg[
 &(a-2\lambda \log z_{21})\Big(\kappa^{\omega\psi\psi}-\lambda\, \kappa^{\psi\psi}\log\big(\tfrac{z_{32}z_{42}}{z_{43}}\big)\Big) 
   \nonumber\\[0.15cm]
   & + (b+a \lambda \log z_{21} - \lambda^2 \log^2 z_{21}) \kappa^{\psi\psi}  \bigg].
\end{alignat}
We obtain the same correlation function on $\mathbb V$ using the conformal mapping. We divide by $\langle \psi_\alpha(x_3) \psi_\alpha(x_4)\rangle_{\mathbb V}$, set $x_4 = x_3^*$, send $x_3 \to \ir \infty$ and find
\be
C_{\arcsdownfour}(x_{21}) \xrightarrow{\tau \to 0^+} \big(a-2\lambda \log(n \tau)\big) \Big(\frac{\kappa^{\omega\psi\psi}}{\kappa^{\psi\psi}}+ \lambda \log(-2\pi \ir \tau)\Big) + b + \lambda^2\log^2(n \tau).
\ee
We compare this result with \eqref{eq:Cf.asy.0+} and equate separately the terms in $\log^2 n$, $\log^2 \tau$, $\log n$, $\log \tau$ and the constant term. This confirms the values obtained in \eqref{eq:lambda.and.ratio} and yields
\be
a = -\frac{2}\pi \big(2 \cos(\tfrac \phi 2)\big)^{1/2} (\gamma + \log 2), \qquad b = -\frac{2 \cos (\frac \phi 2)} {\pi^2} (\gamma + \log 2)^2 = - \frac{a^2}4.
\ee
Interestingly, comparing with the results of \cite{PR04}, we see that the values of $a$ and $b$ are different, but that the relation $b = - \frac{a^2}4$ is the same, leading us to wonder whether this is a universal feature of logarithmic CFTs at $c=-2$.

\section{Discussion and conclusion}\label{sec:conclusion}

In this paper, we computed correlation functions for the model of critical dense polymers on a semi-infinite cylinder. These were obtained from an exact lattice derivation using the XX spin-chain representation of the enlarged periodic Temperley-Lieb algebra. The asymptotic behaviour as the system size $n$ grows to infinity was obtained in terms of integral formulae involving the scaling parameter $\tau = \frac{x-1}n \in (0,1)$. For small $\tau$, the leading behaviour of these correlators are proportional to $\tau^{1/4}$, $\tau^{1/4}\log \tau$, $\log \tau$ and $\log^2 \tau$. This logarithmic dependence of the correlation functions upon the position of the fields is the defining feature of logarithmic conformal field theories. This behaviour involving the square of the logarithm of the distance is rather uncommon, but has previously been observed in the Potts model for certain Fortuin-Kasteleyn probabilities \cite{VJ14}.

In the conformal interpretation, these lattice observables were understood to be ratios of conformal correlation functions: a four-point function divided by a two-point function. Using conformal invariance, we derived differential equations for the four-point functions and solved these equations. We found a perfect agreement with the lattice results, with the logarithmic behaviour arising in two ways: (i) from degenerate solutions to the hypergeometric differential equation, and (ii) from the logarithmic generalisation of the differential equations for highest-weight fields in rank-two Jordan cells that have null descendants.

Admittedly, our conformal derivation is really a hybrid approach, as the constants that arose in the solutions of the differential equations were fixed using the asymptotics of the lattice results. It should in fact be possible to fix these constants directly using the conformal bootstrap \cite{BPZ84,DF84}, and thus to avoid any input coming from the lattice. It would certainly be interesting to understand how this method applies to the case at hand.

We also used the lattice results to compute the structure constants that appear in the operator product expansions of the boundary fields of the model. An intriguing feature that we found pertains to the role played by the field $\psi_\alpha(z)$. Indeed, the fields $\phi_{\halfarcright}(z)$, $\phi_{\halfarcleft}(z)$, $\mu_{\twohalfarcright}(z)$, $\mu_{\twohalfarcleft}(z)$ and $\omega_{\arcdownout}(z)$ are boundary fields that are expected not to depend on $\alpha$, but the calculations of \cref{sec:OPE} show that the structure constants appearing in their OPEs do in fact depend on $\alpha$. These are of the form $\alpha^\iota$ times constants, with $\iota \in \frac12 \mathbb Z$. We interpret this unexpected dependence on $\alpha$ as a consequence of the non-locality of the field $\psi_\alpha(z)$, which appears to modify the conformal behaviour of the other fields that are present in the theory. It is currently uncertain whether this feature extends to other values of $\beta$, and in particular if it occurs only for values of $\beta$ where the boundary CFT has non-trivial indecomposable modules and Jordan cells. There was also another logarithmic field that appeared in the OPEs: $\omega_{\arcsdownsix}(z)$. Our calculation of the structure constants did not allow us to determine whether it coincides with $\omega_{\arcdownout}(z)$, and left open the possibility that the boundary CFT for critical dense polymers has an infinite number of such logarithmic fields with dimension $\Delta = 0$.
(It certainly does have an infinity of rank-two Jordan cells, and some of the corresponding logarithmic couplings have been computed in \cite{VJS11}.)

We believe there is more to be learned about the field $\psi_\alpha(z)$. Notably, our lattice results have only allowed us to put constraints on ratios of structure constants involving the field $\psi_\alpha(z)$, see \eqref{eq:lambda.and.ratio} and \eqref{eq:some.constraint}. This can be traced back to the fact that, in the ratios of conformal correlation functions that we considered, the field $\psi_\alpha(z)$ appears in both the numerator and denominator. For $\beta = 0$, we believe that the fusion of this field with itself is of the form $\psi_{\alpha} \times \psi_{\alpha} = \varphi + \bar \omega + \dots$ where $\bar \omega$ is a logarithmic bulk field with dimension $\Delta =  0$. Indeed, the two-point function $\langle\psi_{\alpha}(z_1) \psi_{\alpha}(z_2)\rangle_{\mathbb C}$ is non-vanishing, and because $\langle \varphi(z)\rangle_{\mathbb C} = 0$, the fusion of $\psi_{\alpha}$ with itself must include a second field with a vanishing conformal dimension. There is, in fact, a simple geometrical interpretation for the fields $\psi_{\alpha}(z)$ and $\bar \omega(z)$. When two fields $\psi_{\alpha}(z_1)$ and $\psi_{\alpha}(z_2)$ are inserted, the loops that encircle only $z_1$ or only $z_2$ have a fugacity~$\alpha$, whereas the loops that encircle neither or both have a fugacity $\beta$. On the lattice, one can bring $z_1$ close to $z_2$ until these points are one lattice spacing apart. In this case, there is a unique loop that passes between the two insertion points, and for $\beta = 0$, this is the only loop in the configuration. This loop is space-filling and its fugacity $\alpha$ appears as an overall prefactor. This is consistent with a field $\bar \omega$ whose conformal dimension is independent of $\alpha$.

There also remains some light to be shed upon the fields $\mu_{\twohalfarcright}(z)$ and $\mu_{\twohalfarcleft}(z)$. Although the consistency of the lattice and conformal results is quite convincing, we struggle to find a proper justification for treating these two fields in an asymmetric way, with one of them belonging to a Kac module and the other one to a staggered module. Moreover, the structure constant $\kappa^{\mu\psi\psi}_1$ that appears in \eqref{eq:CtwoCsix} belongs to a three-point function $\langle\mu_{\twohalfarcright}(z_1)\psi_\alpha(z_2)\psi_\alpha(z_3) \rangle_{\mathbb C}$ for which we have no lattice interpretation. One possibility is that our identification of $\mu_{\twohalfarcright}(z)$ and $\mu_{\twohalfarcleft}(z)$ as primary fields is incorrect, and that the correct identification should be made in terms of specific linear combinations of link states inserted in the four marked nodes, instead of only $\arcsdowntwo$. One attempt at resolving this issue would be to compute the five-point correlator $\langle\mu_{\twohalfarcright}(z_1)\phi_{\halfarcleft}(z_2)\phi_{\halfarcleft}(z_3)\psi_\alpha(z_4)\psi_\alpha(z_5) \rangle_{\mathbb C}$ from the lattice and investigate its conformal behaviour.

It is by now well-known that the six-vertex model at $q=\ir$ and the model of critical dense polymers are intimately related. The model of critical dense polymers has the central charge $c=-2$, whereas the six-vertex model has the central charge $c=1$.  On the torus, the two models share the same modular invariant partition function \cite{dFSZ87,MDPR13}, and many questions that arise in one model have a natural interpretation in the other. For instance, assigning a fugacity $\alpha$ to non-contractible loops in the loop model corresponds to inserting the field $\psi_\alpha(z,z^*)$ at $\ir \infty$. In the vertex model, this corresponds to inserting two electric operators of opposite charges, one at $\ir \infty$ and another one on the boundary of the cylinder, with a twist line connecting the two. In the related Coulomb gas formalism \cite{N84,dFSZ87}, these electric operators correspond to vertex operators. Their scaling dimensions are to be measured relative to the corresponding groundstate in which non-contractible and contractible loops get the same weight, namely $\alpha = \beta = 0$, and this results in the value \eqref{eq:Delta.solution} for $\Delta$. Thus, the insertion of the vertex operators in the six-vertex model changes the central charge from $c=1$ to $c=-2$ and shifts all the scaling dimensions. The indecomposable structures of the representations, which eventually lead to the logarithmic behaviour of the correlation functions of critical dense polymers, are key features of the conformal description with $c=-2$.

The boundary fields that we studied also have an elegant geometric interpretation in terms of the $Q$-state Potts model and its Fortuin-Kasteleyn high-temperature expansion. The loop fugacity is related to the number of states in the Potts model via the relation $\beta = \sqrt Q$. The Potts spins occupy one half of the lattice (for instance the odd sublattice) and the closed loops draw the contours of the Fortuin-Kasteleyn clusters \cite{BKW76}. On the boundary, free boundary conditions for the Potts spins correspond to having a segment of the boundary decorated with simple half-arcs,
$
\psset{unit=0.6cm}
\begin{pspicture}[shift=-0.2cm](0,-0.5)(3.2,0)
\psarc[fillstyle=solid,fillcolor=red,linecolor=white](0.4,0){0.1}{0}{360}
\psarc[fillstyle=solid,fillcolor=red,linecolor=white](1.2,0){0.1}{0}{360}
\psarc[fillstyle=solid,fillcolor=red,linecolor=white](2.8,0){0.1}{0}{360}
\psline[linewidth=0.5pt]{-}(0,0)(3.2,0)
\psarc[linecolor=blue,linewidth=\moyen]{-}(0.4,0){0.2}{180}{0}
\psarc[linecolor=blue,linewidth=\moyen]{-}(1.2,0){0.2}{180}{0}
\rput(2.0,-0.15){...}
\psarc[linecolor=blue,linewidth=\moyen]{-}(2.8,0){0.2}{180}{0}
\end{pspicture}
$\,, where the red circles mark the positions of the Potts spins. One can also choose a boundary where the Potts spins are fixed to one of the $Q$ values. The Potts model has $S_Q$ symmetry, so in computing the partition function on a domain with this boundary, one can equivalently impose that the boundary spins take the same value, which can be any of the $Q$ spin values, and then divide by $Q$. In the loop model, the corresponding boundary condition also consists of simple arcs, but with the spins shifted with respect to the half-arcs:
$
\psset{unit=0.6cm}
\begin{pspicture}[shift=-0.2cm](-0.2,-0.5)(3.4,0)
\psarc[fillstyle=solid,fillcolor=red,linecolor=white](0.0,0){0.1}{0}{360}
\psarc[fillstyle=solid,fillcolor=red,linecolor=white](0.8,0){0.1}{0}{360}
\psarc[fillstyle=solid,fillcolor=red,linecolor=white](1.6,0){0.1}{0}{360}
\psarc[fillstyle=solid,fillcolor=red,linecolor=white](2.4,0){0.1}{0}{360}
\psarc[fillstyle=solid,fillcolor=red,linecolor=white](3.2,0){0.1}{0}{360}
\psline[linewidth=0.5pt]{-}(-0.2,0)(3.4,0)
\psarc[linecolor=blue,linewidth=\moyen]{-}(0.4,0){0.2}{180}{0}
\psarc[linecolor=blue,linewidth=\moyen]{-}(1.2,0){0.2}{180}{0}
\rput(2.0,-0.15){...}
\psarc[linecolor=blue,linewidth=\moyen]{-}(2.8,0){0.2}{180}{0}
\end{pspicture}
$\,. This is usually referred to as {\it wired boundary conditions}.
As a result, the fields $\phi_{\halfarcright}(z)$ and $\phi_{\halfarcleft}(z)$ mark transitions between free and wired (or free and fixed, up to a factor of $Q$) boundary conditions for the Potts spins. Such boundary operators that change the boundary condition from free to wired were previously studied in \cite{C92,JS08}, and found to have the conformal weight~$\Delta_{1,2}$.  Likewise, $\omega_{\arcdownout}(z)$ marks a transition between two adjacent large connected clusters of boundary Potts spins. Finally, the fields $\mu_{\twohalfarcright}(z)$ and $\mu_{\twohalfarcleft}(z)$ correspond to having two spins on the boundary that are in the same cluster, but where all the other adjacent spins are free. Their conformal weight is $\Delta_{1,3}$. These five boundary operators thus appear to be refined versions of the similar operators that insert $d$ adjacent straight defects in the boundary (for $d=1,2$), and whose conformal weights are known \cite{SB89} to equal $\Delta_{1,d+1}$. 
The logarithmic nature of the fields $\mu$ and $\omega$ is special to the value $\beta = \sqrt Q = 0$, as in this case there is a {\it collision} \cite{C99} of the values of the conformal weights: $\Delta_{1,3} = \Delta_{1,1}$.

To conclude, we expect that it will be possible to generalise our results to other values of $\beta$ using the algebraic Bethe ansatz, and similarly to cases where the boundary condition at the edge of the cylinder has blobbed half-arcs \cite{MS94,MW00}. We believe that both cases are accessible using the methods of logarithmic conformal field theory. For other values of $\beta$ of the form $2 \cos\big(\frac{\pi(p'-p)}{p'}\big)$ with $p,p'$ coprime integers, the nature of the logarithmic fields will differ compared to what occurs for $\beta = 0$, as the coincidences of the conformal weights are different. The blobbed case is in particular interesting because it gives access to weights $\Delta_{r,r \pm s}$ for $r\in \mathbb R$ \cite{JS08}, thus offering more possibilities for the formation of indecomposable structures.

\subsection*{Acknowledgments}

AMD is an FNRS Postdoctoral Researcher under the project CR28075116. He acknowledges support from the EOS-contract O013018F. JLJ is grateful for support from the European Research Council under the Advanced Grant NuQFT. The authors thank the staff and members of the School of Mathematics and Statistics at the University of Melbourne for their kind hospitality during a simultaneous visit in 2018. The authors thank Yacine Ikhlef, Philippe Ruelle and David Ridout for helpful discussions, and Gilles Parez for proofreading the manuscript.

%
%

%
%
%



\begin{thebibliography}{10}

\bibitem{O44}
L.~Onsager.
\newblock {Crystal statistics. I. A two-dimensional model with an
  order-disorder transition}.
\newblock {\em Phys.~Rev.}, 65:117, 1944.

\bibitem{WK74}
K.W. Wilson and J.~Kogut.
\newblock {The renormalization group and the $\epsilon$ expansion}.
\newblock {\em Phys.~Rep.}, 12:75--199, 1974.

\bibitem{BPZ84}
A.A. Belavin, A.M. Polyakov, and A.B. Zamolodchikov.
\newblock {Infinite conformal symmetry in two-dimensional quantum field
  theory}.
\newblock {\em Nucl.~Phys.~B}, 241:333--380, 1984.

\bibitem{dFMS97}
P.~Di Francesco, P.~Mathieu, and D.~S\'en\'echal.
\newblock {\em {Conformal Field Theory}}.
\newblock Springer, 1997.

\bibitem{FST79}
L.D. Faddeev, E.K. Sklyanin, and L.A. Takhtajan.
\newblock {The quantum inverse problem method}.
\newblock {\em Teor.~Mat.~Fiz.}, 40:194--220, 1979.

\bibitem{B82}
R.J. Baxter.
\newblock {\em {Exactly solved models in statistical mechanics}}.
\newblock Academic Press, 1982.

\bibitem{KBI93}
V.E. Korepin, N.M. Bogoliubov, and A.G. Izergin.
\newblock {\em {Quantum inverse scattering method and correlation functions}}.
\newblock Cambridge University Press, 1993.

\bibitem{FQS84}
D.~Friedan, Z.~Qiu, and S.~Shenker.
\newblock {Conformal invariance, unitarity, and critical exponents in two
  dimensions}.
\newblock {\em Phys.~Rev.~Lett.}, 52:1575--1578, 1984.

\bibitem{KRR88}
V.~Kac, A.~Raina, and N.~Rozhkovskaya.
\newblock {\em {Bombay} lectures on highest weight representations of infinite
  dimensional {Lie} algebras}, volume~29 of {\em Advanced Series in
  Mathematical Physics}.
\newblock World Scientific, Singapore, 1988.

\bibitem{IK11}
K.~Iohara and Y.~Koga.
\newblock {\em Representation theory of the {Virasoro} algebra}.
\newblock Springer Monographs in Mathematics. Springer-Verlag, London, 2011.

\bibitem{RS92}
L.~Rozansky and H.~Saleur.
\newblock {Quantum field theory for the multi-variable Alexander-Conway
  polynomial}.
\newblock {\em Nucl.~Phys.~B}, 376:461--509, 1992.

\bibitem{G93}
V.~Gurarie.
\newblock Logarithmic operators in conformal field theory.
\newblock {\em Nucl.~Phys.~B}, 410:535--549, 1993.
\newblock
  \href{http://arxiv.org/abs/hep-th/9303160}{\textsf{arXiv:hep-th/9303160}}.

\bibitem{C99}
J.L. Cardy.
\newblock {Logarithmic correlations in quenched random magnets and polymers},
  1999.
\newblock
  \href{https://arxiv.org/abs/cond-mat/9911024}{\textsf{arXiv:cond-mat/9911024}}.

\bibitem{GL02}
V.~Gurarie and A.W.W. Ludwig.
\newblock {Conformal algebras of two-dimensional disordered systems}.
\newblock {\em J.~Phys.~A:~Math.~Gen.}, 35:L377, 2002.
\newblock
  \href{https://arxiv.org/abs/cond-mat/9911392}{\textsf{arXiv:cond-mat/9911392}}.

\bibitem{DJS10}
J.~Dubail, J.L. Jacobsen, and H.~Saleur.
\newblock {Conformal field theory at central charge \mbox{$c=0$}: A measure of
  the indecomposability ($b$) parameters}.
\newblock {\em Nucl.~Phys.~B}, 834:399--422, 2010.
\newblock \href{https://arxiv.org/abs/1001.1151}{\textsf{arXiv:1001.1151
  [math-ph]}}.

\bibitem{VJS11}
R.~Vasseur, J.L. Jacobsen, and H.~Saleur.
\newblock {Indecomposability parameters in chiral logarithmic conformal field
  theory}.
\newblock {\em Nucl.~Phys.~B}, 851:314--345, 2011.
\newblock \href{https://arxiv.org/abs/1103.3134}{\textsf{arXiv:1103.3134
  [math-ph]}}.

\bibitem{RSA14}
D.~Ridout and Y.~Saint-Aubin.
\newblock {Standard modules, induction and the structure of the Temperley-Lieb
  algebra}.
\newblock {\em Adv.~Theor.~Math.~Phys.}, 18:957--1041, 2014.
\newblock \href{https://arxiv.org/abs/1204.4505}{\textsf{arXiv:1204.4505
  [math-ph]}}.

\bibitem{R96}
F.~Rohsiepe.
\newblock {On reducible but indecomposable representations of the Virasoro
  algebra}, 1996.
\newblock
  \href{https://arxiv.org/abs/hep-th/9611160}{\textsf{arXiv:hep-th/9611160}}.

\bibitem{KG96}
H.~Kausch and M.~Gaberdiel.
\newblock {Indecomposable fusion products}.
\newblock {\em Nucl.~Phys.~B}, 477:293--318, 1996.
\newblock
  \href{https://arxiv.org/abs/hep-th/9604026}{\textsf{arXiv:hep-th/9604026}}.

\bibitem{MR07}
P.~Mathieu and D.~Ridout.
\newblock {From percolation to logarithmic conformal field theory}.
\newblock {\em Phys.~Lett.~B}, 657:120--129, 2007.
\newblock \href{https://arxiv.org/abs/0708.0802}{\textsf{arXiv:0708.0802
  [hep-th]}}.

\bibitem{MR08}
P.~Mathieu and D.~Ridout.
\newblock {Logarithmic $M(2, p)$ minimal models, their logarithmic couplings,
  and duality}.
\newblock {\em Nucl.~Phys.~B}, 801:268--295, 2008.
\newblock \href{https://arxiv.org/abs/0711.3541}{\textsf{arXiv:0711.3541
  [hep-th]}}.

\bibitem{KR09}
K.~Kyt\"ol\"a and D.~Ridout.
\newblock {On staggered indecomposable Virasoro modules}.
\newblock {\em J.~Math.~Phys.}, 50:123503, 2009.
\newblock \href{https://arxiv.org/abs/0905.0108}{\textsf{arXiv:0905.0108
  [math-ph]}}.

\bibitem{GRR13}
A.~Gainutdinov, D.~Ridout, and I.~Runkel~(Guest Editors).
\newblock {Special issue on logarithmic conformal field theory}.
\newblock {\em J.~Phys.~A: Math.~Theor.}, 46 (49), 2013.

\bibitem{GJSV13}
A.M. Gainutdinov, J.L. Jacobsen, H.~Saleur, and R.~Vasseur.
\newblock {A physical approach to the classification of indecomposable Virasoro
  representations from the blob algebra}.
\newblock {\em Nucl.~Phys.~B}, 873:614--681, 2013.
\newblock \href{https://arxiv.org/abs/1212.0093}{\textsf{arXiv:1212.0093
  [hep-th]}}.

\bibitem{PR07}
P.A. Pearce and J.~Rasmussen.
\newblock Solvable critical dense polymers.
\newblock {\em J.~Stat.~Mech.}, 0702:P02015, 2007.
\newblock
  \href{http://arxiv.org/abs/hep-th/0610273}{\textsf{arXiv:hep-th/0610273}}.

\bibitem{PRZ06}
P.A. Pearce, J.~Rasmussen, and J.-B. Zuber.
\newblock {Logarithmic Minimal Models}.
\newblock {\em J.~Stat.~Mech.}, 0611:P11017, 2006.
\newblock
  \href{http://arxiv.org/abs/hep-th/0607232}{\textsf{arXiv:\mbox{hep-th}/0607232}}.

\bibitem{K95}
H.G. Kausch.
\newblock {Curiosities at $c = -2$}.
\newblock
  \href{https://arxiv.org/abs/hep-th/9510149}{\textsf{arXiv:hep-th/9510149}}.

\bibitem{GK99}
M.R. Gaberdiel and H.G. Kausch.
\newblock {A local logarithmic conformal field theory}.
\newblock {\em Nucl.~Phys.~B}, 538:631--658, 1999.
\newblock
  \href{https://arxiv.org/abs/hep-th/9807091}{\textsf{arXiv:hep-th/9807091}}.

\bibitem{K00}
H.G. Kausch.
\newblock {Symplectic fermions}.
\newblock {\em Nucl.~Phys.~B}, B583:513--541, 2000.
\newblock
  \href{https://arxiv.org/abs/hep-th/0003029}{\textsf{arXiv:hep-th/0003029}}.

\bibitem{KW01}
S.~Kawai and J.F. Wheater.
\newblock {Modular transformation and boundary states in logarithmic conformal
  field theory}.
\newblock {\em Phys.~Lett.~B}, 508:203--210, 2001.
\newblock
  \href{https://arxiv.org/abs/hep-th/0103197}{\textsf{arXiv:hep-th/0103197}}.

\bibitem{BF02}
A.~Bredthauer and M.~Flohr.
\newblock {Boundary states in $c = -2$ logarithmic conformal field theory}.
\newblock {\em Nucl.~Phys.~B}, 639:450--470, 2002.
\newblock
  \href{https://arxiv.org/abs/hep-th/0204154}{\textsf{arXiv:hep-th/0204154}}.

\bibitem{GL96}
J.J. Graham and G.I. Lehrer.
\newblock {Cellular Algebras}.
\newblock {\em Invent.~Math.}, 123:1--34, 1996.

\bibitem{PR04}
G.~Piroux and P.~Ruelle.
\newblock {Pre-logarithmic and logarithmic fields in a sandpile model}.
\newblock {\em J.~Stat.~Mech.}, page P10006, 2004.
\newblock
  \href{https://arxiv.org/abs/hep-th/0407143}{\textsf{arXiv:hep-th/0407143}}.

\bibitem{JPR06}
M.~Jeng, G.~Piroux, and P.~Ruelle.
\newblock {Height variables in the Abelian sandpile model: scaling fields and
  correlations}.
\newblock {\em J.~Stat.~Mech.}, page P10015, 2006.
\newblock
  \href{https://arxiv.org/abs/cond-mat/0609284}{\textsf{arXiv:cond-mat/0609284
  [cond-mat.stat-mech]}}.

\bibitem{R13}
P.~Ruelle.
\newblock {Logarithmic conformal invariance in the Abelian sandpile model}.
\newblock {\em J.~Phys.~A: Math.~Theor.}, 46:494014, 2013.
\newblock \href{https://arxiv.org/abs/1303.4310}{\textsf{arXiv:1303.4310
  [hep-th]}}.

\bibitem{I99}
E.V. Ivashkevich.
\newblock {Correlation functions of dense polymers and $c = -2$ conformal field
  theory}.
\newblock {\em J.~Phys.~A: Math.~Gen.}, 32:1691, 1999.
\newblock
  \href{https://arxiv.org/abs/cond-mat/9801183}{\textsf{arXiv:cond-mat/9801183
  [cond-mat.stat-mech]}}.

\bibitem{IH05}
E.V. Ivashkevich and C.K. Hu.
\newblock {Exact multileg correlation functions for the dense phase of
  branching polymers in two dimensions}.
\newblock {\em Phys.~Rev.~E}, 71:015104, 2005.

\bibitem{MDJ18}
A.~Morin-Duchesne and J.L. Jacobsen.
\newblock {Two-point boundary correlation functions of dense loop models}.
\newblock {\em SciPost~Phys.}, 4:034, 2018.
\newblock \href{https://arxiv.org/abs/1712.08657}{\textsf{arXiv:1712.08657
  [cond-mat.stat-mech]}}.

\bibitem{VJS12}
R.~Vasseur, J.L. Jacobsen, and H.~Saleur.
\newblock {Logarithmic observables in critical percolation}.
\newblock {\em J.~Stat.~Mech.}, page L07001, 2012.
\newblock \href{https://arxiv.org/abs/1206.2312}{\textsf{arXiv:1206.2312
  [cond-mat.stat-mech]}}.

\bibitem{TCDJ19}
X.~Tan, R.~Couvreur, Y.~Deng, and J.L. Jacobsen.
\newblock {Observation of non-scalar and logarithmic correlations in 2D and 3D
  percolation}.
\newblock {\em Phys.~Rev.~E}, 99:050103, 2019.
\newblock \href{https://arxiv.org/abs/1809.06650}{\textsf{arXiv:1809.06650
  [cond-mat.stat-mech]}}.

\bibitem{VJ14}
R.~Vasseur and J.L. Jacobsen.
\newblock {Operator content of the critical Potts model in $d$ dimensions and
  logarithmic correlations}.
\newblock {\em Nucl.~Phys.~B}, 880:435--475, 2014.
\newblock \href{https://arxiv.org/abs/1311.6143}{\textsf{arXiv:1311.6143
  [cond-mat.stat-mech]}}.

\bibitem{CJV17}
R.~Couvreur, J.L. Jacobsen, and R.~Vasseur.
\newblock {Non-scalar operators for the Potts model in arbitrary dimension}.
\newblock {\em J.~Phys.~A:~Math.~Theor.}, 50:474001, 2017.
\newblock \href{https://arxiv.org/abs/1704.02186}{\textsf{arXiv:1704.02186
  [cond-mat.stat-mech]}}.

\bibitem{F02}
M.~Flohr.
\newblock {Bits and pieces in logarithmic conformal field theory}.
\newblock {\em Int.~J.~Mod.~Phys.~A}, 18:4497--4591, 2003.
\newblock
  \href{https://arxiv.org/abs/hep-th/0111228}{\textsf{arXiv:hep-th/0111228}}.

\bibitem{C89}
J.L. Cardy.
\newblock {Boundary conditions, fusion rules and the Verlinde formula}.
\newblock {\em Nucl.~Phys.~B}, 324:581--596, 1989.

\bibitem{C04}
J.L. Cardy.
\newblock {Boundary conformal field theory}.
\newblock
  \href{https://arxiv.org/abs/hep-th/0411189}{\textsf{arXiv:hep-th/0411189}}.

\bibitem{DJS09}
J.~Dubail, J.L. Jacobsen, and H.~Saleur.
\newblock {Conformal two-boundary loop model on the annulus}.
\newblock {\em Nucl.~Phys.~B}, 813:430--459, 2009.
\newblock \href{https://arxiv.org/abs/0812.2746}{\textsf{arXiv:0812.2746
  [math-ph]}}.

\bibitem{GJP16}
J.~de~Gier, J.L. Jacobsen, and A.~Ponsaing.
\newblock {Finite-size corrections for universal boundary entropy in bond
  percolation}.
\newblock {\em SciPost~Phys.}, 1:012, 2016.
\newblock \href{https://arxiv.org/abs/1610.04006}{\textsf{arXiv:1610.04006
  [math-ph]}}.

\bibitem{GRS13a}
A.M. Gainutdinov, N.~Read, and H.~Saleur.
\newblock {Continuum limit and symmetries of the periodic $g\ell(1|1)$ spin
  chain}.
\newblock {\em Nucl.~Phys.~B}, 871:245--288, 2013.
\newblock \href{https://arxiv.org/abs/1112.3403}{\textsf{arXiv:1112.3403
  [hep-th]}}.

\bibitem{GRS13b}
A.M. Gainutdinov, N.~Read, and H.~Saleur.
\newblock {Bimodule structure in the periodic $g\ell(1|1)$ spin chain}.
\newblock {\em Nucl.~Phys.~B}, 871:289--329, 2013.
\newblock \href{https://arxiv.org/abs/1112.3407}{\textsf{arXiv:1112.3407
  [hep-th]}}.

\bibitem{MDSA13}
A.~Morin-Duchesne and Y.~Saint-Aubin.
\newblock {Jordan cells of periodic loop models}.
\newblock {\em J.~Phys.~A: Math.~Theor.}, 46:494013, 2013.
\newblock \href{https://arxiv.org/abs/1302.5483}{\textsf{arXiv:1302.5483
  [math-ph]}}.

\bibitem{GRSV15}
A.M. Gainutdinov, N.~Read, H.~Saleur, and R.~Vasseur.
\newblock {The periodic $s\ell(2|1)$ alternating spin chain and its continuum
  limit as a bulk logarithmic conformal field theory at $c = 0$}.
\newblock {\em J.~High~Energ.~Phys.}, page 114, 2015.
\newblock \href{https://arxiv.org/abs/1409.0167}{\textsf{arXiv:1409.0167
  [hep-th]}}.

\bibitem{L91}
D.~Levy.
\newblock {Algebraic structure of translation-invariant spin-$\frac12$ XXZ and
  $q$-Potts quantum chains}.
\newblock {\em Phys.~Rev.~Lett.}, 67:1971--1974, 91.

\bibitem{MS93}
P.~Martin and H.~Saleur.
\newblock {On an algebraic approach to higher dimensional statistical
  mechanics}.
\newblock {\em Commun.~Math.~Phys.}, 158:155--190, 1993.
\newblock
  \href{https://arxiv.org/abs/hep-th/9208061}{\textsf{arXiv:hep-th/9208061}}.

\bibitem{GL98}
J.J. Graham and G.I. Lehrer.
\newblock {The representation theory of affine Temperley-Lieb algebras}.
\newblock {\em Enseign.~Math.}, 44:173--218, 1998.

\bibitem{G98}
R.M. Green.
\newblock {On representations of affine Temperley-Lieb algebras}.
\newblock {\em CMS~Conf.~Proc.}, 24:245--261, 1998.

\bibitem{EG99}
K.~Erdmann and R.M. Green.
\newblock {On representations of affine Temperley-Lieb algebras, II}.
\newblock {\em Pac.~J.~Math.}, 191:243--274, 1999.
\newblock \href{https://arxiv.org/abs/math/9811017}{\textsf{arXiv:math/9811017
  [math.RT]}}.

\bibitem{PRV10}
P.A. Pearce, J.~Rasmussen, and S.P. Villani.
\newblock {Solvable critical dense polymers on the cylinder}.
\newblock {\em J.~Stat.~Mech.}, page P02010, 2010.
\newblock \href{https://arxiv.org/abs/0910.4444}{\textsf{arXiv:0910.4444
  [hep-th]}}.

\bibitem{MDPR13}
A.~Morin-Duchesne, P.A. Pearce, and J.~Rasmussen.
\newblock {Modular invariant partition function of critical dense polymers}.
\newblock {\em Nucl.~Phys.~B}, 874:312--357, 2013.
\newblock \href{https://arxiv.org/abs/1303.4895}{\textsf{arXiv:1303.4895
  [hep-th]}}.

\bibitem{OLBC10}
F.W.J. Olver, D.W. Lozier, R.F. Boisvert, and C.W. Clark.
\newblock {\em {NIST Handbook of Mathematical Functions}}.
\newblock Cambridge University Press, 2010.

\bibitem{N84}
B.~Nienhuis.
\newblock {Critical behavior of two-dimensional spin models and charge
  asymmetry in the Coulomb gas}.
\newblock {\em J.~Stat.~Phys.}, 34:731--761, 1984.

\bibitem{PS90}
V.~Pasquier and H.~Saleur.
\newblock {Common structures between finite systems and conformal field
  theories through quantum groups}.
\newblock {\em Nucl.~Phys.~B}, 330:523--556, 1990.

\bibitem{ZF86}
A.B. Zamolodchikov and V.A. Fateev.
\newblock {Disorder fields in two-dimensional conformal quantum field theory
  and $N=2$ extended supersymmetry}.
\newblock {\em Sov.~Phys.~JETP}, 63:913, 1986.

\bibitem{DF84}
V.S. Dotsenko and V.A. Fateev.
\newblock {Conformal algebra and multipoint correlation functions in $2d$
  statistical models}.
\newblock {\em Nucl.~Phys.~B}, 240:1984, 312--348.

\bibitem{dFSZ87}
P.~Di Francesco, H.~Saleur, and J.-B. Zuber.
\newblock {Relations between the Coulomb gas picture and conformal invariance
  of two-dimensional critical models}.
\newblock {\em J.~Stat.~Phys.}, 49:57--79, 1987.

\bibitem{BKW76}
R.J. Baxter, S.B. Kelland, and F.Y. Wu.
\newblock {Equivalence of the Potts model or Whitney polynomial with an
  ice-type model}.
\newblock {\em J.~Phys.~A:~Math.~Gen.}, 9:397--406, 1976.

\bibitem{C92}
J.L. Cardy.
\newblock {Critical percolation in finite geometries}.
\newblock {\em J.~Phys.~A}, 25:1992, L201.
\newblock
  \href{https://arxiv.org/abs/hep-th/9111026}{\textsf{arXiv:hep-th/9111026}}.

\bibitem{JS08}
J.L. Jacobsen and H.~Saleur.
\newblock {Conformal boundary loop models}.
\newblock {\em Nucl.~Phys.~B}, 788:2008, 137--166.
\newblock
  \href{https://arxiv.org/abs/math-ph/0611078}{\textsf{arXiv:math-ph/0611078}}.

\bibitem{SB89}
H.~Saleur and M.~Bauer.
\newblock {On some relations between local height probabilities and conformal
  invariance}.
\newblock {\em Nucl.~Phys.~B}, 320:591--624, 1989.

\bibitem{MS94}
P.~Martin and H.~Saleur.
\newblock {The blob algebra and the periodic Temperley-Lieb algebra}.
\newblock {\em Lett.~Math.~Phys.}, 30:189--206, 1994.
\newblock
  \href{https://arxiv.org/abs/hep-th/9302094}{\textsf{arXiv:hep-th/9302094}}.

\bibitem{MW00}
P.~Martin and D.~Woodcock.
\newblock {On the structure of the blob algebra}.
\newblock {\em J.~Alg.}, 225:957--988, 2000.

\end{thebibliography}

\end{document}